\documentclass[modern]{aastex61}

\received{...}
\revised{...}
\accepted{......}
\submitjournal{ApJ}

\shorttitle{{\it Spitzer} SNe}
\shortauthors{Szalai et al.}

\usepackage{graphicx}

\begin{document} 

 \title{A Comprehensive Analysis of {\it Spitzer} Supernovae}

\correspondingauthor{Tam\'as Szalai}
\email{szaszi@titan.physx.u-szeged.hu}

\author[0000-0003-4610-1117]{Tam\'as Szalai}
\affiliation{Department of Optics and Quantum Electronics, University of Szeged, H-6720 Szeged, D\'om t\'er 9., Hungary}   
\affiliation{Konkoly Observatory, MTA CSFK, Konkoly-Thege M. \'ut 15-17, Budapest, 1121, Hungary}
   
\author{Szanna Zs\'iros} 
\affiliation{Department of Optics and Quantum Electronics, University of Szeged, H-6720 Szeged, D\'om t\'er 9., Hungary} 

\author{Ori D. Fox}
\affiliation{Space Telescope Science Institute, 3700 San Martin Drive, Baltimore, MD 21218, USA}
	     
\author{Ond\v{r}ej Pejcha}   
\affiliation{Institute of Theoretical Physics, Faculty of Mathematics and Physics, Charles University in Prague, Czech Republic}
\affiliation{Lyman Spitzer Jr. Fellow, Department of Astrophysical Sciences, Princeton University, 4 Ivy Lane, Princeton, NJ 08540, USA}

\author{Tom\'as M\"uller}
\affiliation{Millennium Institute of Astrophysics, Santiago, Chile}
\affiliation{Instituto de Astrof\'isica, Pontificia Universidad Cat\'olica de Chile, Av. Vicu\~na Mackenna 4860, 782-0436 Macul, Santiago, Chile}
\affiliation{Department of Physics and Astronomy, University of Southampton, Southampton, Hampshire, SO17 1BJ, UK}
\affiliation{LSSTC Data Science Fellow}

\begin{abstract}

The mid-infrared (mid-IR) wavelength regime offers several advantages for following the late-time evolution of supernovae (SNe).  First, the peaks of the SN spectral energy distributions shift toward longer wavelengths following the photospheric phase.  Second, mid-IR observations suffer less from effects of interstellar extinction. Third, and perhaps most important, the mid-IR traces dust formation and circumstellar interaction at late-times ($>$100 days) after the radioactive ejecta component fades. The {\it Spitzer Space Telescope} has provided substantial mid-IR observations of SNe since its launch in 2003.  More than 200 SNe have been targeted, but there are even more SNe that have been observed serendipitously.  Here we present the results of a comprehensive study based on archival {\it Spitzer}/IRAC images of more than 1100 SN positions; from this sample, 119 SNe of various subclasses have been detected, including 45 SNe with previously unpublished mid-IR photometry. The photometry reveal significant amounts of warm dust in some cases. We perform an in-depth analysis to constrain the origin and heating mechanism of the dust, and present the resulting statistics.

\end{abstract}

\keywords{editorials, notices --- 
miscellaneous --- catalogs --- surveys}
   \keywords{supernovae: general --
                infrared: stars --
                circumstellar matter
               }

%

\section{Introduction}\label{intro}

Tracing the multi-wavelength evolution of supernovae (SNe) over many years, and even decades, can provide important clues about the shock physics, circumstellar environment, and dust production.  The current ground-based transient surveys ensure the optical follow-up of hundreds of SNe per year, but these observations are typically at early times during the photospheric phase.  Late-time optical spectra and/or non-optical observations are more rare because they require large apertures or space telescopes. 

The {\it Spitzer Space Telescope} (hereafter {\it Spitzer}) has been the primary source of mid-infrared (mid-IR) observations of many SNe.  Between 2003 and 2009, in the cryogenic (or Cold Mission) phase, only a moderate number ($<$50) of nearby SNe were targeted by {\it Spitzer}. Since 2009, even with post-cryo (Warm Mission) {\it Spitzer}, more than 150 more SNe have been targeted. Two surveys, in particular, contributed to this surge: a program aimed to observe a large sample of Type IIn SNe \citep[73 observed SN sites, 13 detected targets, see][]{Fox11,Fox13}, and SPitzer InfraRed Intensive Transients Survey (SPIRITS), a systematic mid-IR study of nearby galaxies \citep{Kasliwal17}. SPIRITS has resulted in the detection of 44 objects of various types of SNe \citep[observing 141 sites,][]{Tinyanont16}, three obscured SNe missed by previous optical surveys \citep{Jencson17,Jencson18}, and a large number of other variables and transients including ones with unusual infrared behavior \citep{Kasliwal17}.

These mid-IR observations have several advantages over optical observations, including increased sensitivity to the ejecta as it expands and cools, less impact by interstellar extinction, and coverage of atomic and molecular emission lines generated by shocked gas as it cools \citep[see e.g.][]{Reach06a}.  Most of the mid-IR observations are sensitive to warm dust in the SN environment.  The origin and heating mechanism of the dust, however, is not always obvious as the dust may be newly formed or pre-existing in the circumstellar medium (CSM).  Newly-condensed dust may form in either the ejecta or in a cool dense shell (CDS) produced by the interaction of the ejecta forward shock with a dense shell of CSM \citep[see e.g][]{Pozzo04,Mattila08,Smith09}. Pre-existing dust may be radiatively heated by the peak SN luminosity or by X-rays generated by late-time CSM interaction, thereby forming an IR echo \citep[see e.g.][]{Bode80,Dwek83,Graham86,Sugerman03,Kotak09}.  In this case, the dust is a useful probe of the CSM characteristics and the pre-SN mass loss from either the progenitor or companion star \citep[see e.g.][for a review]{Gall11}.

Based on theoretical expectations \citep[see e.g.][]{Kozasa09,Gall11}, Type II-P explosions are likely the best candidates for dust formation among SNe.  Some of these objects were targets of {\it Spitzer} observations in the early years of the mission. These data typically trace dust formation $\sim$1-3 yr after explosion and estimate the physical parameters of newly-formed dust. In addition to several detailed studies of single objects \citep[e.g.][]{Meikle06,Meikle07,Meikle11,Sugerman06,Kotak09,Andrews10,Fabbri11,Szalai11}, \citet{Szalai13} presented an analysis of twelve Type II-P SNe, yielding nine detections and three upper limits. The results do not support the theoretical prediction of significant($>>$0.001 $M_{\odot}$) dust production in SNe or the large dust masses observed in some old SN remnants and/or high-redshift galaxies. Several ways to reconcile this inconsistency include imperfections of grain condensation models, the probability of clumping dust formation, or significant grain growth in the interstellar matter (ISM) (see \citet{Gall11} for a review, as well as \citet{Szalai13}). Another possibility is that a significant amount of dust may be present in the form of very cold ($<$50 K) grains in the ejecta, but to date, far-IR and sub-mm observations have only been able to detect such dust in the very nearby case of SN 1987A \citep{Matsuura11,Matsuura15,Indebetouw14,Wesson15}.

Type IIn SNe exhibit signatures of interaction between the ejecta and dense CSM. This shock interaction may lead to either heating of pre-existing circumstellar grains or dust condensation in the CDS that can form between the forward and reverse shock. Papers on individual objects \citep[e.g.][]{Gerardy02,Fox10,Andrews11a,Gall14}, together with the comprehensive {\it Spitzer} study of SNe IIn mentioned above \citep{Fox11,Fox13} show how the mid-IR evolution can be used to the trace the mass-loss history of the progenitor in the years leading up to the SN. 

In contrast with the relatively large number of Type II-P and IIn SNe with published {\it Spitzer} data, there are fewer published mid-IR observations of thermonuclear explosions of C/O white dwarfs (Type Ia SNe) or stripped-envelope core-collapse SNe (SE CCSNe, including Type Ib/c, Ibn, and IIb ones).  Historically, these SN subclasses are less likely to form new dust due to their high ejecta velocities and less likely to have pre-existing, dense CSMs.  For example, \citet{Chomiuk16} and \citet{Maeda15a} use radio and near-IR observations, respectively, to place strict upper limits on the amount of material surrounding SNe Ia.

In recent years, however, many SNe within the SNe Ia and stripped-envelope subclasses have shown signs of a dense CSM and/or warm dust. For example, SNe Ia-CSM, which are thought to be thermonuclear explosions exploding in dense, H-rich shells of ambient CSM \citep[producing IIn-like emission features in their late-time spectra, see e.g.][]{Silverman13,Fox15,Inserra16} and are very bright in mid-IR, even 3-4 years after explosion \citep{FF13,Graham17}. The subluminous thermonuclear Type Iax SN~2014dt showed an excess of mid-IR emission (over the expected fluxes of more normal SNe Ia) at $\sim$1 yr after explosion \citep[][see also in Section \ref{res_ev}]{Fox16}, and an excess of near-IR emission was observed by circumstellar dust around the super-Chandrasekhar candidate SN~2012dn \citep{Yamanaka16,Nagao17}.

Some stripped-enveloped SNe show mid-IR emission at late-times, too. For example, the Type Ic SN 2014C showed a excess of mid-IR emission develop $\sim$1 year post-explosion \citep{Tinyanont16}, as did several SNe IIb, including SN 2013df \citep{Szalai16,Tinyanont16} and SN~2011dh \citep{Helou13}. The Type Ibn SN subclass, which shows narrow helium lines, do not typically show a late-time mid-IR excess (respect to the expected flux level originates from the cooling ejecta). However, the Type Ibn SN~2006jc was bright in early-time {\it Spitzer} images \citep{Mattila08}.

Despite the relatively high number of SNe with reported {\it Spitzer} observations, most of the analysis consists of single object papers. There have been some broader studies on SNe IIn \citep{Fox11,Fox13}, SNe IIP \citep{Szalai13}, and SNe Ia \citep{Johansson17}, and a SPIRITS summary by \citet{Tinyanont16}, which includes observations of $\sim$140 core-collapse SNe within 20 Mpc.

The motivation of the current work, however, is to provide a complete review of all SNe currently in the {\it Spitzer} archive to compare the mid-IR properties of different SN subclasses.  This paper includes mid-IR observations of more than 1100 SN positions, from which 119 objects have been detected.  Within this detected sample, many observations were previously unpublished and 45 targets were observed serendipitously during other science programs.

In Section \ref{obs}, we describe the steps of data collection and photometry of {\it Spitzer}/IRAC (Infrared Array Camera) data. We present our results in Section \ref{res}, including a statistical analysis of the mid-IR evolution of the different SN subclasses and simple models fits to the spectral energy distributions (SEDs). Finally, the conclusions of our study are presented in Section \ref{conc}.

\section{Observations and data analysis}\label{obs}

\subsection{Collection of supernova data from the {\it Spitzer} Heritage Archive}\label{obs_coll}

Using the list of SNe on the website of Central Bureau for Astronomical Telegrams (CBAT) \footnote{http://www.cbat.eps.harvard.edu/lists/Supernovae.html}, and the website of All-Sky Automated Survey for Supernovae \citep[ASAS-SN][]{Shappee14,Holoien17a,Holoien17b,Holoien17c}\footnote{http://www.astronomy.ohio-state.edu/$\sim$assassin}, we selected all SNe that were discovered before 2015 and have been spectroscopically classified. We also selected additional nearby ($z \lesssim$ 0.05) SNe listed in the Open Supernova Catalog\footnote{https://sne.space} \citep{Guillochon17}. This search returned $\sim$4500 objects, all of which have had their positions searched in the {\it Spitzer} Heritage Archive (SHA)\footnote{http://sha.ipac.caltech.edu} (using a 100$\arcsec$ environment for the queries). We found 1142 SN sites that have been observed post-explosion with {\it Spitzer}. For these SNe, we downloaded the available IRAC data for further analysis whether or not the data had been previously published. We note that although MIPS (Multiband Imaging Photometer) and IRS (Infrared Spectrograph) data can also contribute to the understanding of the mid-IR behavior of SNe \citep[see e.g.][]{Kotak06,Kotak09,Gerardy07,Fabbri11,Szalai11,Szalai13}, only a few objects observed with these instruments exist so we focus only on IRAC data in this article.

\subsection{Object identification and photometry on {\it Spitzer}/IRAC images}\label{obs_phot}

\begin{figure}
\centering
\includegraphics[width=10cm]{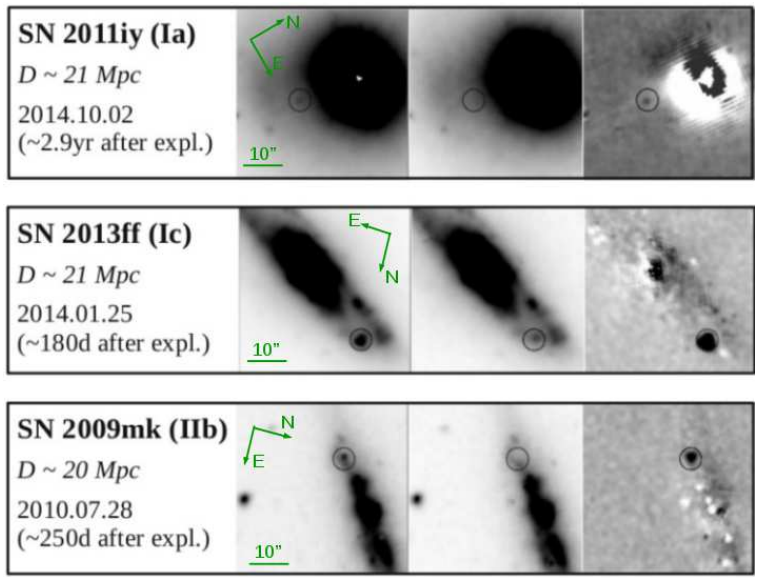}
\caption{HOTPANTS template subtraction of our {\it Spitzer} data.  For each SN, the three panels show (left) the most recent {\it Spitzer}/IRAC 4.5 $\mu$m image, (center) template, and (right) differenced image.
\label{fig:sne_new}}
\end{figure}

We collected and analyzed all available IRAC post-basic calibrated data (PBCD). The scale of these images is 0.6$\arcsec$/pixel. Identifying a point source at the position of an SN explosion can be difficult at the large distances to some of these galaxies, where compact \ion{H}{2} regions or the host clusters of SNe may also appear as point-like sources on {\it Spitzer}/IRAC images. Furthermore, the target can be faint or on top of a complex background. We therefore performed image subtraction with HOTPANTS\footnote{http://www.astro.washington.edu/users/becker/hotpants.html} whenever a template exists (Fig. \ref{fig:sne_new}). This procedure achieved a good match between the background levels of target and template frames, resulting in net background levels close to zero in the subtracted images.
However, not all targets have templates. In these cases, local background was estimated by measuring actual flux fluctuations via placing apertures covering the region of the SN site. 
In all cases (including either image-subtracted or non-subtracted images), we defined the source as a positive detection if i) the source showed epoch-to-epoch flux changes, and ii) its flux was {\it above} the local background by at least 5 $\mu$Jy and 15 $\mu$Jy at 3.6 and 4.5 $\mu$m, respectively (according to point-source sensitivities in Table 2.10 of the IRAC Instrument Handbook version 2).

Moreover, in some cases, only a single-epoch set of {\it Spitzer} observations is available, thus, epoch-to-epoch flux changes cannot be used as indicators of the presence of SNe.  
In these cases, as a first step, we used archival pre-explosion 2MASS JHK$_s$ images in order to exclude the potential false-positive detections (compact \ion{H}{2} regions etc.). For the precise astrometric comparison, we collected the absolute coordinates of the concerned SNe from the Open Supernova Catalog and derived their (x,y) coordinates in the {\it Spitzer}/IRAC images (note that the uncertainties of the absolute SN coordinates have not been reported in the most cases). {\it Spitzer}/IRAC post-BCD images have a pointing to 2MASS with an accuracy of 0.15\arcsec (see IRAC Instrument Handbook\footnote{https://irsa.ipac.caltech.edu/data/SPITZER/docs/irac/iracinstrumenthandbook/}); an additional limit is the 0.6\arcsec/pixel resolution of {\it Spitzer}/IRAC PBCD images. The basic astrometric criterion of a potential positive detection was an agreement between the absolute SN coordinates and the position of the photometric center of the mid-IR point source within two IRAC pixels (1.2\arcsec).
In the second step, we carried out aperture photometry on the pre-explosion 2MASS JHKs images (using the same aperture and annulus/dannulus parameters as during the {\it Spitzer} photometry). Since, in most cases, there are no detectable point sources on the 2MASS images at the positions of the SNe, it was not possible to estimate reliable photometric errors based on photon statistics; instead, we have used a $\pm$0.4 mag value as a general photometric error, based on the upper limit of 2MASS photometric uncertainties reported in \citet{Skrutskie06}.
In order to reveal the presence of any possibly real mid-IR excess at post-explosion {\it Spitzer}/IRAC images,  we have fitted simple blackbodies to the SEDs consist of the upper limits of pre-explosion 2MASS photometry (assuming a general uncertainty of 0.4 mag mentioned above). The photometric criterion of a positive detection was to  find {\it Spitzer}/IRAC fluxes being above the fitted SED with a 3-$\sigma$ photometric error in at least one IRAC channel.   
Conclusively, we labeled in total 7 SNe with single-epoch {\it Spitzer} data as positive detections; we note that all of these SNe are expected to show strong mid-IR radiation at the given epoch (strongly interacting SNe IIn, or early-caught SNe of other types). We present all the pairs of images ({\it Spitzer}/IRAC + pre-explosion 2MASS K$_s$) and SED fittings lead us to select the single-epoch positive detections, together with an example for negative detections, in Appendix B.



We performed a photometric analysis for all positive detections at every epoch. For isolated sources, we implemented aperture photometry on the PBCD frames using the \texttt{phot} task of IRAF\footnote{IRAF is distributed by the National Optical Astronomy Observatories, which are operated by the Association of Universities for Research in Astronomy, Inc., under cooperative agreement with the National Science Foundation.} as a first step. We generally used an aperture radius of 2\arcsec~and a background annulus from 2\arcsec~to 6\arcsec~(2$-$2$-$6 configuration), applied aperture corrections of 1.213, 1.234, 1.379, and 1.584 for the four IRAC channels (3.6, 4.5, 5.8, and 8.0 $\mu$m, respectively) as given in IRAC Data Handbook, but sometimes used the 3$-$3$-$7 configuration (aperture corrections: 1.124, 1.127, 1.143, and 1.234, respectively) or the 5$-$12$-$20 configuration (aperture corrections: 1.049, 1.050, 1.058, and 1.068, respectively). For targets with templates, we compared the results before and after template subtraction to test for consistency. We generally found good agreement between the two methods ($\lesssim$10\% difference in fluxes, which is within the approximated uncertainty of {\it Spitzer}/IRAC photometry). In the few cases where the difference between the two methods was more than 10\%, we preferred the results of image subtraction photometry.

For sources on top of complex backgrounds without a corresponding template, we implemented the photometric method described by \citet{Fox11} (called hereafter as ``Fox+11 method''). This method applies a set of single apertures with a fixed radius to estimate both the SN and average background flux. This technique allows us to visually identify only local background associated with the SN, as opposed to the annuli regarding the aperture configurations mentioned above.

We compared our results to any previously published {\it Spitzer}/IRAC SN-photometry. In general, we found good agreements with the published values ($\lesssim$10\% difference in fluxes). In a few cases, the flux differences are larger, but each of these cases consist of either a very faint target and/or complex sky background.

The target details and resulting mid-IR photometry of all SNe with previously unpublished {\it Spitzer} photometry are listed in Tables \ref{tab:sn} and \ref{tab:phot}, respectively. We clearly highlight SNe identified on a single-epoch set of {\it Spitzer} images, as well as all the other SNe where image subtraction can not be applied; in all these cases, measured fluxes are strictly handled as upper limits. Flux uncertainties in Table \ref{tab:phot} generally based on photon statistics provided by \texttt{phot}, but, where photometry was carried out on subtracted images, increment of the noise level by $\sqrt{2}$ is also taken into account.



\section{Results}\label{res}

\subsection{Demographics}\label{res_stat}

\begin{figure*}
\includegraphics[width=5cm]{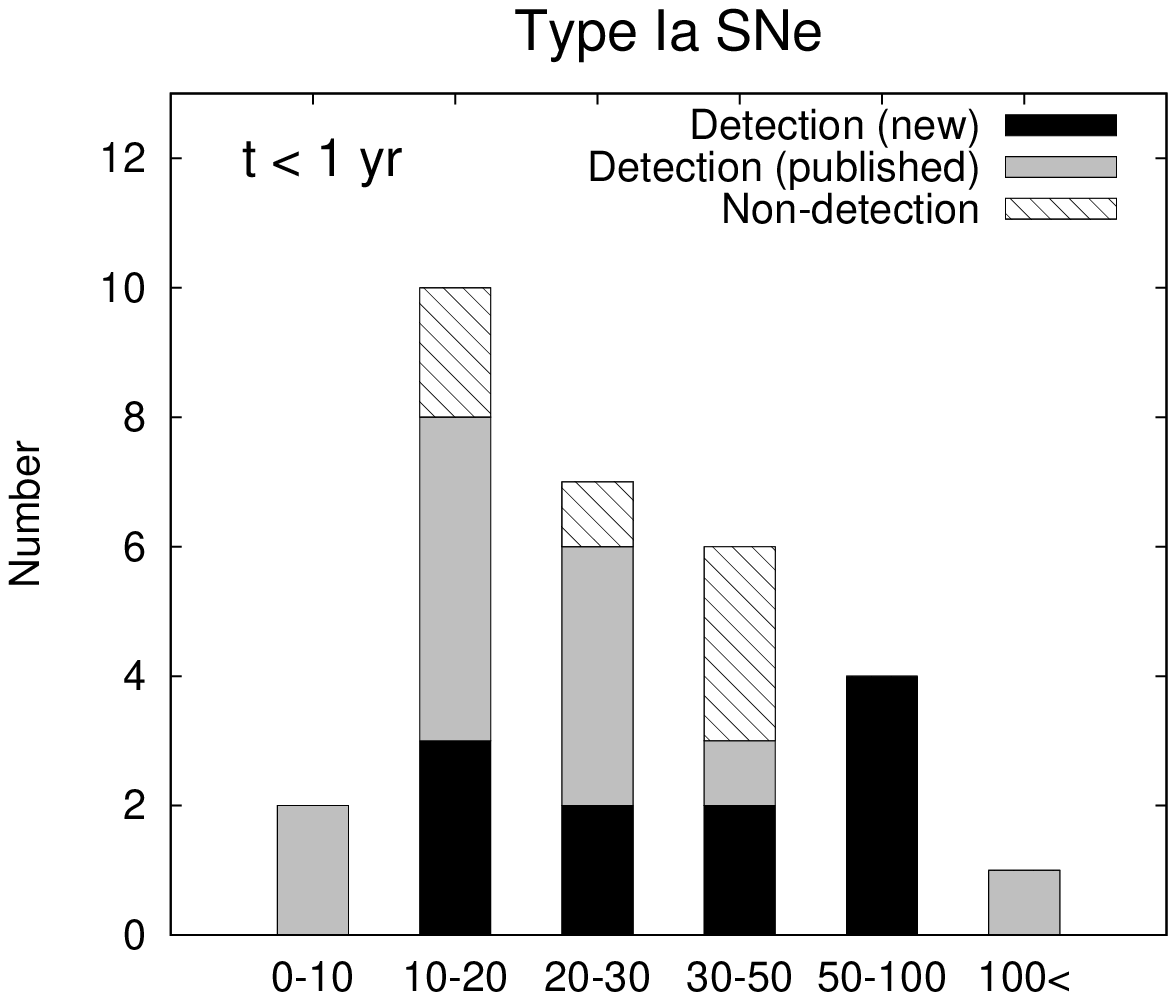}
\includegraphics[width=5cm]{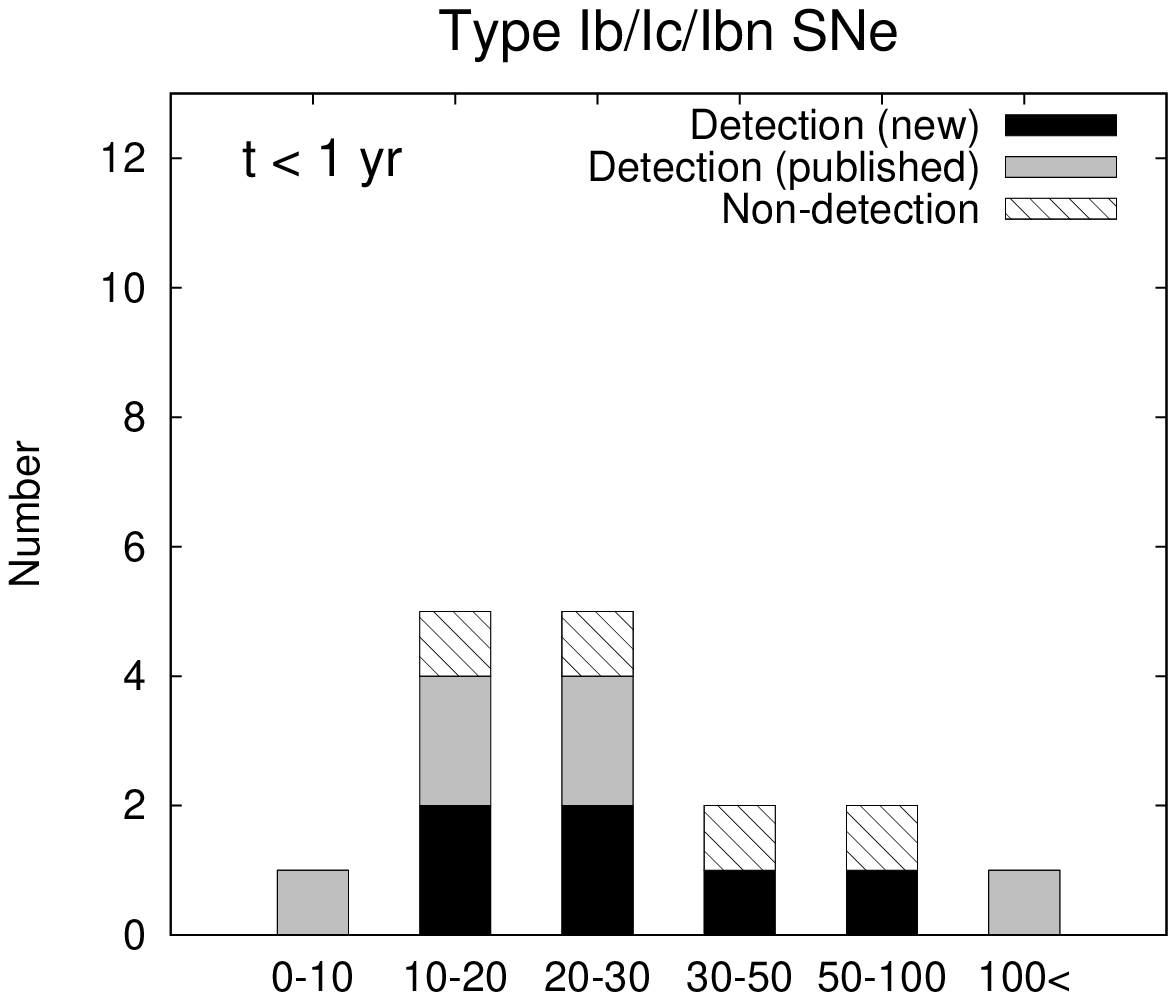}
\includegraphics[width=5cm]{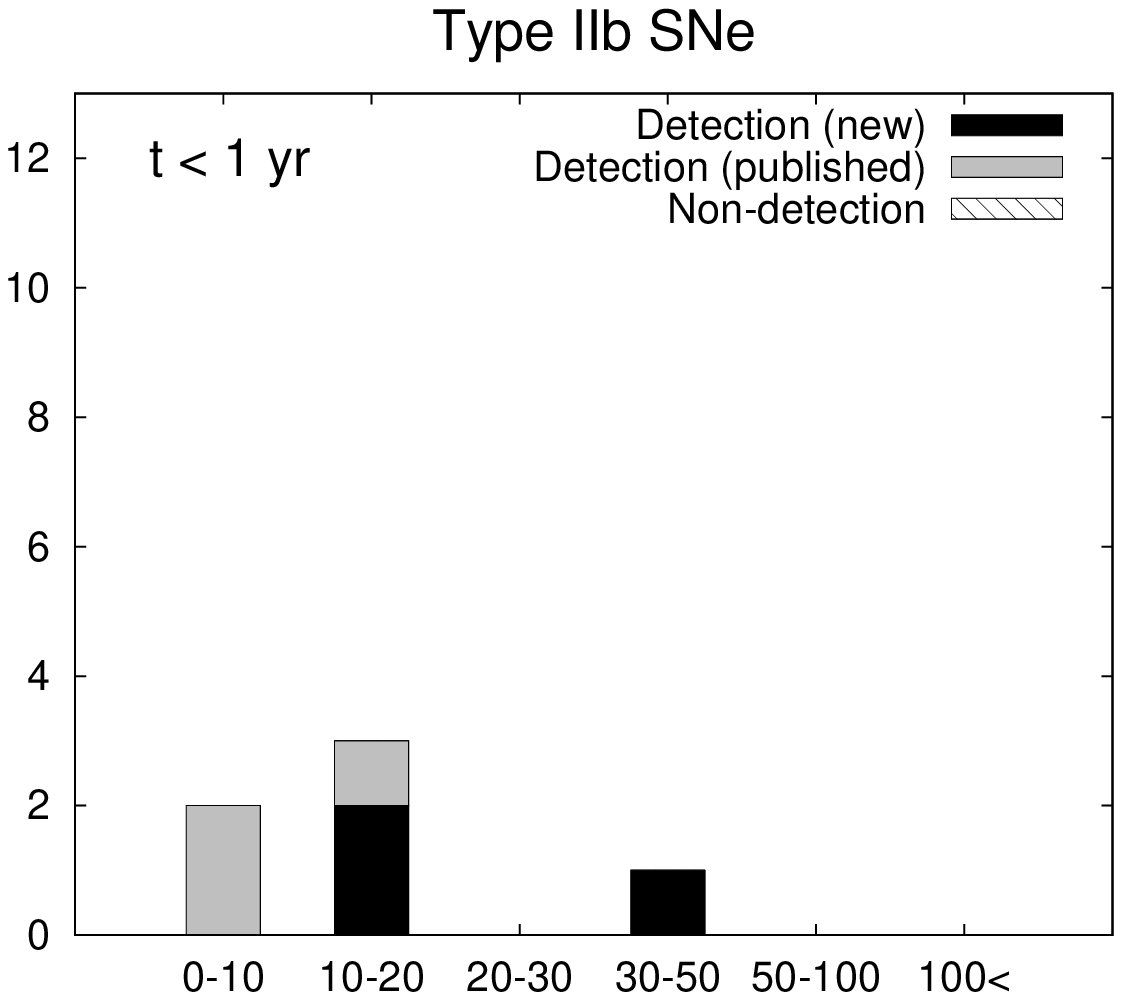}
\includegraphics[width=5cm]{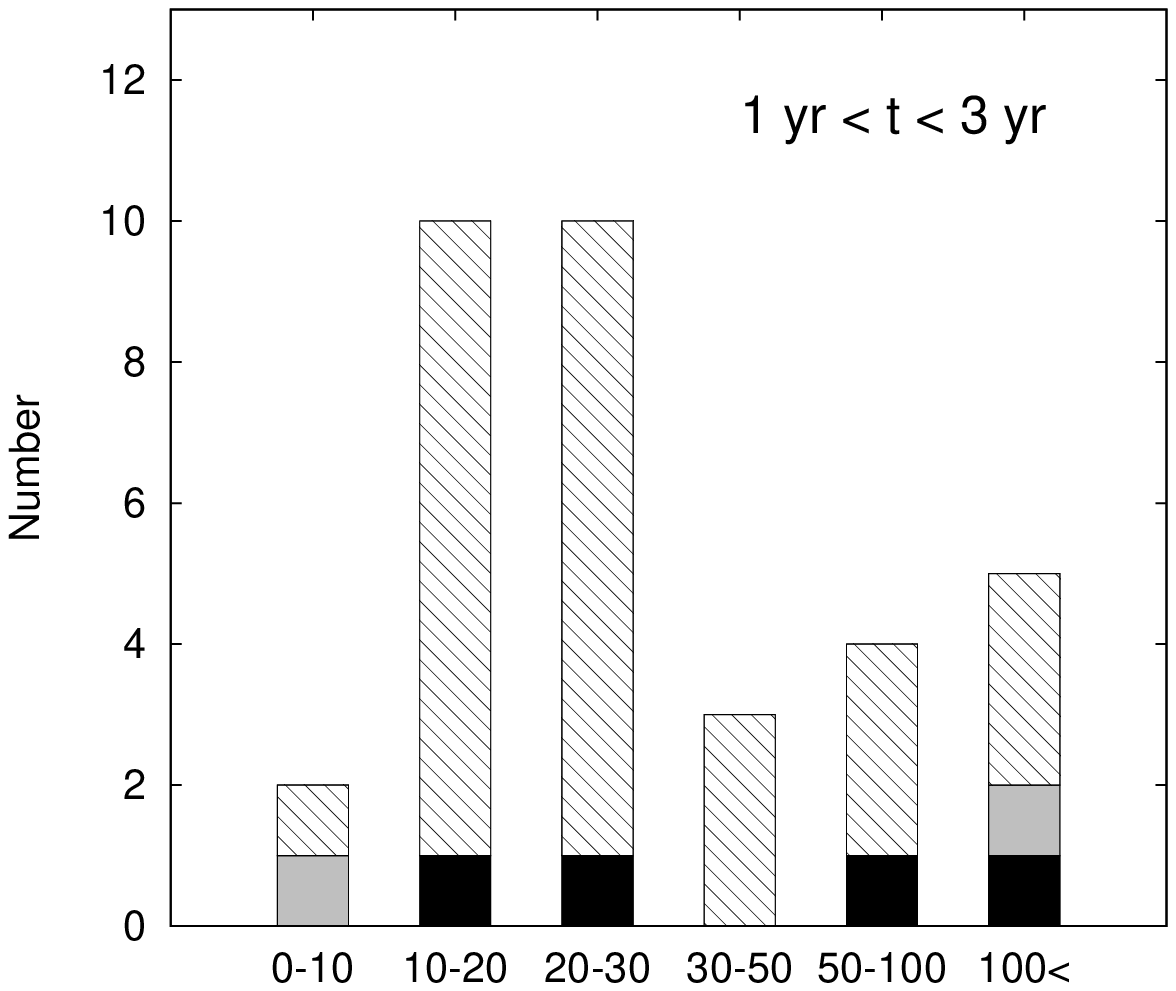}
\includegraphics[width=5cm]{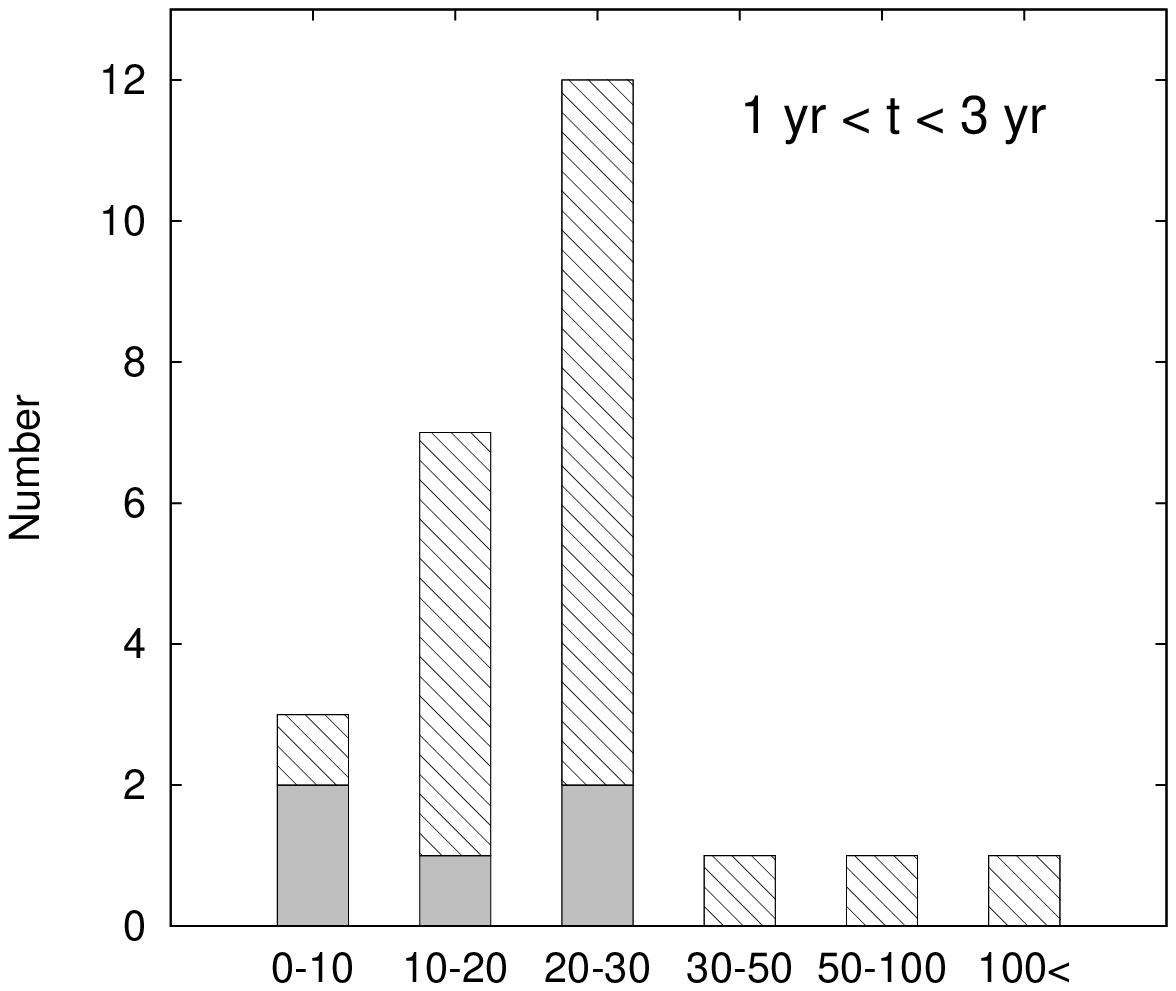}
\includegraphics[width=5cm]{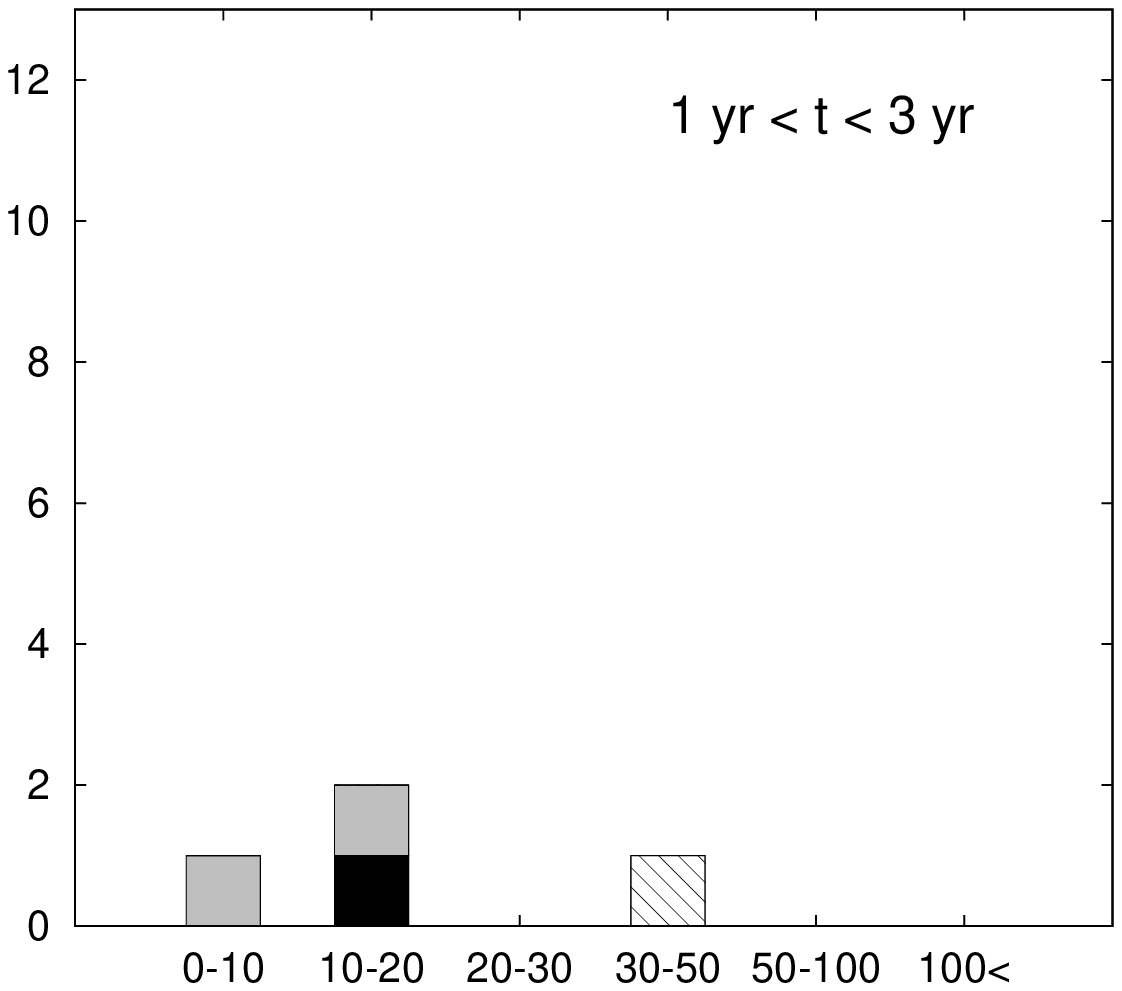}
\includegraphics[width=5cm]{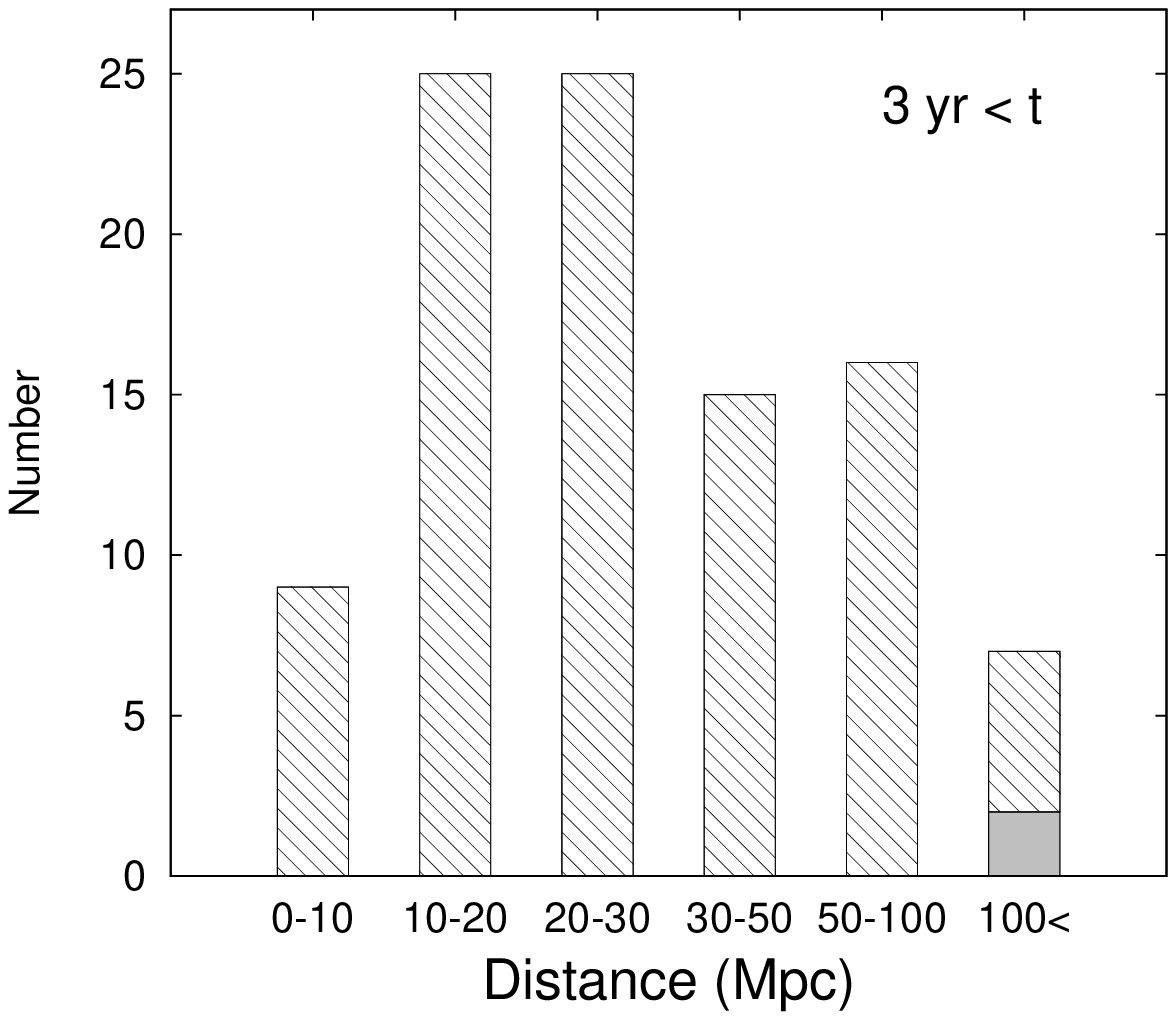}
\includegraphics[width=5cm]{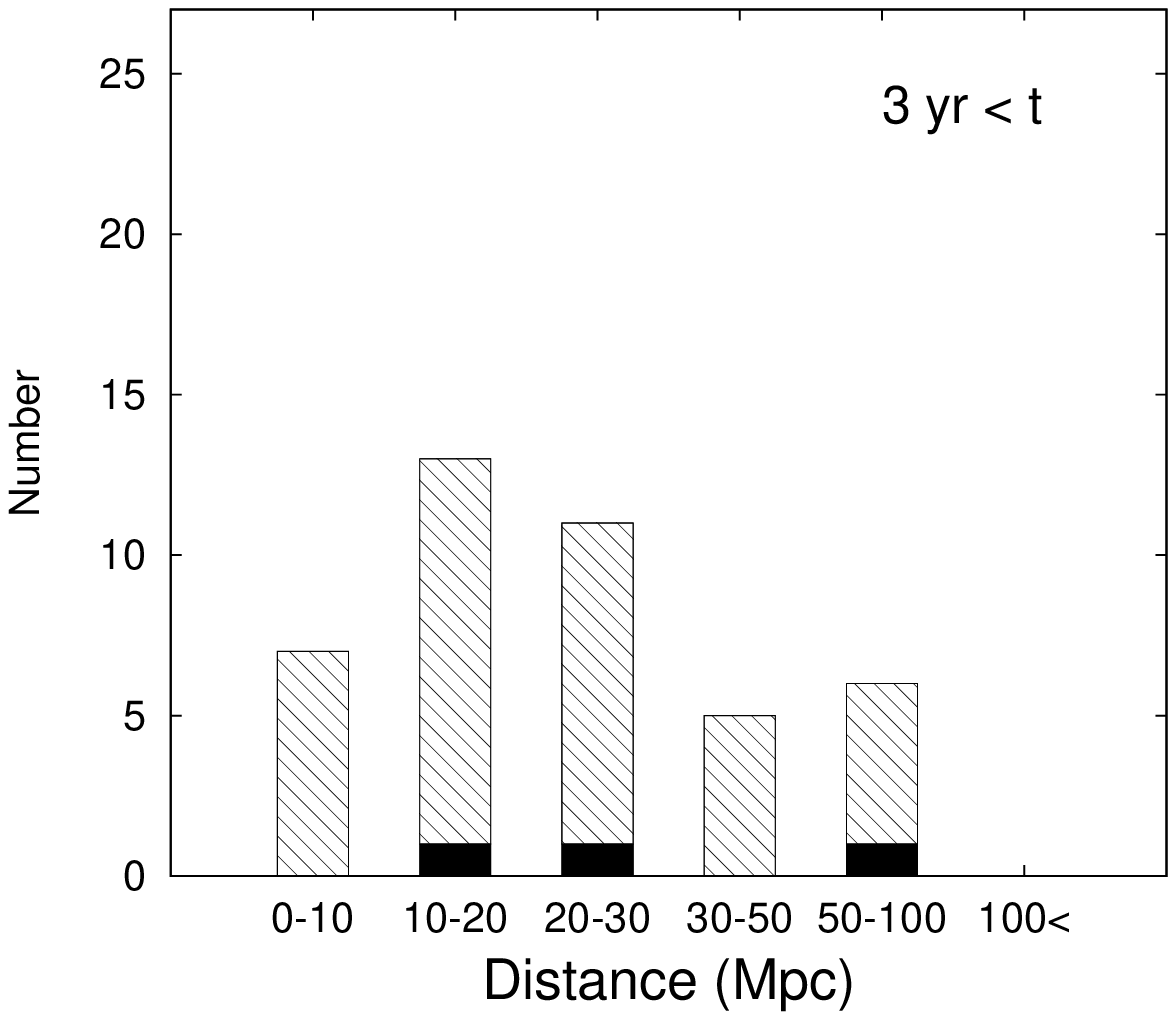}
\includegraphics[width=5cm]{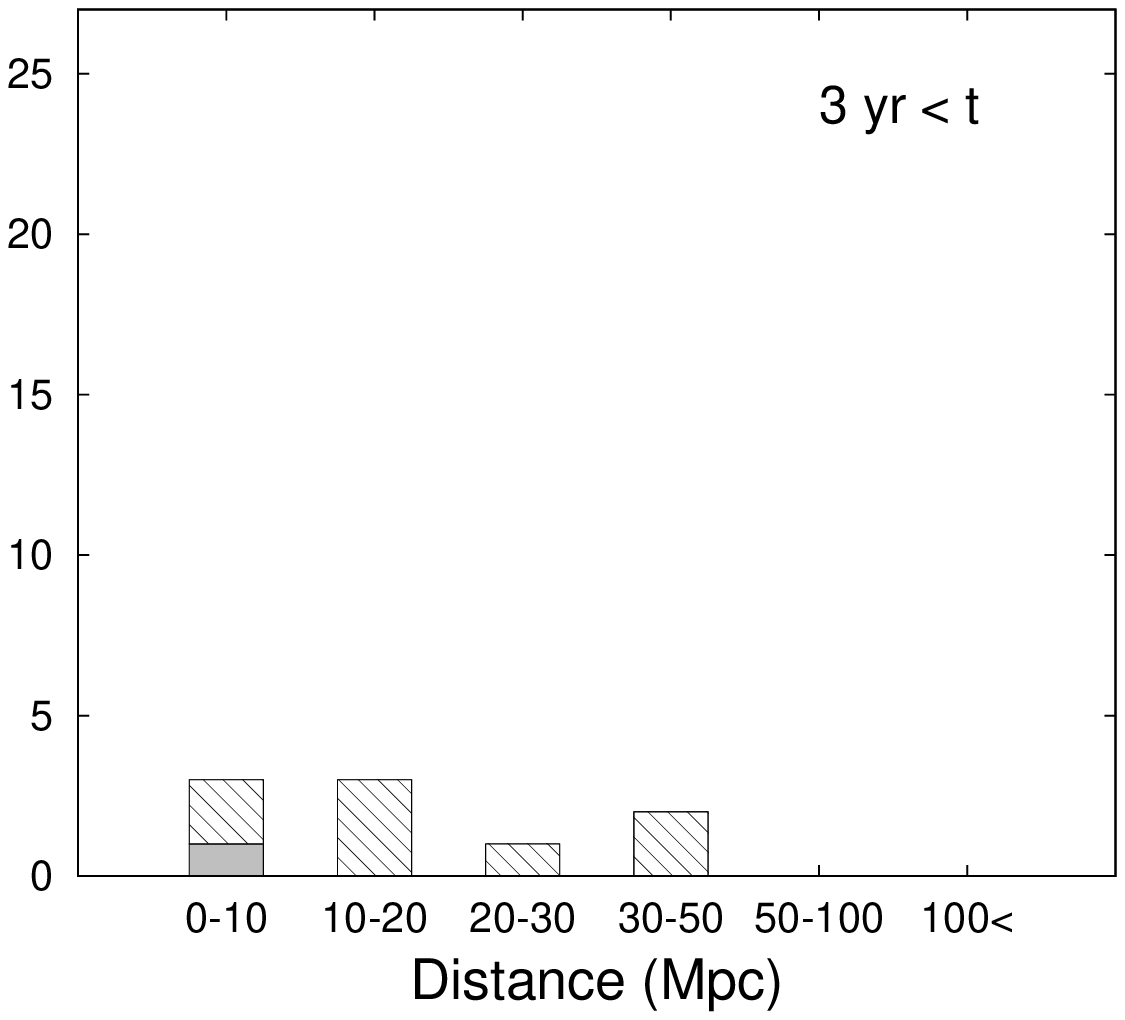}
\caption{The statistics of detected Type Ia and stripped-envelope CC SNe in our {\it Spitzer}/IRAC sample.  The statistics are divided by type and epoch.  The number of detections is plotted as a function of distance in each case.  We do not include SNe located in distant ($z\gtrsim$0.05), anonymous galaxies and/or too close to the center of their hosts.  We also exclude most SNe with only a single-epoch {\it Spitzer}/IRAC observations.}
\label{fig:stat}
\end{figure*}

\begin{figure*}
\includegraphics[width=5cm]{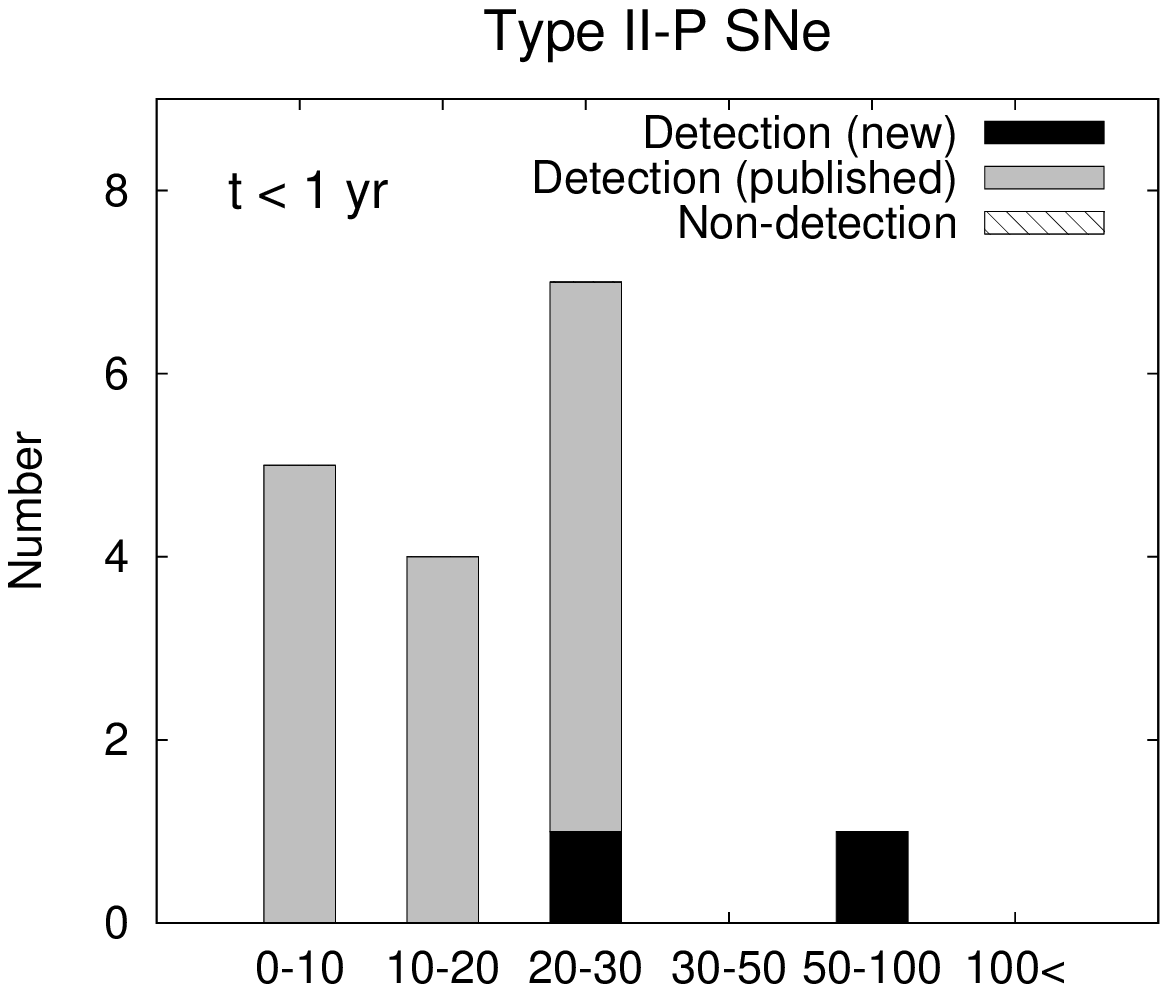}
\includegraphics[width=5cm]{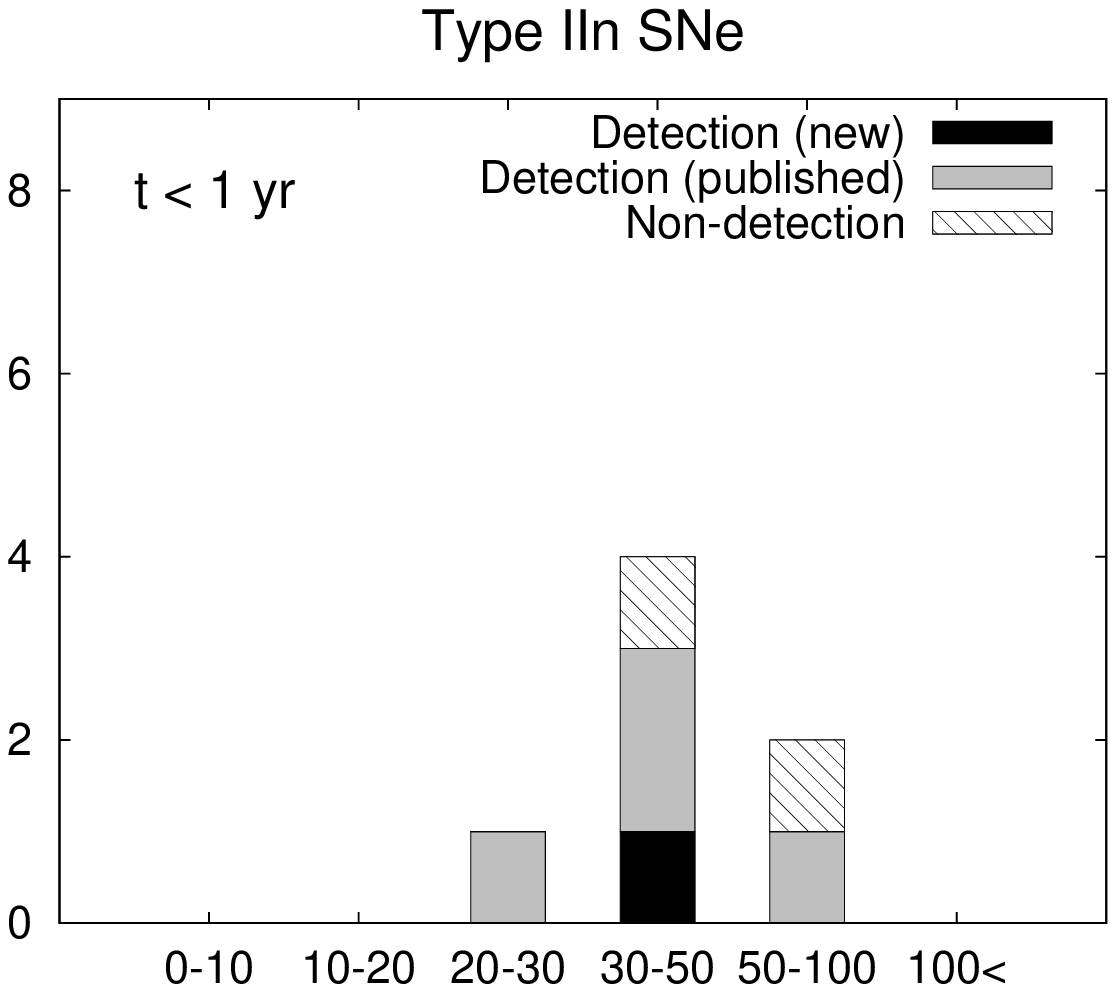}
\includegraphics[width=5cm]{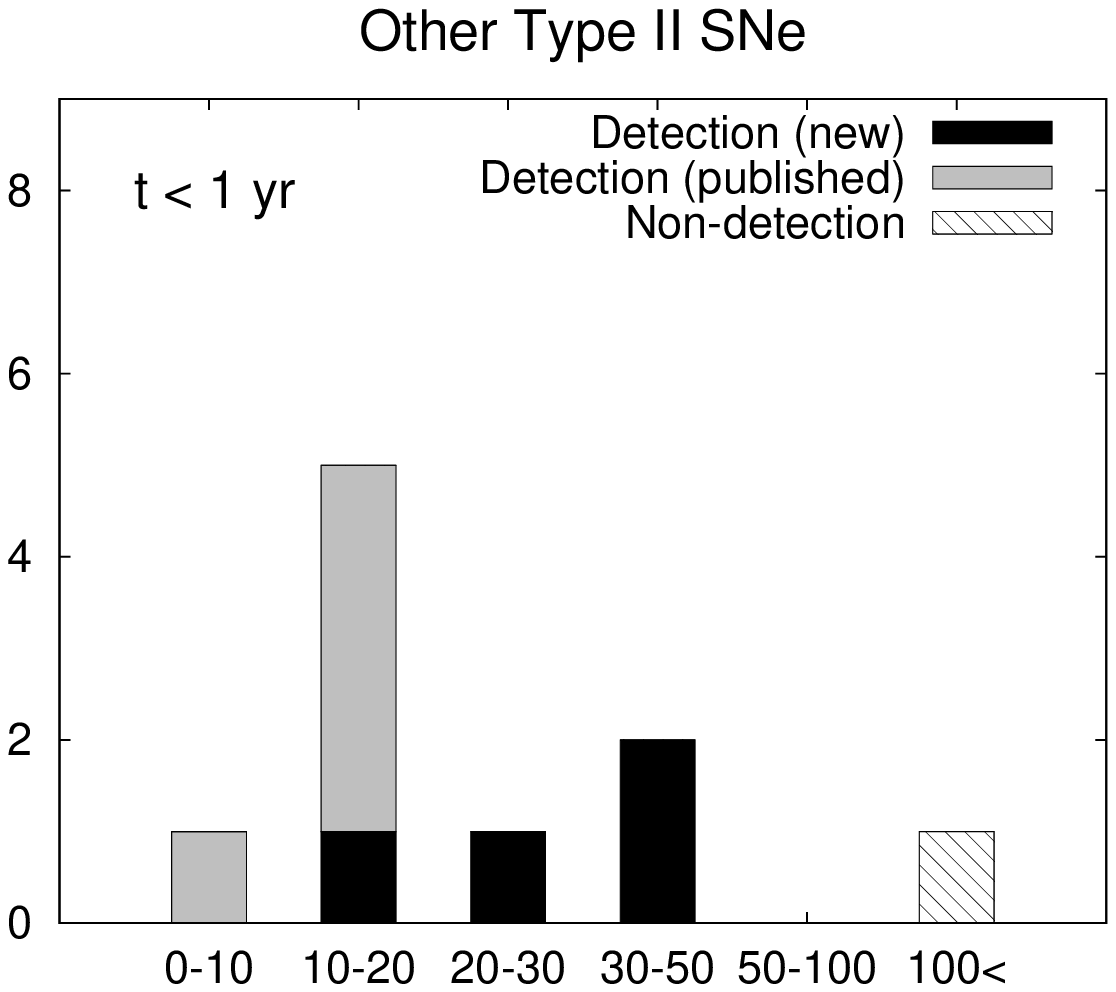}
\includegraphics[width=5cm]{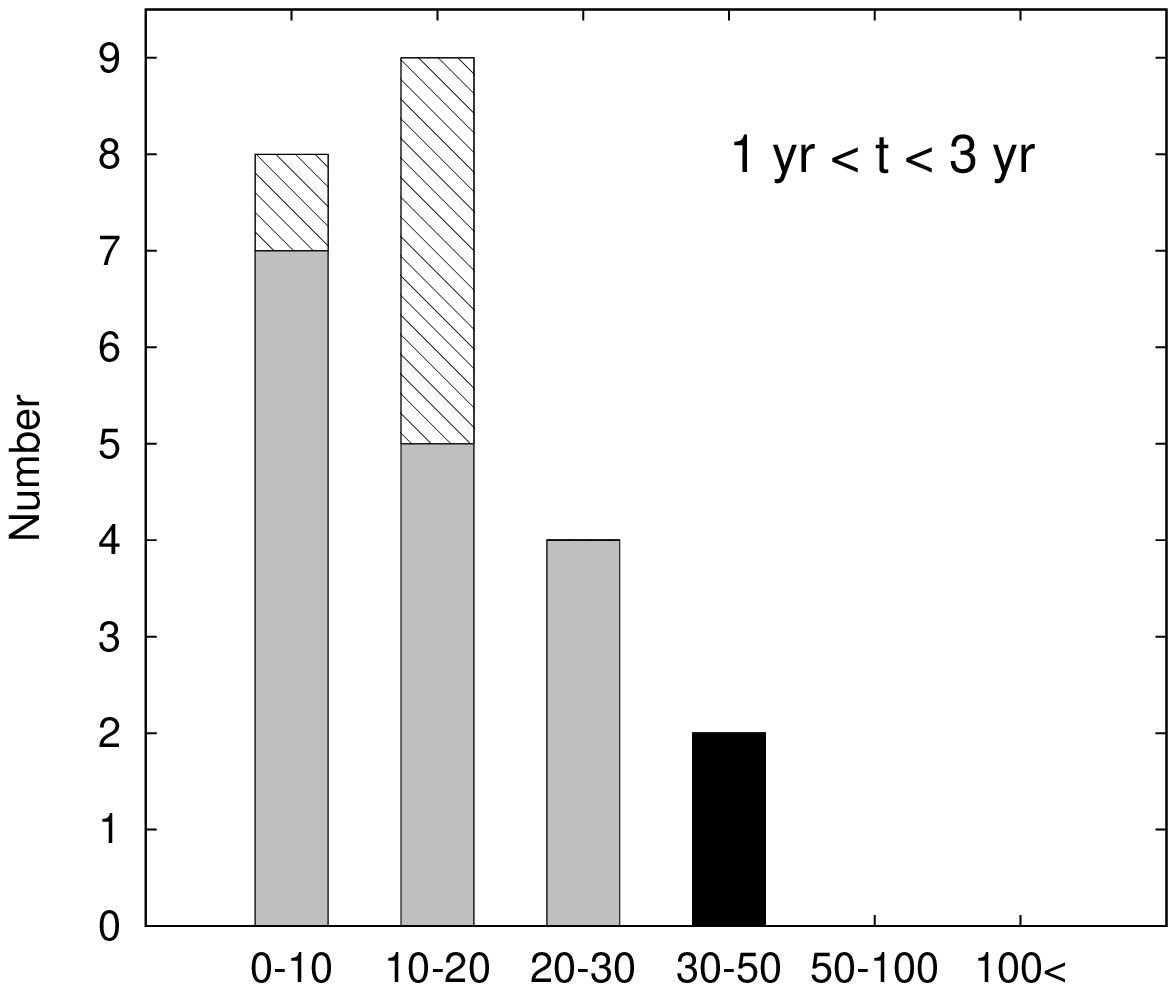}
\includegraphics[width=5cm]{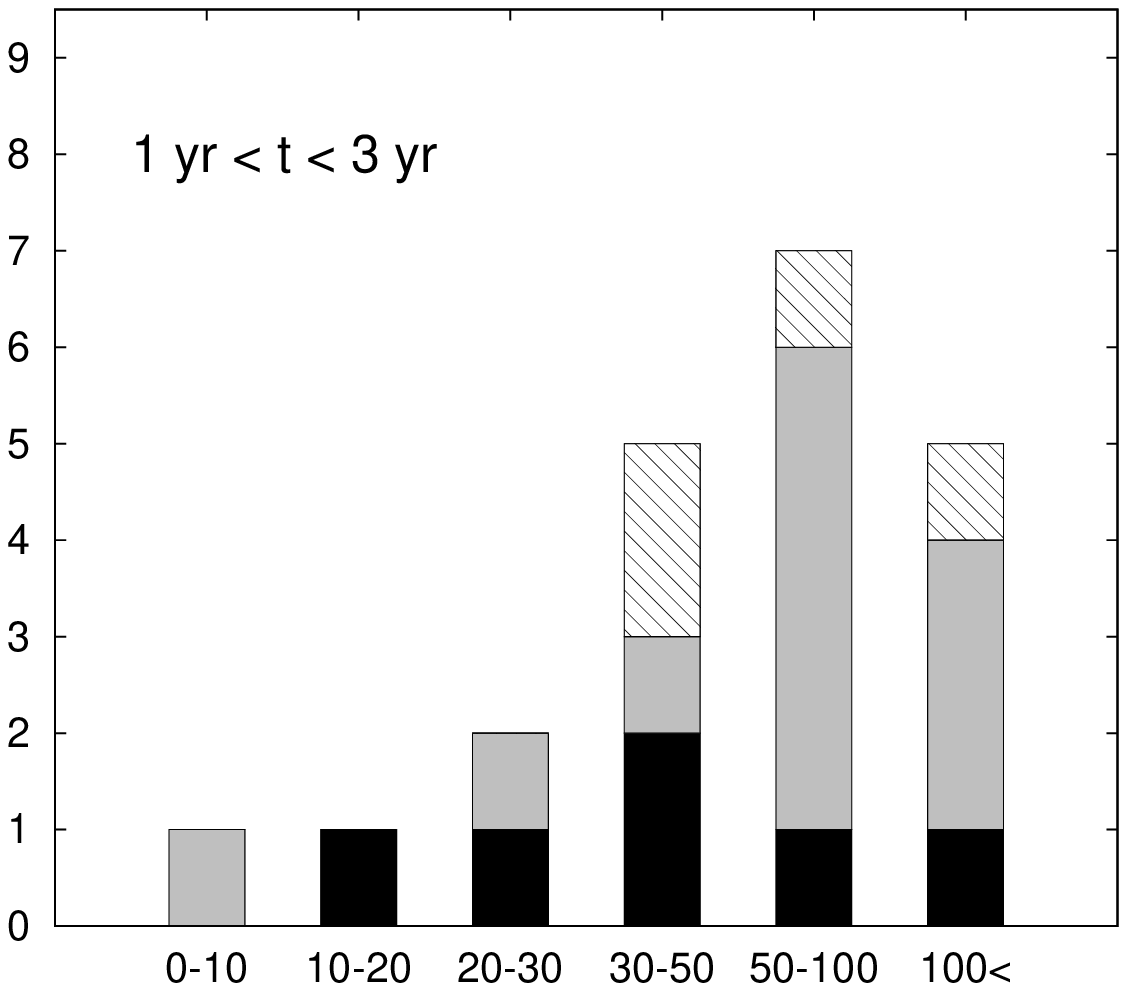}
\includegraphics[width=5cm]{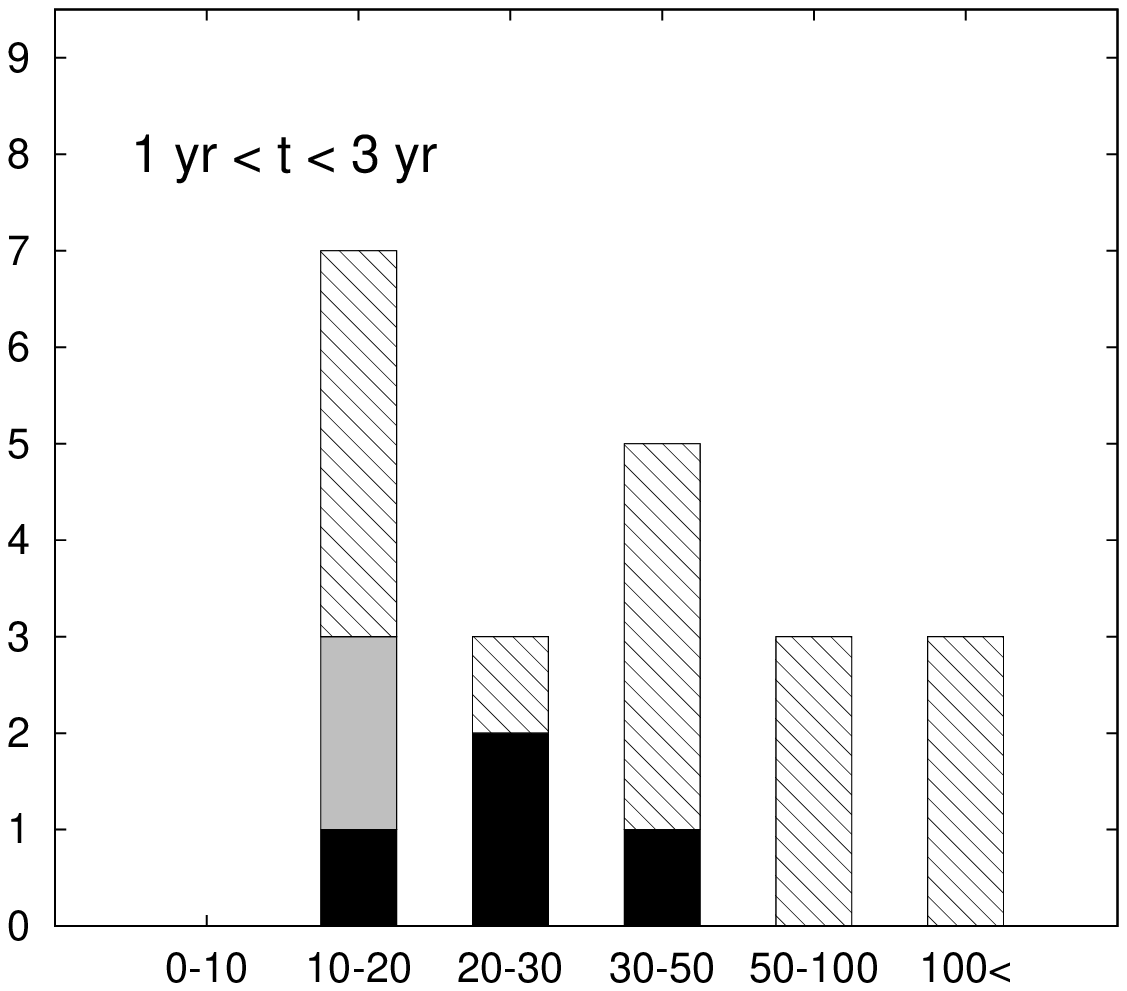}
\includegraphics[width=5cm]{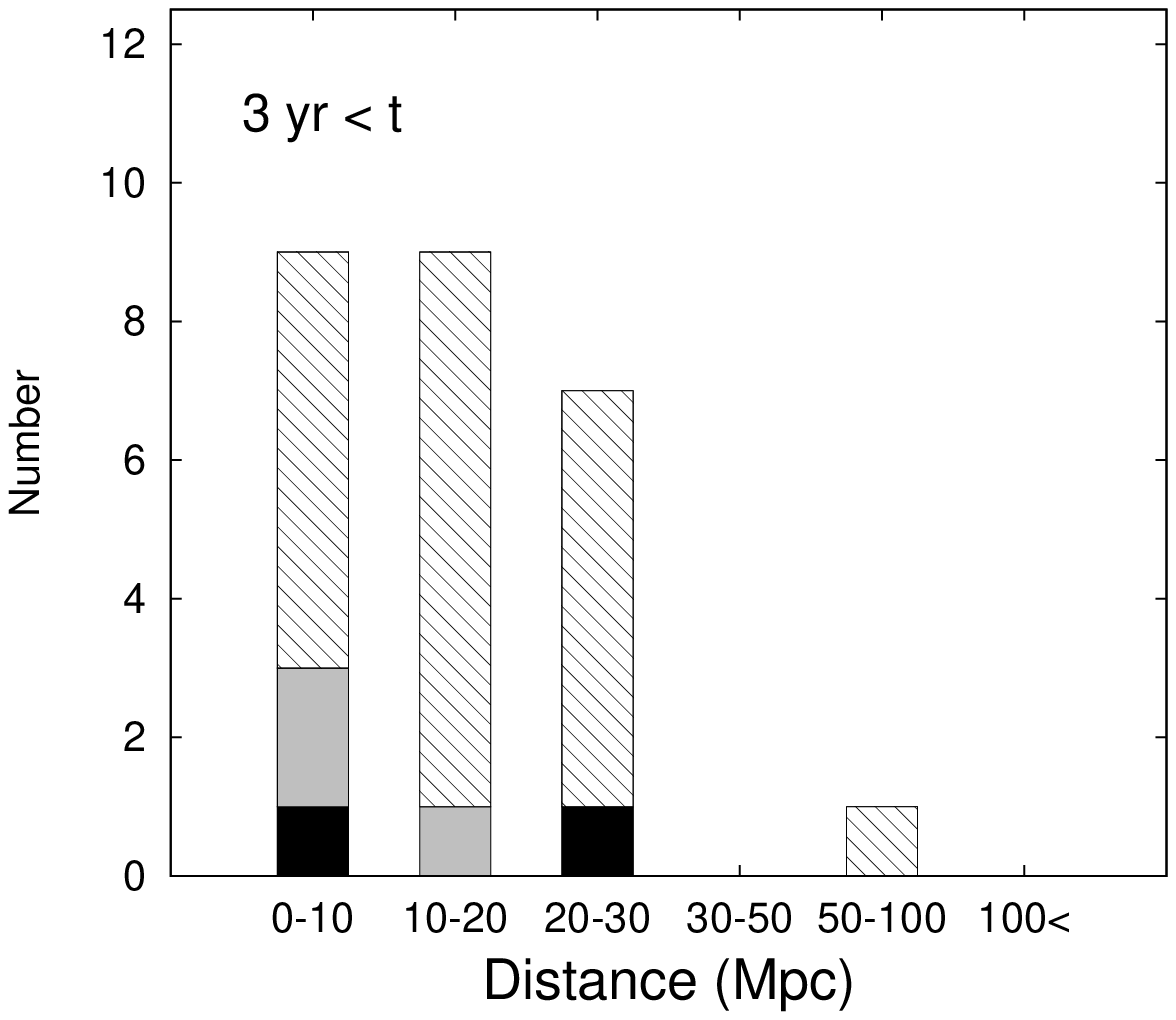}
\includegraphics[width=5cm]{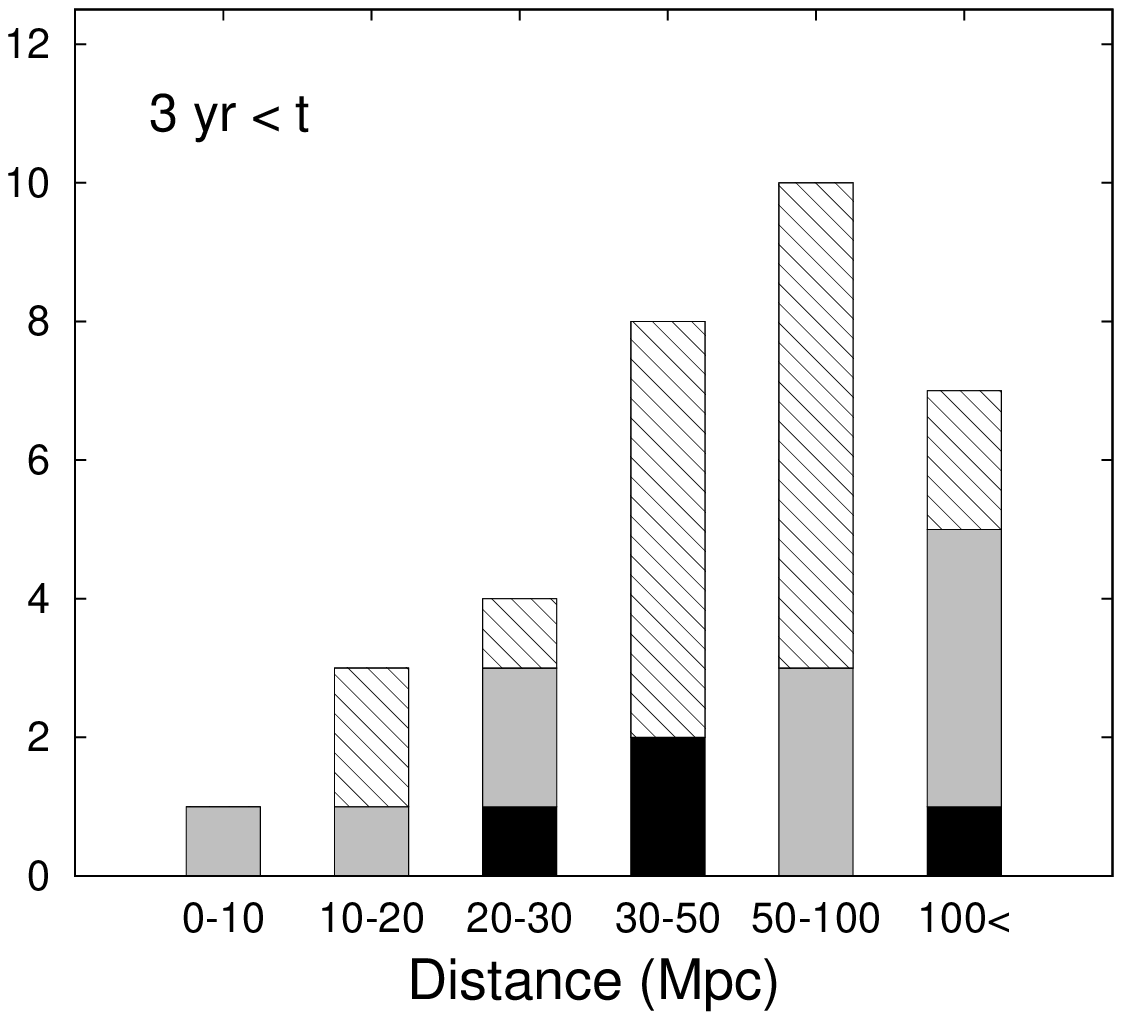}
\includegraphics[width=5cm]{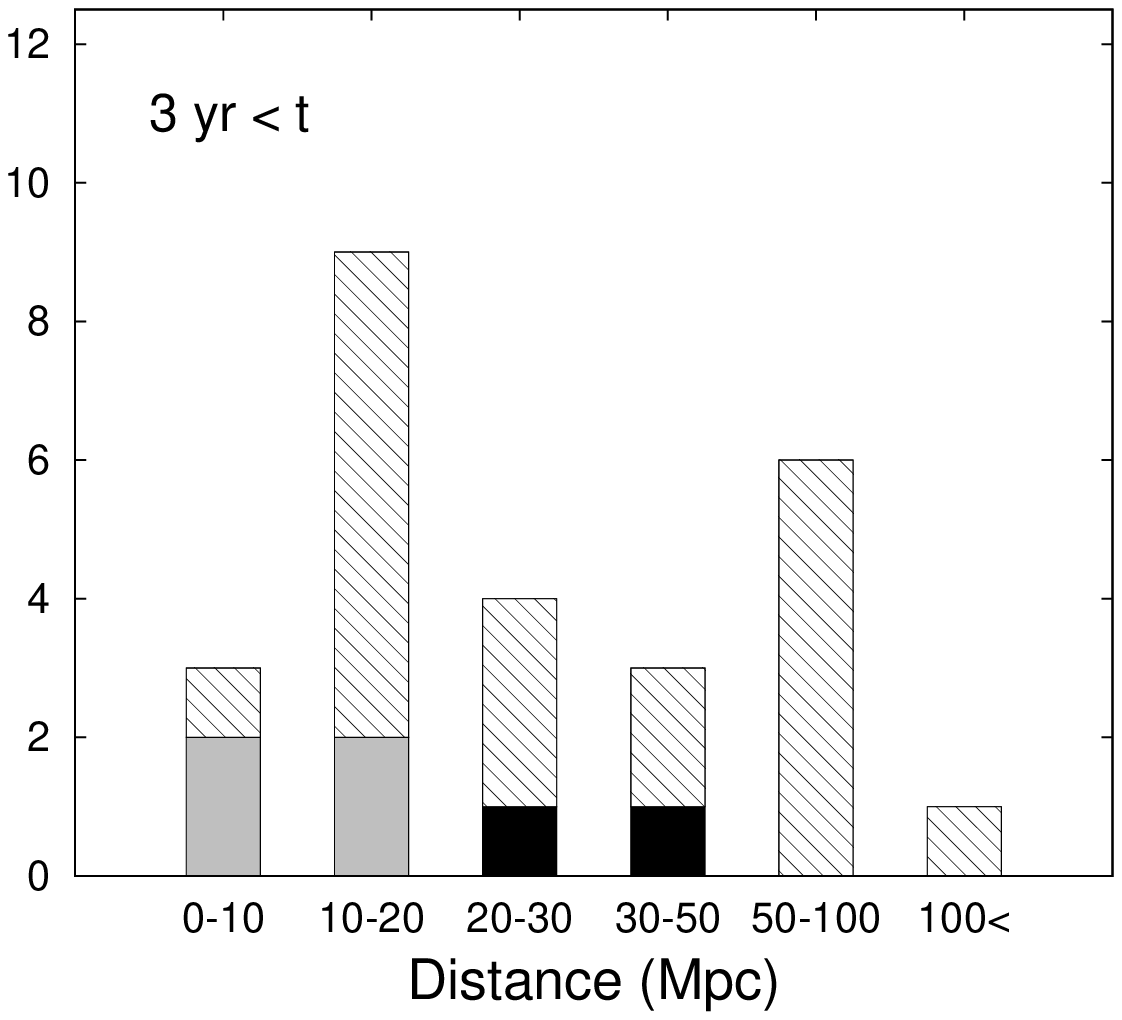}
\caption{Same as Figure \ref{fig:stat}, except in this case for SNe II-P, Type IIn, and other (unclassified) Type II SNe.}
\label{fig:stat_II}
\end{figure*}

The total number of the observed SN positions is over 1100. The majority of SNe are nearby (z$<$0.05). We detect 119 SNe, including 45 objects that have no previously published {\it Spitzer} photometry. Only $\sim$12\% of the SN sites were observed pre-explosion. We also highlight three specific targets (SNe 2012aw, 2012fh, and 2013ee), which have been noted by \citet{Tinyanont16} as positive {\it Spitzer} detections, but without any corresponding photometry.  We summarize the statistics of our SN sample in Table \ref{tab:stat} and Figs. \ref{fig:stat} and \ref{fig:stat_II}.

About 40$\%$ of the objects (mostly Type Ia) are located in distant, anonymous galaxies, but these observations did not yield any SN detections.  There are also $\sim$60 SNe that are also located in complex regions of the galaxy, typically very close to the galaxy nuclei.  In these cases, even template subtraction is not effective due to the asymmetric profile of the IRAC point-spread function (PSF). We do not include any of these SNe in this analysis.





Following the methods presented by \citet{Tinyanont16}, we present the detection rates separated in three time bins after discovery: less than one year, one to three years, and more than three years. If a SN is observed with at least one detection in a bin, it is considered detected, even though it might fade away later in the same bin. 


\begin{longrotatetable}
\begin{deluxetable}{c|cccc|ccccccc|ccccc}
\tabletypesize{\scriptsize}
\tablecaption{\label{tab:stat} Statistics of the {\it Spitzer}/IRAC data regarding the sample of studied SNe.}
\tablehead{}
\startdata
~ & \multicolumn{16}{c}{{\bf Total number of observed SN sites: 1142/693$^{\dagger}$}} \\
Total number of & \multicolumn{4}{c}{Thermonuclear SNe} & \multicolumn{7}{c}{Stripped-envelope CC SNe} & \multicolumn{5}{c}{Type II SNe}  \\
observed SN sites & Ia & Ia-pec & Iax & Ia-CSM & Ib & Ib-pec & Ibn & Ib/c & Ic & Ic-pec & IIb & II-P & II-P pec. & IIn & II-L & Unclass. SN II \\
~ & 723/294$^{\dagger}$ & 25/23$^{\dagger}$ & 8 & 5 & 59/53$^{\dagger}$ & 1 & 2 & 1 & 73/63$^{\dagger}$ & 5/4$^{\dagger}$ & 25 & 36 & 2 & 101 & 4 & 72 \\
\hline
~ & \multicolumn{16}{c}{{\bf SN sites with multiple observations: 553/334$^{\dagger}$}} \\
SN sites with & \multicolumn{4}{c}{Thermonuclear SNe} & \multicolumn{7}{c}{Stripped-envelope CC SNe} & \multicolumn{5}{c}{Type II SNe}  \\
multiple observations & Ia & Ia-pec & Iax & Ia-CSM & Ib & Ib-pec & Ibn & Ib/c & Ic & Ic-pec & IIb & II-P & II-P pec. & IIn & II-L & Unclass. SN II \\
~ & 325/112$^{\dagger}$ & 9 & 4 & 5 & 27/25$^{\dagger}$ & 1 & 1 & -- & 35/33$^{\dagger}$ & 3 & 14 & 32 & 2 & 38 & 4 & 53 \\
\hline
~ & \multicolumn{16}{c}{{\bf SN sites with pre-explosion images: 111/87$^{\dagger}$}} \\
SN sites with& \multicolumn{4}{c}{Thermonuclear SNe} & \multicolumn{7}{c}{Stripped-envelope CC SNe} & \multicolumn{5}{c}{Type II SNe}  \\
pre-explosion images & Ia & Ia-pec & Iax & Ia-CSM & Ib & Ib-pec & Ibn & Ib/c & Ic & Ic-pec & IIb & II-P & II-P pec. & IIn & II-L & Unclass. SN II \\
~ & 43/20$^{\dagger}$  & 3 & 2 & -- & 10 & -- & 1 & -- & 9/8$^{\dagger}$  & -- & 4 & 10 & 2 & 9 & 1 & 17 \\
\hline
~ & \multicolumn{16}{c}{{\bf Total number of positive detections: 119}} \\
Total number of & \multicolumn{4}{c}{Thermonuclear SNe} & \multicolumn{7}{c}{Stripped-envelope CC SNe} & \multicolumn{5}{c}{Type II SNe}  \\
positive detections & Ia & Ia-pec & Iax & Ia-CSM & Ib & Ib-pec & Ibn & Ib/c & Ic & Ic-pec & IIb & II-P & II-P pec. & IIn & II-L & Unclass. SN II \\
~ & {\bf 24} & 1 & 2 & 5 & {\bf 5} & -- & {\bf 1} & 1 & 7 & 1 & 7 & 22 & 1 & {\bf 25} & 2 & {\bf 15}  \\
\hline
~ & \multicolumn{16}{c}{{\bf Unpublished positive detections: 45}} \\
Unpublished & \multicolumn{4}{c}{Thermonuclear SNe} & \multicolumn{7}{c}{Stripped-envelope CC SNe} & \multicolumn{5}{c}{Type II SNe}  \\
positive detections & Ia & Ia-pec & Iax & Ia-CSM & Ib & Ib-pec & Ibn & Ib/c & Ic & Ic-pec & IIb & II-P & II-P pec. & IIn & II-L & Unclass. SN II \\
~ & {\bf 13} & 1 & 1 & 2 & {\bf 3} & -- & -- & 1 & 2 & -- & 4 & 4 & -- & {\bf 6} & 1 & {\bf 7} \\
\hline
\enddata
\tablecomments{$^{\dagger}$Total number of objects / Number of objects excluding SNe in distant, anonymous galaxies}
\end{deluxetable}
\end{longrotatetable}

\subsection{Mid-IR evolution: trends and outliers}\label{res_ev}

\begin{figure*}
\includegraphics[width=15cm]{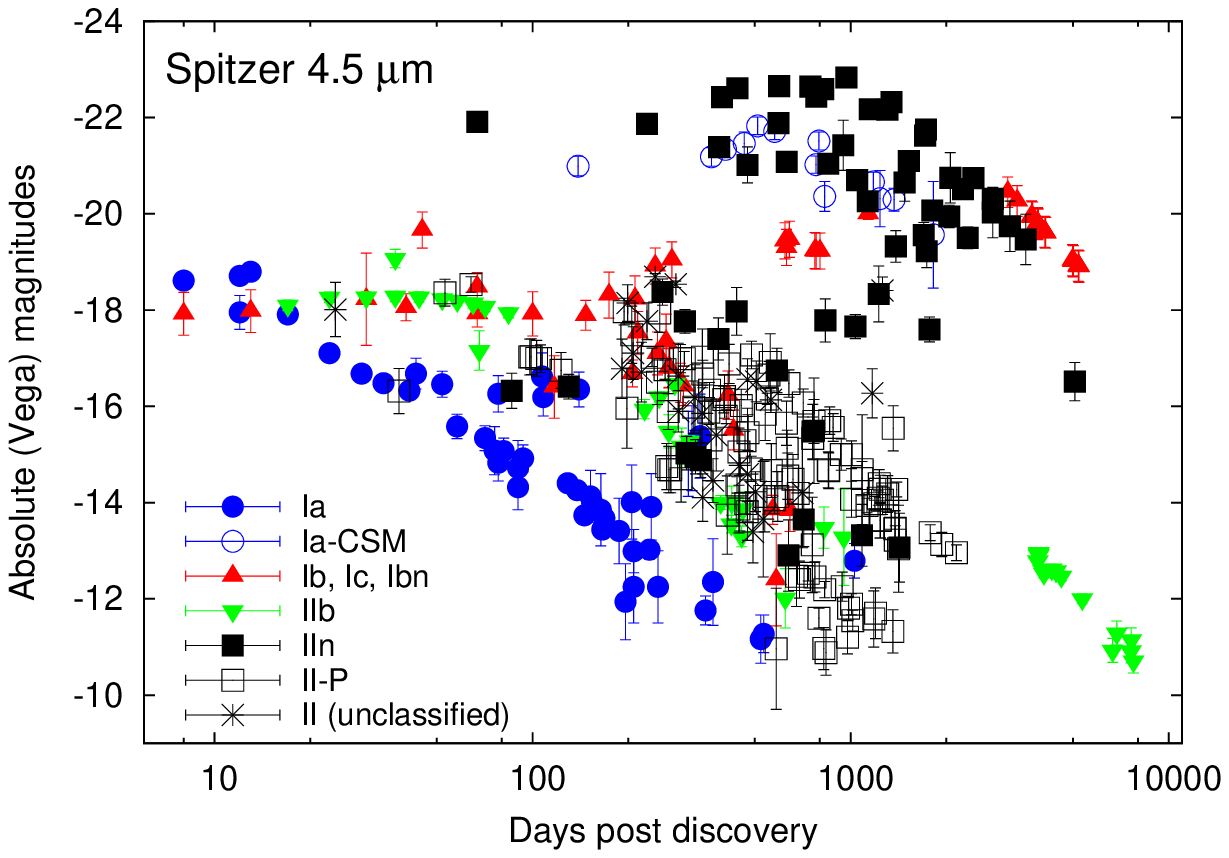}
\caption{4.5 $\mu$m absolute Vega magnitudes of all SNe identified as point sources in {\it Spitzer}/IRAC images. Values and sources of data are shown in Table \ref{tab:absmag_all}.}
\label{fig:absmag}
\end{figure*}

Fig. \ref{fig:absmag} plots the mid-IR photometry of all SNe with positive {\it Spitzer} detections.  Table \ref{tab:absmag_all}
lists all corresponding Vega magnitudes, distances, and $E(B-V)$ values.  For plotting purposes, Fig. \ref{fig:absmag} excludes some objects with decade-long mid-IR datasets -- e.g. Type II-pec SN~1987A \citep{Dwek10}, or Type II-L SN~1979C \citep{Tinyanont16} --, but these SNe are included in our statistical analysis. Figs. \ref{fig:absmag_Ia}-\ref{fig:absmag_II} highlight the SN subclasses so that individual SNe can be identified and photometric details can be ascertained. Tables \ref{tab:sn} and \ref{tab:absmag_all} contain all the sources of previously published {\it Spitzer} data we used for constructing Figs. \ref{fig:stat}-\ref{fig:absmag_II} and for the analysis we present below.


\subsubsection{Thermonuclear SNe}

\begin{figure}
\includegraphics[width=15cm]{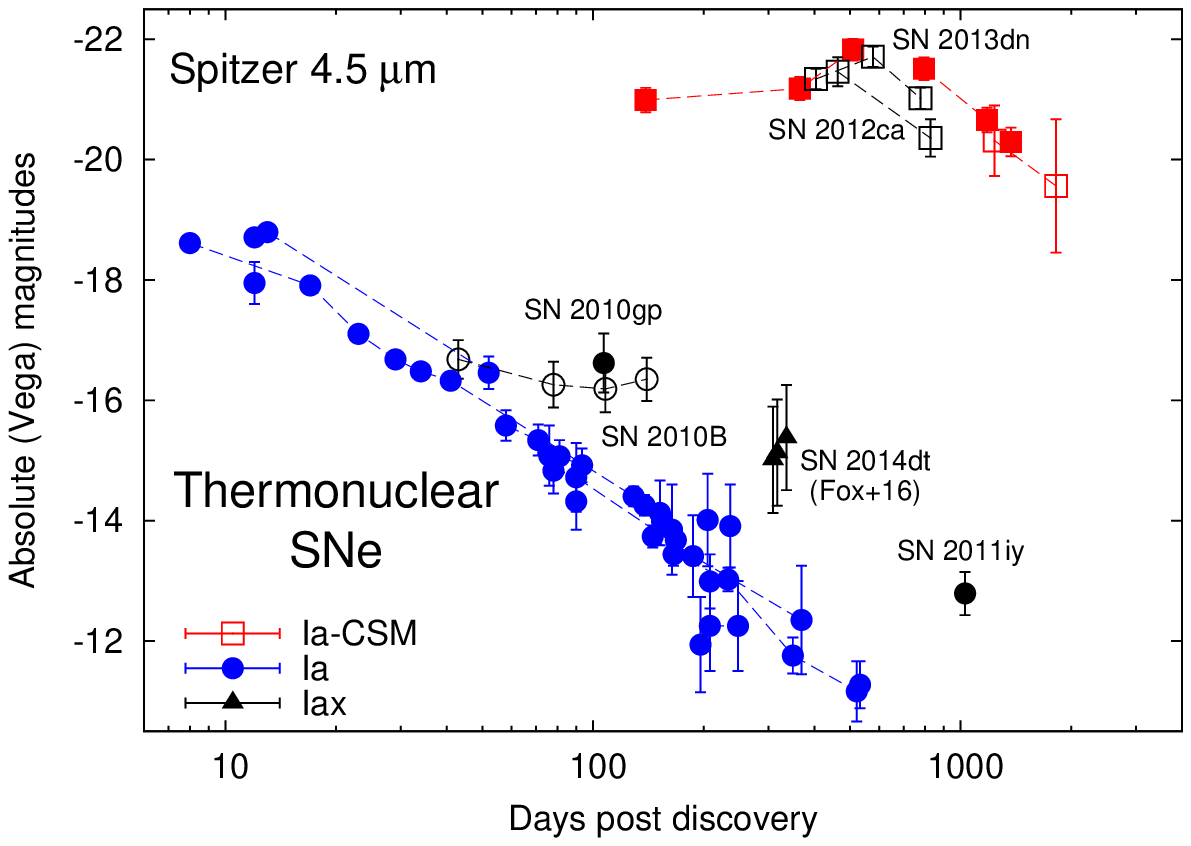}
\caption{Mid-IR evolution of thermonuclear SNe; highlighted objects are marked with black symbols, while filled and empty symbols denote SNe whose absolute magnitudes were determined with or without template subtraction, respectively. Values and sources of data are shown in Table \ref{tab:absmag_all}.}
\label{fig:absmag_Ia}
\end{figure}

This work more than doubles the number of SNe Ia with positive mid-IR detection (33 vs. 15).  Fig. \ref{fig:absmag_Ia} shows that most Type Ia SNe have a relatively well-defined evolution compared to the other SN subclasses, consistent with previous results \citep{Tinyanont16, Johansson17}. The handful of Type Ia-CSM SNe, however, are extremely bright at mid-IR wavelengths.  This sample is small, so it is difficult to draw any definitive conclusions about the overall trend.  For example, PTF11kx ($D \sim$200 Mpc) is still detectable at $\sim$1800 days, while SN~2002ic ($D \sim$280 Mpc) faded at a similar age.

We do not find any previously unpublished SNe Ia with mid-IR fluxes comparable to those of the known SNe Ia-CSM.  This result suggests that a dense CSM is rare in the environments of thermonuclear SNe (which may also hint that SNe Ia-CSM arise from different progenitor systems than the majority of SNe Ia), or, that CSM shells may be far away from the explosion sites. Based on the existing (rough) estimations, SN Ia-CSM objects may contribute between 1\% and 5\% of all SNe Ia \citep[see e.g.][]{MP18}, which seems to be supported by our results; however, future systematic surveys are necessary for the thorough study of this problem.

We do find some other SNe Ia that deviate from the expected mid-IR evolution. SNe 2010B and 2010gp, observed at early times, are noticeably brighter. Note, however, that SN 2010B has a complex background that may be contributing additional mid-IR flux.  On the other hand, SN 2010gp has a template for subtraction, so these results are quite robust.  SN~2011iy, which is relatively nearby ($d \sim$20 Mpc), also appears as a point source at 4.5 $\mu$m after image subtraction at $\sim$1030 days after explosion (see Fig. \ref{fig:sne_new}). This SN, however, is not detectable at 3.6 $\mu$m.

SN~2014dt, classified as a Type Iax SN, should be also highlighted here: this object shows a clear and even growing mid-IR excess $\sim$1 yr after explosion, which has been explained with the presence of newly-formed dust, pre-existing dust, or possibly a bound remnant \citep{Fox16,Foley16}. The only other SN Iax we identified as a mid-IR source on {\it Spitzer} images is SN~2005P. While SN 2005P seems to have a slight 8.0 $\mu$m~detection at $\sim$180 days, there is no comparable data with SN 2014dt. By $\sim$1 year post-explosion, SN 2005P has faded below the detection threshold.

\subsubsection{Stripped-envelope CC SNe}

\begin{figure}
\includegraphics[width=15cm]{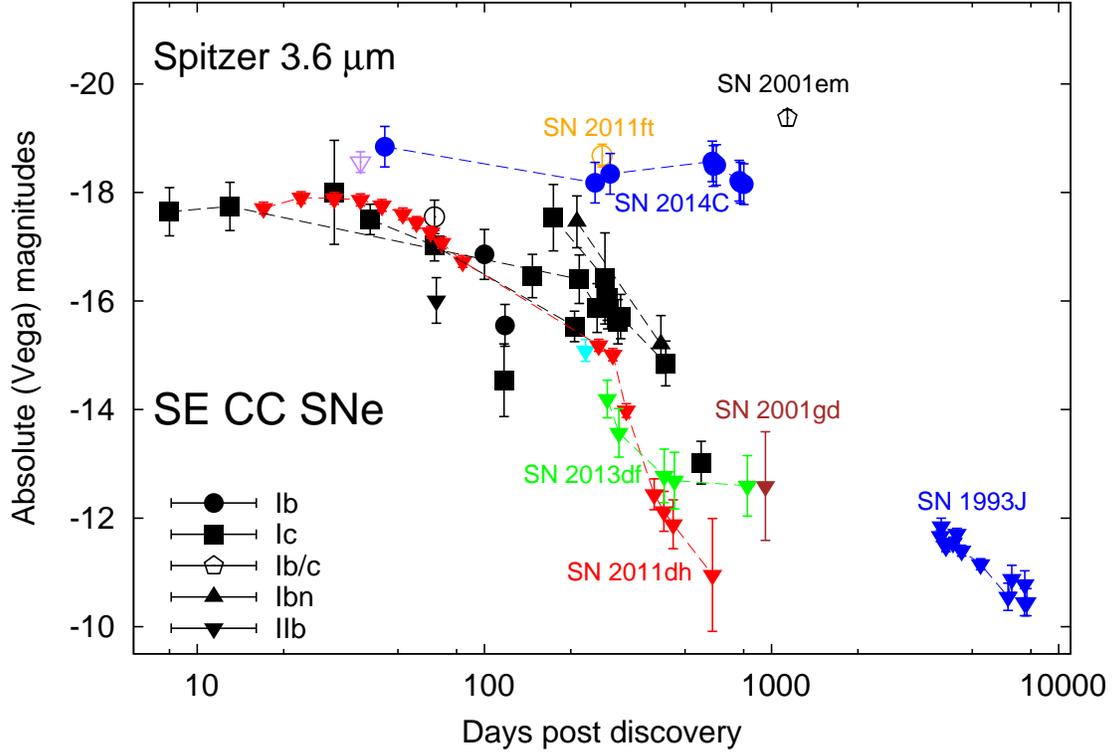}
\caption{Mid-IR evolution of stripped-envelope core-collapse SNe; highlighted objects are marked with labels, while filled and empty symbols denote SNe whose absolute magnitudes were determined with or without image subtraction, respectively. Values and sources of data are shown in Table \ref{tab:absmag_all}.}
\label{fig:absmag_Ibc}
\end{figure}

Fig. \ref{fig:absmag_Ibc} plots the mid-IR absolute magnitudes of SE CCSNe.  The stripped-envelope designation encompasses various subclasses, including SNe IIb, Ib, and Ic, so their mid-IR evolution exhibits a bit of heterogeneity, particularly at later times.  

The mid-IR evolution of ``normal'' Type Ib/Ic SNe seem to be fastest amongst SE CCSNe.  SN~2014C presents itself as a special case in which the explosion transforms from a ``normal'' Type Ib into a strongly-interacting, Type IIn-like SN \citep{Milisavljevic15,Margutti17}. SN~2014C is located within NGC 7331, which has been followed extensively as part of the SPIRITS program. \citet{Tinyanont16} show a roughly constant IR-luminosity in the first $\sim$800 days and a unique re-brightening at $\sim$250 days as the CSM interaction begins. 

Another interesting object is SN~2001em, a strongly-interacting Type Ib/c object, which generated strong X-ray, radio and optical emission for $\sim$3 years post-explosion \citep[see][]{Stockdale04,Pooley04,Soderberg04,Chugai06}. Unlike SN 2014C, however, the transformational process was not observed by {\it Spitzer}, making a direct comparison impossible. SN~2001em was observed by {\it Spitzer} only once. Fig. \ref{fig:absmag_Ibc} shows SN~2001em is even brighter than SN~2014C, although background contributions have not been removed. We present a more detailed analysis of SN~2001em in Section \ref{res_sed}.

Finally, it is worth mentioning the one observation of SN~2011ft, a distant ($d \sim$ 100 Mpc) Type Ib SN that is as bright as SN~2014C at $\sim$250 days after explosion. With only a single 3.6 $\mu$m, more observations are planned.

Observations exist for only two SNe Ibn: SN~2006jc (4 epochs, but only one from the first year) and PS1-12sk (1 epoch). These two events are bright in mid-IR during the early-time CSM interaction, but the brightness declines quickly. Following the interpretation of \citet{Mattila08}, the early mid-IR radiation may arise from newly formed dust in the CDS, while the source of the later-time mid-IR flux is probably an IR echo from pre-existing dust in the CSM.

Among SNe IIb, the moderately interacting SN~2013df \citep{Kamble16,Maeda15b,Szalai16} produces a slowly-declining mid-IR light curve between $\sim$270-820 days \citep{Tinyanont16,Szalai16}. SN 2001gd shows a similar brightness at $\sim$950 days. SN~2011dh, one of the best-sampled SN in mid-IR, has been also detectable up to almost two years after explosion. The Type IIb SN 1993J ($D$ $\sim$3.7 Mpc) is detected at $>$24 years post-explosion in mid-IR \citep{Tinyanont16}, while the Type IIb SN~2008ax ($D$ $\sim$7.8 Mpc) is not detected at even $\sim$4 years after explosion.


The differences between SNe IIb seem to correlate with the assumed sizes of the progenitors of SE CCSNe. SNe 1993J, 2001gd and 2013df, which are detected by {\it Spitzer} at later epochs, have been classified as Type eIIb objects \citep{Chevalier10,Szalai16}, which denotes that these explosions originate from extended progenitors (yellow or red giants). SN~2008ax is known as a representative of Type cIIb objects, which are defined to have more compact progenitors, similar to those of SNe Ib/c. SN~2011dh seems to be an intermediate case in both its progenitor radius ($R \sim$ a few tens of $R{_\odot}$) and mid-IR evolution.


\subsubsection{Type II-P SNe}\label{res_ev_IIP}

\begin{figure}
\includegraphics[width=15cm]{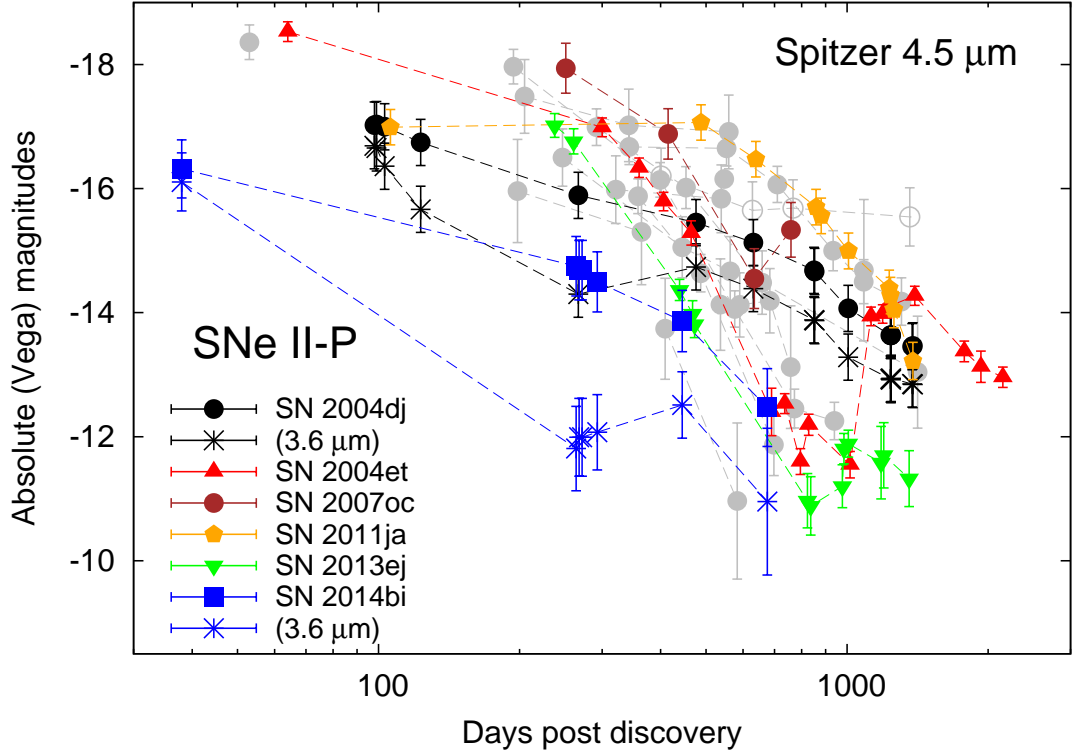}
\caption{4.5 $\mu$m absolute magnitudes of Type II-P explosions. Colored symbols denote objects where mid-IR rebrightening occurred. Filled and empty symbols denote SNe whose absolute magnitudes were determined with or without image subtraction, respectively. In two cases (SNe 2004dj and 2014bi), re-brightening can be only observed at 3.6 $\mu$m (see details in text), which curves are also shown (marked with asterisks). All data are shown in Table \ref{tab:absmag_all}.}
\label{fig:absmag_IIP}
\end{figure}

Fig. \ref{fig:absmag_IIP} plots the mid-IR absolute magnitudes of Type IIP SNe, which show a relatively homogeneous mid-IR evolution. Theoretical models suggest that the ejecta of most of Type II-P SNe may form dust between $\sim$300-600 days due to the slow expansion velocities and high densities. Only a few SNe show evidence for a rebrightening in the mid-IR between $\sim$300-600 days: SNe~2004dj \citep{Szalai11,Meikle11}, 2011ja \citep{Andrews16,Tinyanont16}, and 2014bi \citep{Tinyanont16}. This unexpectedly low rate may be influenced by the poor sampling of the other observed Type IIP SNe. Furthermore, while both SNe~2004dj and 2014bi show the rebrightening effect at 3.6 $\mu$m \,\citep[the first object even at 5.8 and 8.0 \micron, see][]{Szalai11}, it is not detectable at 4.5 \micron \,(there is a linear flux decline instead). \citet{Szalai11} suggest that additional flux at 4.5 $\mu$m arises from the 1$-$0 vibrational band of CO at 4.65 $\mu$m \citep[see][]{Kotak05} during the declining phase, but disappears at $\sim$500d \citep{Szalai11,Szalai13}, thereby making 4.5 \micron~light-curves difficult to interpret for SNe II-P.

Two other Type IIP SNe, 2004et \citep{Kotak09,Fabbri11} and 2007oc \citep{Szalai13}, as well as Type IIP/IIL SN~2013ej \citep{Tinyanont16,Mauerhan17} also show mid-IR rebrightening, but it occurred between $\sim$700-1000 days. This rebrightening is detected at both 3.6 and 4.5 $\mu$m (at least in the cases of SNe 2004et and 2007oc; SN~2013ej becomes undetectable at 3.6 $\mu$m after $\sim$800 days). The above papers suggest this rebrightening is due to new dust forming in the CDS behind the reverse shock and not within the ejecta.

\subsubsection{Type IIn SNe}\label{res_ev_IIn}

\begin{figure}
\includegraphics[width=15cm]{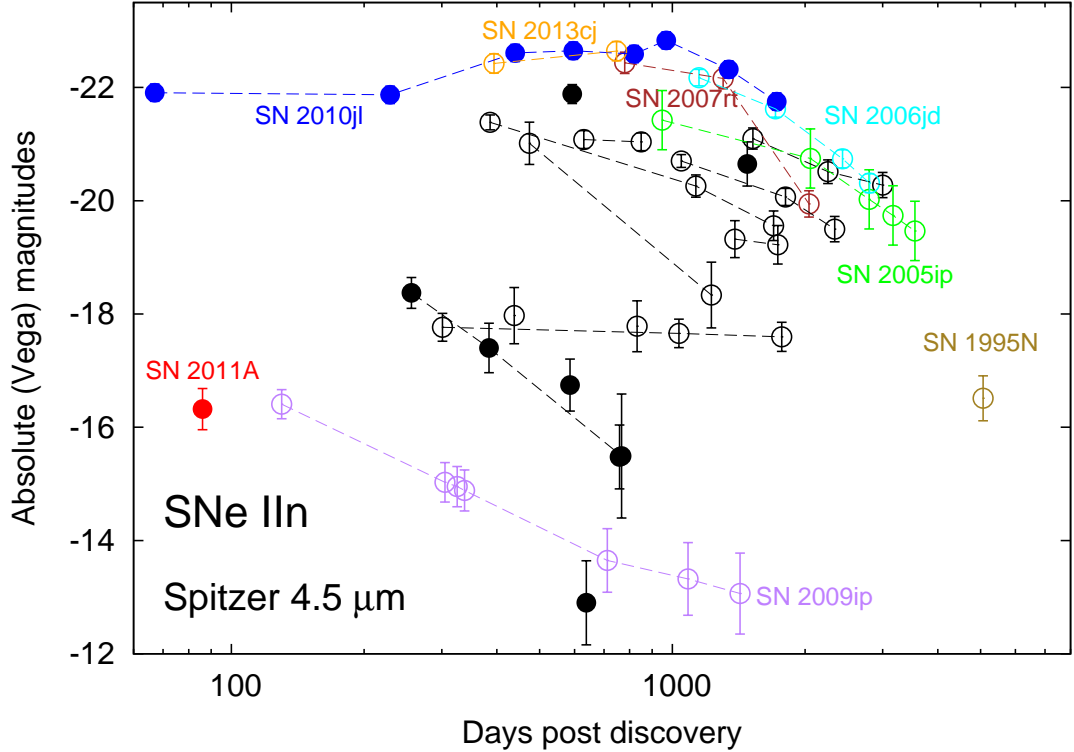}
\caption{Mid-IR evolution of the studied Type IIn explosions.  Highlighted objects are marked with colored symbols (see details in text). Filled and empty symbols denote SNe whose absolute magnitudes were determined with or without image subtraction, respectively. In the case of SN~2009ip, epochs are defined relative to the large outburst occurred in 2012. All the data are shown in Table \ref{tab:absmag_all}.}
\label{fig:absmag_IIn}
\end{figure}

Fig. \ref{fig:absmag_IIn} plots the mid-IR absolute magnitudes of SNe IIn.  For most SNe IIn, \citet{Fox11,Fox13} show that the mid-IR radiation arises from pre-existing dust, which is radiatively heated by optical emission generated by ongoing interaction between the forward shock and CSM. While many SNe IIn show early evidence for CSM interaction (e.g., strong emission in H$\alpha$ / X-ray / radio),  only a hand-full of {\it Spitzer} observations exist in the first few months post-explosion. SNe 2009ip \citep[][and this work]{Fraser15} and 2011A were faint mid-IR sources in the first months, but both of these objects are considered low-luminosity Type IIn events/impostors (see the analyses of e.g. \citealt{Pastorello13}, \citealt{Fraser13}, \citealt{Mauerhan13}, and \citealt{Margutti14}, and \citealt{deJaeger15}). By contrast, SN~2010jl was extremely bright in mid-IR during the first year \citep{Andrews11a,Fox13,Fransson14, Williams15}. The origin of the mid-IR excess has been debated, but is likely a combination of both newly formed and pre-existing dust \citep{Gall14,Fransson14}.

The mid-IR evolution of SNe IIn is heterogeneous. While many SNe IIn remain bright for year post-explosion, the decline rates are not always the same.  Furthermore, many SNe IIn are not even detected \citep[see Fig. \ref{fig:stat_II}, as well as][]{Fox11}. These differences likely correspond to the extent of pre-SN mass-loss, but may also suggest different geometries, shock velocities, and progenitors.
\vspace{3mm}

\begin{figure}
\includegraphics[width=15cm]{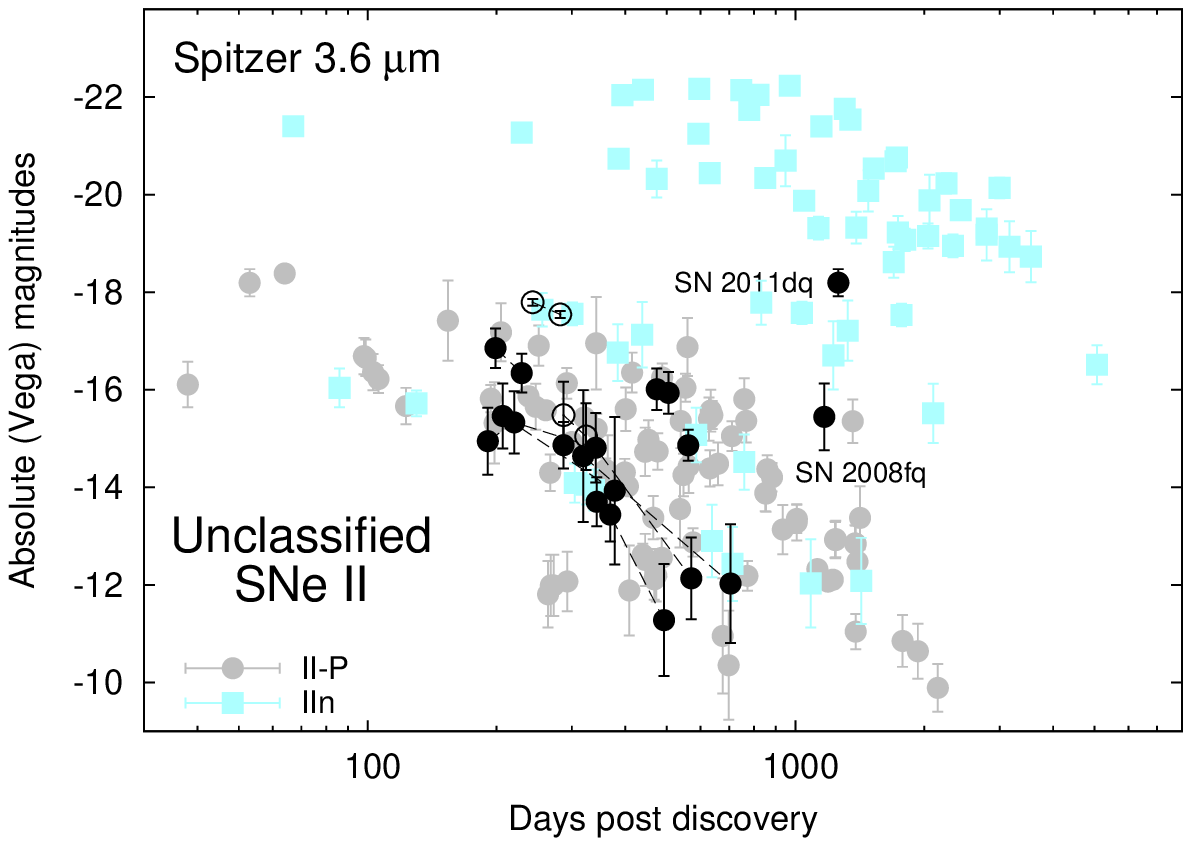}
\caption{Mid-IR evolution of the unclassified Type II SNe in our sample (black symbols) compared to that of known SNe II-P and IIn (gray circles and rectangles, respectively). Filled and empty symbols denote SNe whose absolute magnitudes were determined with or without image subtraction, respectively (absolute magnitudes shown here are calculated from 3.6 $\mu$m fluxes, because several objects have been observed only at this wavelength). Values and sources of data are shown in Table \ref{tab:absmag_all}.}
\label{fig:absmag_II}
\end{figure}

Fig. \ref{fig:absmag} shows that Type II-P and Type IIn SNe have quite distinct late-time mid-IR evolution. This dichotomy serves as a useful classification method for several unclassified targets in our sample (see Table \ref{tab:stat}).  Most of these sources are likely Type IIP, except SNe 2005kd, 2008fq, and 2011dq, which may SNe IIn, as it is shown in Fig. \ref{fig:absmag_II}.

%
%
%
%

\subsection{SED fittings: limitations, methods, consequences}\label{res_sed}

Mid-IR SEDs of SNe span the peak of the thermal emission from warm dust and can place useful constraints on the dust properties \citep[see e.g.][]{Kotak09,Szalai11,Szalai13}. In most of our sample, these fits are limited to only two photometry points. Further challenges exist. During the first several months after explosion, a hot component arising from an optically thick gas in the innermost part of the ejecta may affect the continuum emission at these wavelengths. Moreover, the line emission by CO at 4.65 $\mu$m (described in Section \ref{res_ev_IIP}) may also contribute a significant flux at 4.5 $\mu$m (although this effect has only been observed in some Type II-P SNe before $\sim$500 days after explosion). Most of our sample with previously unpublished {\it Spitzer} data lack the multi-wavelength data that can improve these fits. We also note that while the Galactic extinction (typically at a level below $E(B-V)$=0.1, see Table \ref{tab:sn}) is practically negligible at mid-IR wavelengths, the host galaxy extinction can be more important for both thermonuclear \citep[see e.g.][]{Phillips13} and core-collapse SNe \citep[see e.g.][and references therein]{Jencson18}. Unfortunately, regarding most of the studied SNe, we have no information about the host extinction. As a simple estimation \citep[based on the results of][]{Xue16}, an extreme value of $E(B-V)_{total}$=1.0 mag can attenuate the measured flux by $\sim$20\% at 3.6 or 4.5 $\mu$m.

We illustrate our fitting process using data from the SN IIn 2011A and SN II-P 2014cx. Both of these objects were observed by {\it Spitzer} within 3 months after explosion (at +86 and +53 days, respectively). In the case of SN~2011A, contemporaneous {\it g'r'i'z'} data can be found in \citet{deJaeger15}, and in the case of SN~2014cx, BVRI and {\it g'r'i'} data obtained at the epoch of {\it Spitzer} observations can be found in \citet{Huang16}. The mid-IR fluxes were transformed to $F_{\lambda}$ values and dereddened using the galactic reddening law parametrized by \citet{Fitzpatrick07} assuming R$_V$ = 3.1 and adopting $E(B-V)$ values listed in Table \ref{tab:sn}.  

\begin{figure*}
\includegraphics[width=8cm]{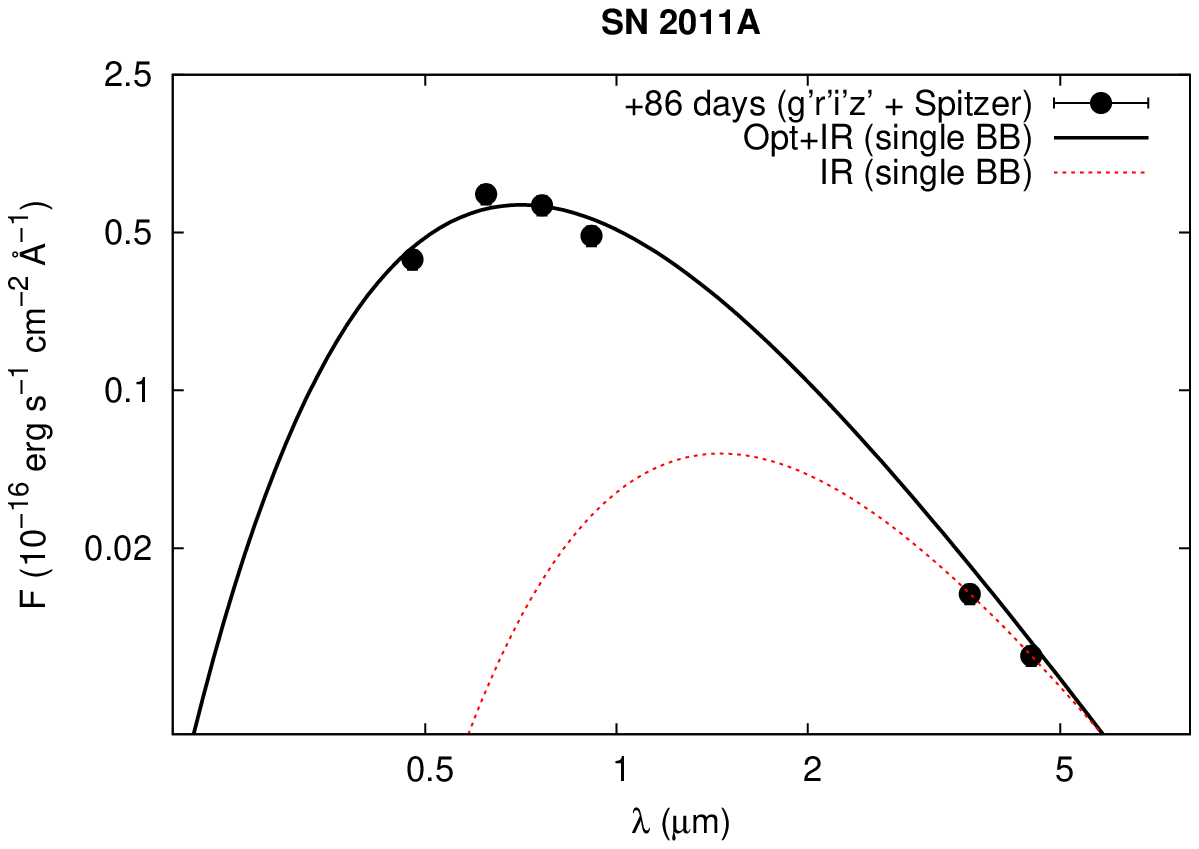}
\includegraphics[width=8cm]{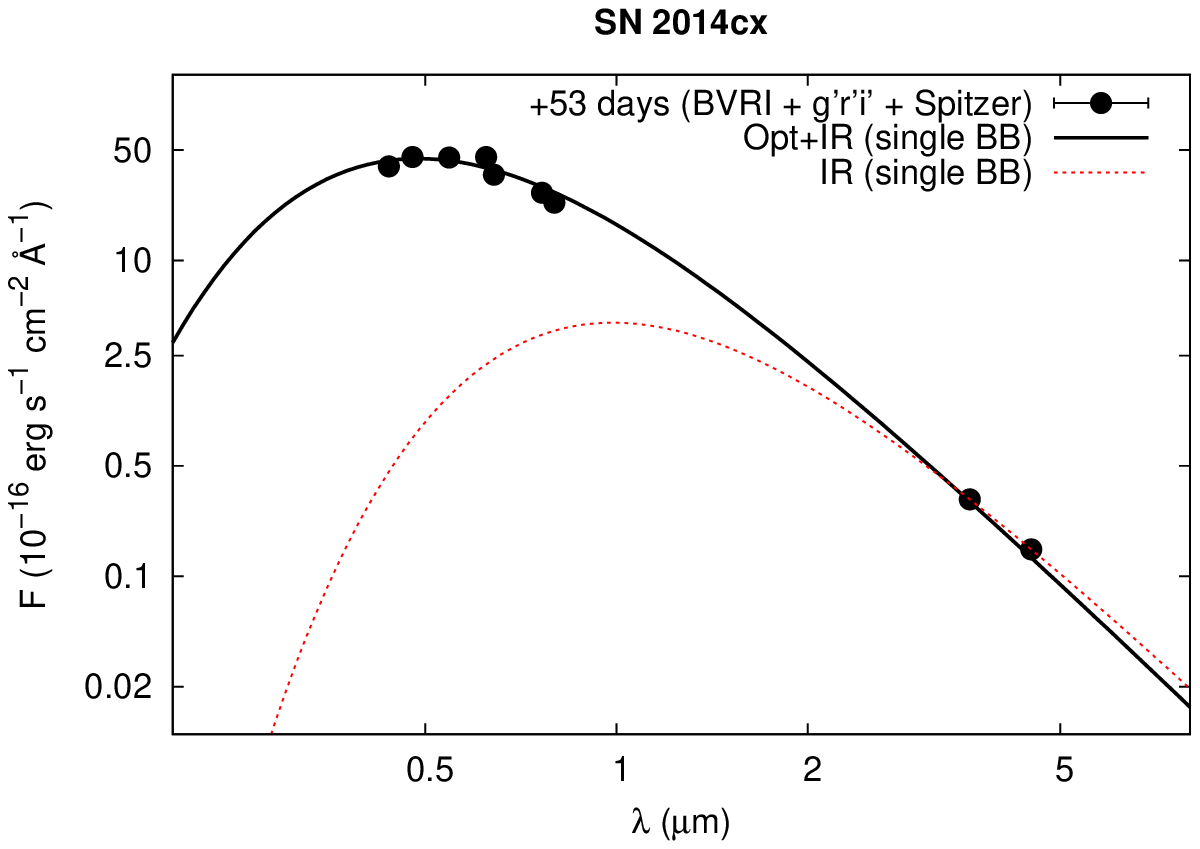}
\caption{Comparison of single blackbody fits to the (left) Type IIn SN 2011A at +86 days, and (right) Type II-P SN 2014cx at +53 days.  Fits are applied to both the combined optical-IR SEDs and only the mid-IR fluxes.}
\label{fig:11A_14cx}
\end{figure*}

Fig. \ref{fig:11A_14cx} shows that single component black bodies (BBs) provide a good fit to the combined optical-IR SEDs of both SNe. Fitting only the mid-IR data yields significantly different parameters (see Table \ref{tab:models}, highlighting the shortcomings of fitting just two data points). Regardless, the SED in this case does not show any evidence for an excess of mid-IR emission above the optically peaked SED.

\begin{figure}
\includegraphics[width=8cm]{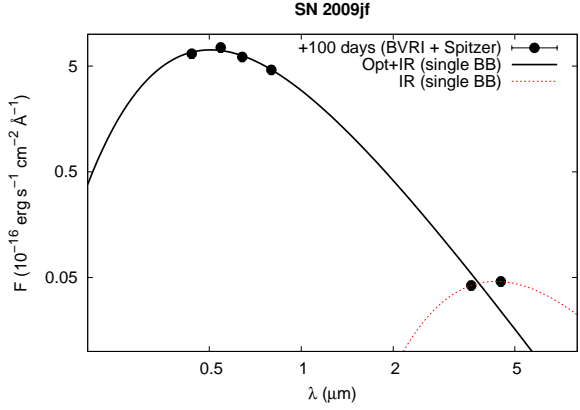}
\caption{Comparison of single blackbody fits to the Type Ib SN 2009jf at +100 days.  A fit is applied to both the combined optical-IR SEDs and only the mid-IR fluxes.}
\label{fig:09jf}
\end{figure}

The Type Ib SN~2009jf also has sufficient data to construct a combined optical-IR SED adopting BVRI measurements from \citet{Sahu11b}. {\it Spitzer} data were obtained at 100 days after explosion, while optical data are from +94 and +105 days. Unlike SNe 2011A and 2014cx, Fig. \ref{fig:09jf} shows that SN 2009jf exhibits an excess at 4.5 $\mu$m, but not at 3.6 $\mu$m. Fitting the mid-IR data with a single BB is difficult.

\begin{figure*}
\includegraphics[width=8cm]{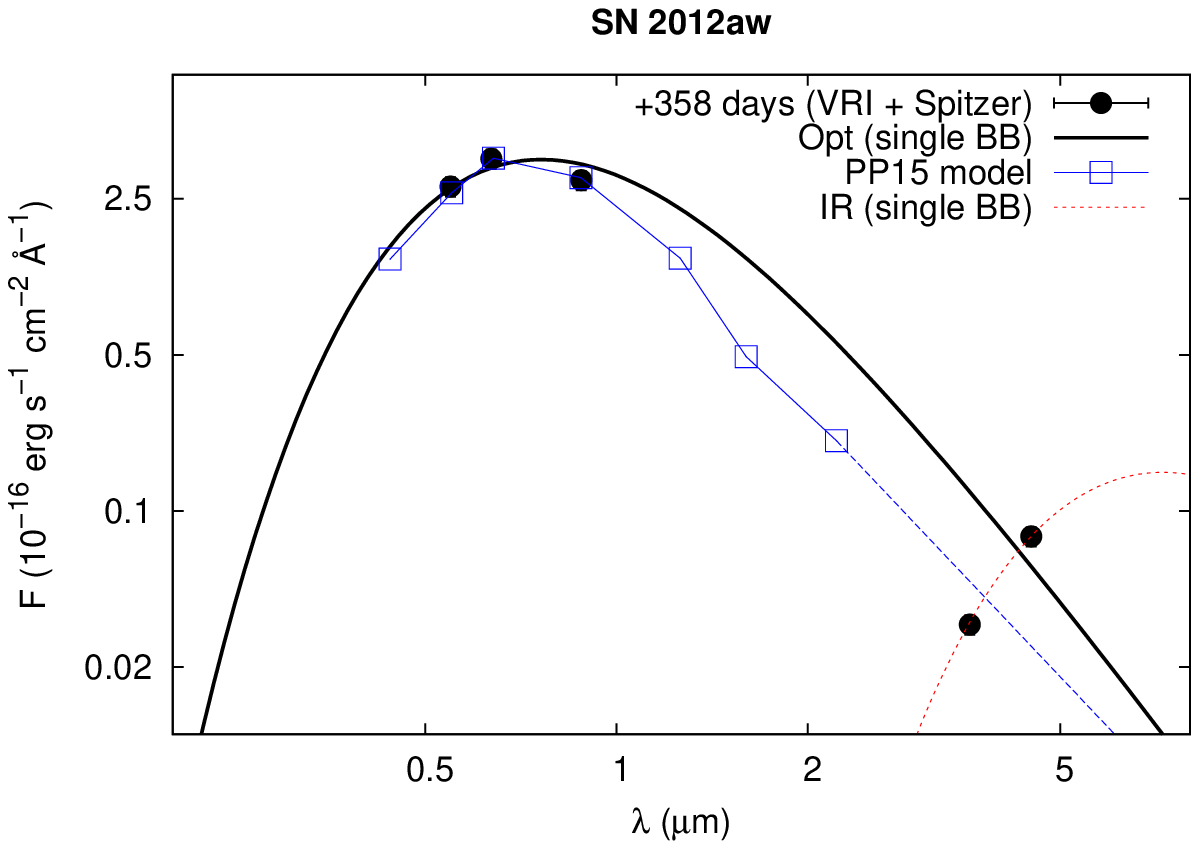}
\includegraphics[width=8cm]{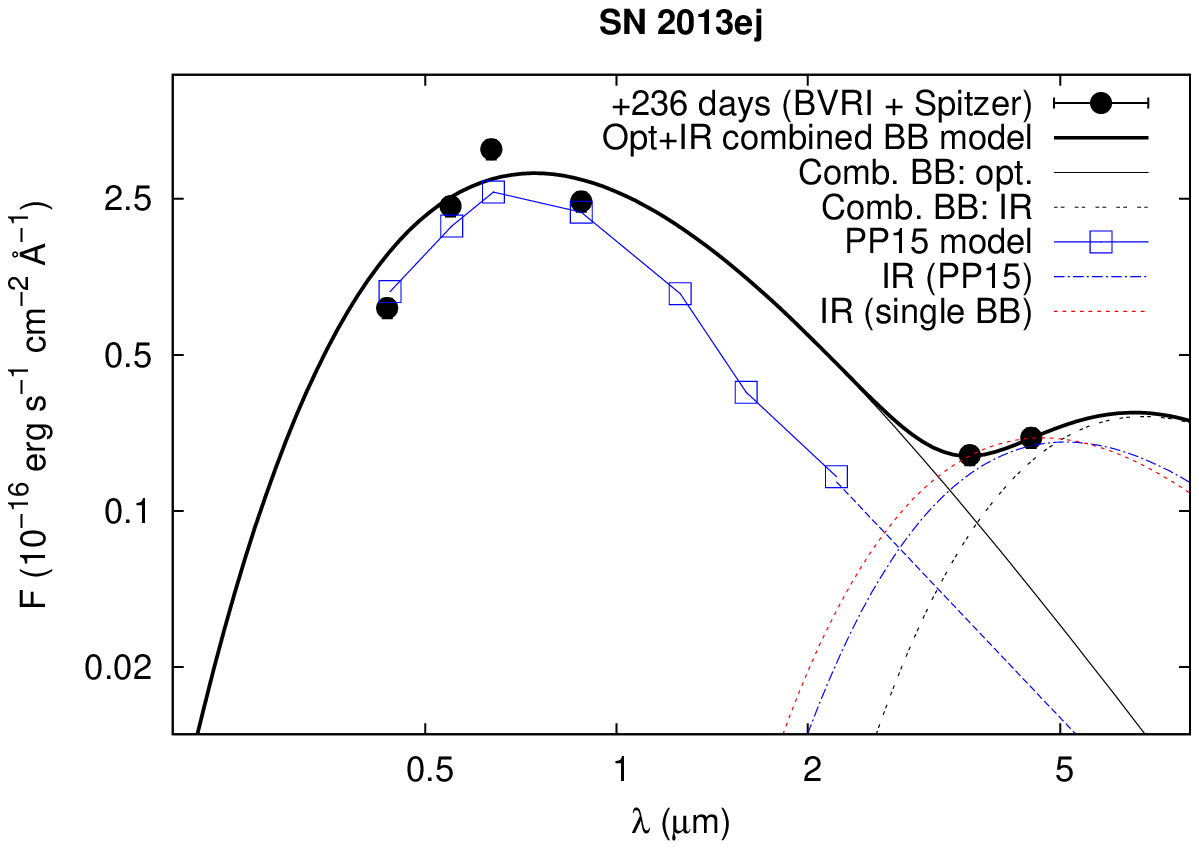}
\caption{Comparison of single blackbody fits to the (left) SN 2012aw at +358 days and (right) SN 2013ej at +236 days.  Fits are applied to both the combined optical-IR SEDs and only the mid-IR fluxes.  SEDs calculated using the PP15 model are marked with open rectangles.}
\label{fig:12aw_13ej}
\end{figure*}

Fig. \ref{fig:12aw_13ej} shows data for the Type II-P SN~2012aw. The earliest {\it Spitzer} observations occur on day 358, and we extrapolate V, R, and I-band data from day $\sim$330 \citep{DallOra14}. The hot component cannot be adequately modeled by a simple BB curve since the optical depth of the continuously expanding ejecta is quite low at this time. Therefore, we applied the global light-curve model of Type II-P SNe \citep[][called hereafter PP15 model]{Pejcha15} to estimate the contribution of the hot component to the mid-IR fluxes. In order to construct the PP15 model SED, we calculated its values at the wavelengths of BVRIJHK filters, while, at longer wavelengths, we used the the Rayleigh$-$Jeans approximation (F$_{\lambda} \propto \lambda^{-4}$). Like SN 2009jf, there is an excess at 4.5 $\mu$m, indicating a warm dust component is present.  Fitting this component is difficult with just two data points and complicated even further by the potential 4.5 $\mu$m line emission in SNe II-P described above.

Finally, Fig. \ref{fig:12aw_13ej} shows a similar analysis for the CSM-interacting Type II-P/II-L SN 2013ej \citep{Leonard13,Bose15,Kumar16,Dhungana16,Chakraborti16,Mauerhan17}. Despite the amount of published data, modeling of the combined (UV)-optical-IR SEDs has not been presented in the literature. Only one epoch (+236d), however, has nearly contemporaneous mid-IR and optical data \citep[][respectively]{Tinyanont16,Bose15}. Like SN 2012aw, we fit the optical data with the PP15 model with parameters given by \citet{Muller17}. We ignore the R-band data in this case, however, given the strong H$\alpha$-emission arising from CSM interaction \citep{Bose15,Huang15,Dhungana16,Mauerhan17}.  The results for each fit are given in Table \ref{tab:models2}.

\begin{table}
\footnotesize
\caption{\label{tab:models} Parameters of single BBs fitted to the optical-IR SEDs of SNe 2011A (IIn), 2014cx (II-P), 2009jf (Ib), and 2012aw (II-P).}
\newcommand\T{\rule{0pt}{3.1ex}}
\newcommand\B{\rule[-1.7ex]{0pt}{0pt}}
\begin{tabular}{c|cc|cc|cc|cc}
\hline
\hline
~ & \multicolumn{2}{c}{SN 2011A} & \multicolumn{2}{c}{SN 2014cx} & \multicolumn{2}{c}{SN 2009jf} & \multicolumn{2}{c}{SN 2012aw}\T \\
~ & \multicolumn{2}{c}{(IIn, +86d)} & \multicolumn{2}{c}{(II-P, +53d)} & \multicolumn{2}{c}{(Ib, +100d)} & \multicolumn{2}{c}{(II-P, +358d)}\B \\
~ & $R$ & $T$ & $R$ & $T$ & $R$ & $T$ & $R$ & $T$\T \\
~ & (10$^{16}$ cm) & (K) & (10$^{16}$ cm) & (K) & (10$^{16}$ cm) & (K) & (10$^{16}$ cm) & (K)\B \\
\hline
Opt. + IR (single BB) & 0.07 & 4100 & 0.13 & 5960 & 0.09 & 5760 & 0.05 & 3810 \T\\\
IR (single BB) & 0.12 & 1990 & 0.23 & 2940 & 1.48 & 670 & 2.79 & 400\B \\
\hline
\end{tabular}
\end{table}

\begin{table*}
\scriptsize
\caption{\label{tab:models2} Parameters of two- and one-component blackbodies fitted to the combined optical-IR SED of the known interacting Type II-P/II-L SN 2013ej, together with dust parameters determined from fitting a simply analytic dust model comparing with previously published results of \citet{Tinyanont16}.}
\newcommand\T{\rule{0pt}{3.1ex}}
\newcommand\B{\rule[-1.7ex]{0pt}{0pt}}
\begin{tabular}{c|cccc|ccc}
\hline
\hline
~ & \multicolumn{7}{c}{SN 2013ej (II-P/II-L, +236d)}\T\B \\
~ & $R_{opt}$ & $T_{opt}$ & $R_{IR}$ & $T_{IR}$ & $T_{dust}$ & $M_{dust}$ & $L_{dust}$\T \\
~ & (10$^{16}$ cm) & (K) & (10$^{16}$ cm) & (K) & (K) & (10$^{-5}$ M$_{\odot}$) & ($10^{6} L_{\odot}$)\B \\
\hline
Two-comp. BBs & 0.05 & 3910 & 3.48 & 430 & 360 & 580 & 5.3\T\\\
Opt. (PP15) + IR BB & -- & -- & 1.51 & 570 & 460 & 98.8 & 3.3 \\
IR (single BB) & -- & -- & 1.25 & 620 & 490 & 69.9 & 2.9 \\
\hline
Tinyanont et al. (2016) - IR & -- & -- & -- & -- & 477 & 75.0 & 2.7 \\
\hline
\end{tabular}
\end{table*}

We performed a similar analysis on the rest of the targets in our sample.   Since we are most interested in late-time emission and want to minimize contributions from the early-time photospheric light-curve, we excluded targets that did not meet certain criteria.  For example, we did not include SNe without late-time observations or only single filter IRAC photometry.  We also exclude ``normal'' Type Ia SNe since their mid-IR photometry do not probe warm dust (but see \citealt{Nozawa11}).

For the SNe we analyze, we follow the method published in a number of papers \citep[see e.g.][]{Meikle07,Fox10,Fox11,Fox16,FF13,Szalai13,Graham17} by assuming a spherically symmetric, optically thin dust shell. We calculate the {\it minimum} shell radius by fitting BBs ($R_{BB}$) to the observed SEDs and, from the radii and the estimated ages, we also constrain the corresponding expansion velocities ($v_{BB}$) by assuming a constant expansion over time (see Table \ref{tab:dustparnew}).  

For comparison, we also fit analytic dust model adopted from \citet{Fox10,Fox11}, assuming only thermal emission of optically thin dust with mass $M_d$, with a particle radius $a$, at a distance $d$ from the observer, thermally emitting at a single equilibrium temperature $T_d$; hence, the flux can be written as

\begin{equation}
F_{\nu} = \frac{M_d B_{\nu}(T_d) \kappa_{\nu}(a)}{d^2} ,   
\end{equation}

\noindent where $B_{\nu}(T_d)$ is the Planck-function, $\kappa_{\nu}$ is the dust mass absorption coefficient as a function of the grain size. We chose pure graphite composition assuming single-sized grains of $a$=0.1 $\mu$m \citep[following][]{Fox10,Fox11}. During the fit, only $T_d$ and $M_d$ are free parameters; $\kappa_{\nu}$ has been determined from Fig. 4 of \citet{Fox10}. In cases of two-point SEDs, we are limited to using one temperature component.



\begin{figure}
\begin{center}
\includegraphics[width=7cm]{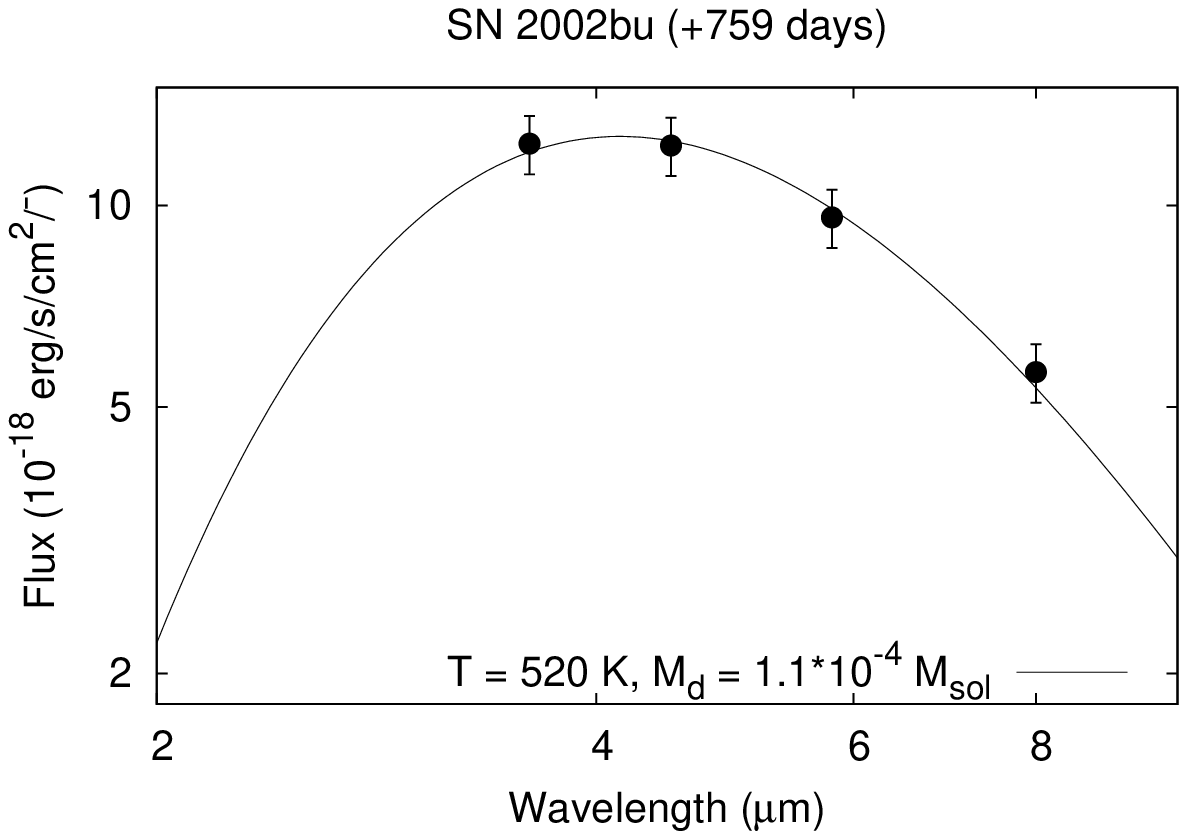}
\includegraphics[width=7cm]{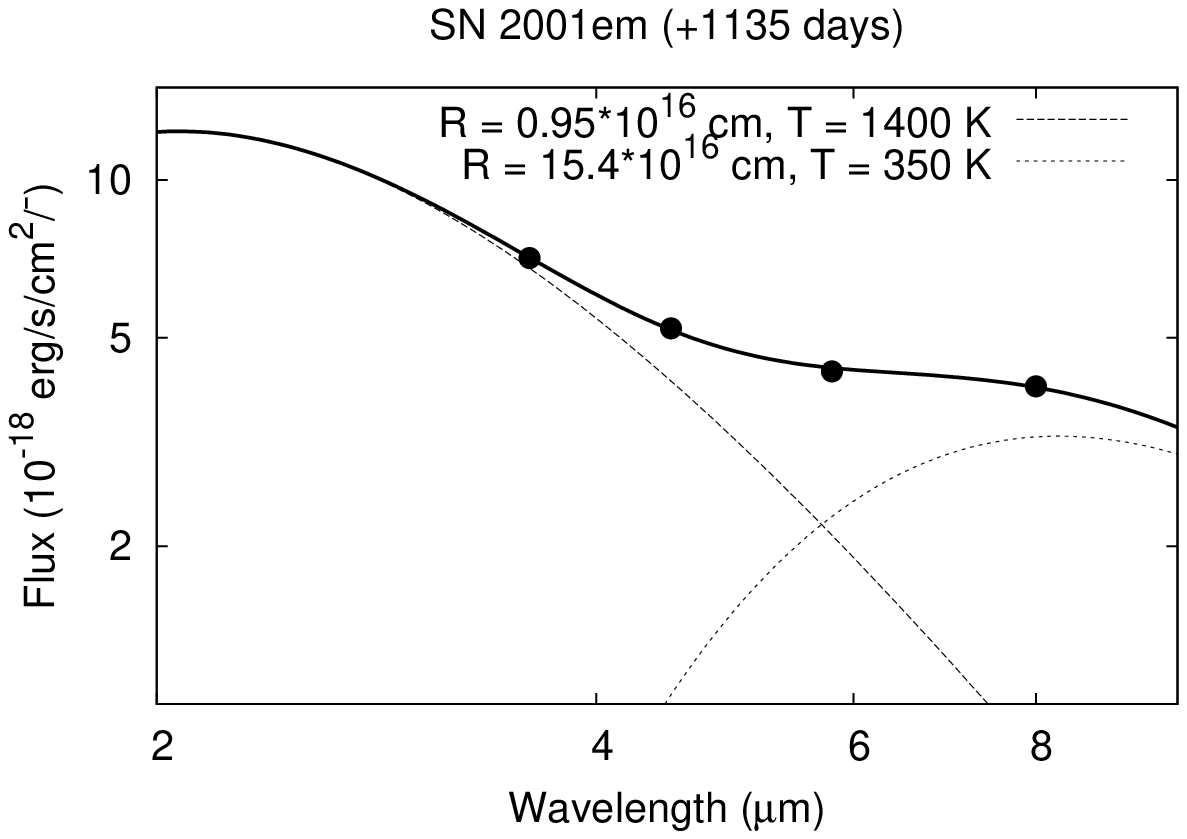}
\caption{{\it Left:} One-component carbonaceous dust model fit to the four-point mid-IR SED of the Type IIn SN 2002bu. {\it Right}: Two-component blackbody model fit to the four-point mid-IR SED of the known interacting Type Ib/c SN 2001em.}
\end{center}
\label{fig:01em}
\end{figure}

Figure \ref{fig:01em} compares the analytical and blackbody fits in two SNe that have data from all four IRAC channels: the Type IIn SN 2002bu and and the Type Ib/c 2001em. SN 2002bu was observed $\sim$2 years post-explosion and can be fit with just a single-component graphite or blackbody dust model. SN~2001em, however, requires a two-component model. If we fit use blackbodies, we get the parameters shown in Figure \ref{fig:01em}. We can compare our results with those of \citet{Chugai06} who constructed a model for the strong late-time
X-ray, radio, and H$\alpha$ emission from SN~2001em and developed a picture in which the SN ejecta collide with a dense massive CS shell.
Our two-component model gives $\sim10^{16}cm$ and $\sim15\times 10^{16}cm$ for the two radii, which is compatible with the estimated size of the single CS shell (r$\sim7 \times 10^{16}cm$) calculated by \citet{Chugai06} from X-ray, radio, and H$\alpha$ data contemporaneous with mid-IR observations. If we change the longer-wavelength blackbody to a graphite dust model, we get $T_{dust}$ = 280K and an upper limit of $M_{dust}\approx$0.2 $M_{\odot}$, which are in a good agreement with the calculations of \citet{Chugai06} who derived -- indirectly -- 300K for dust temperature and 2-3 $M_{\odot}$ for the mass of the CS shell (which gives 0.02-0.03 $M_{\odot}$ dust mass assuming a 0.01 dust-to-gas mass ratio). These results strengthen previous conclusions of CSM interaction with SN 2001em, but further suggest the presence of multiple pre-explosion dust shells.

\begin{figure}
\begin{center}
\includegraphics[width=15cm]{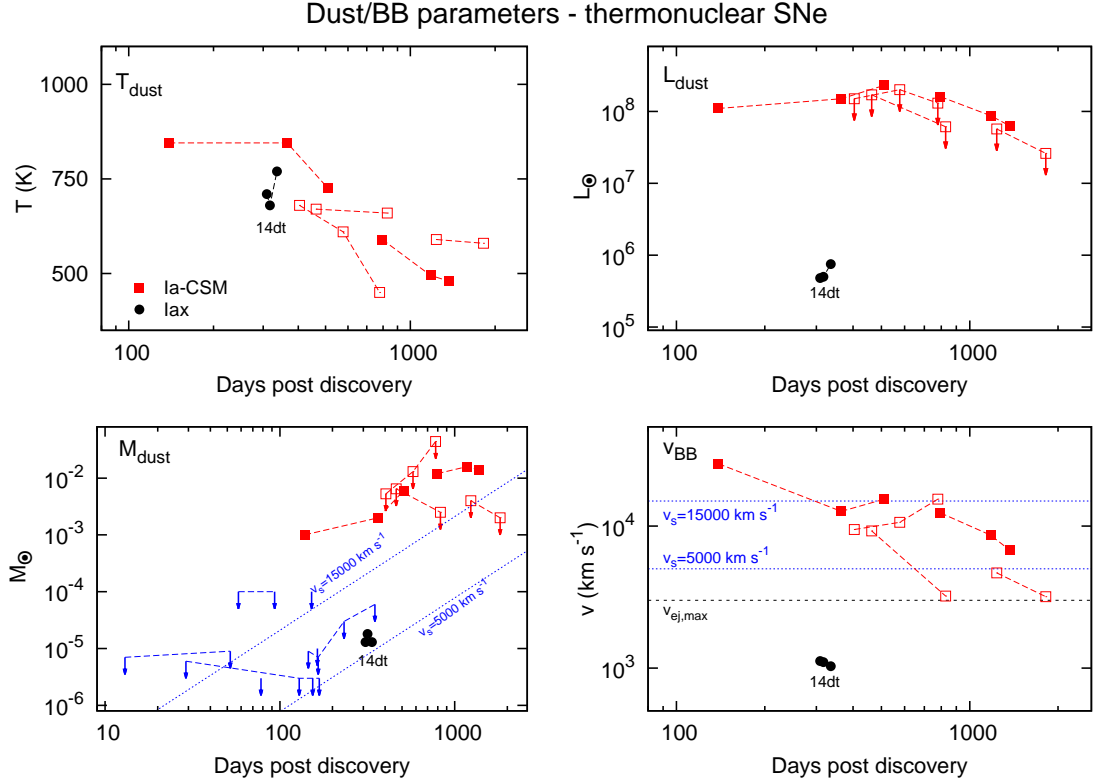}
\caption{Dust parameters (temperatures -- top left, luminosities -- top right, and dust masses -- bottom left) and blackbody velocities (belong to minimum dust radii) of thermonuclear SNe derived from the SED fits. Filled and empty symbols denote SNe whose absolute magnitudes were determined with or without image subtraction, respectively -- in the latter cases, only upper limits can be determined for dust masses and luminosities (marked with arrows on both bottom left and top right panels). Blue arrows denote upper dust mass limits for a group of SNe Ia calculated by \citet{Johansson17}. Dotted lines on the bottom left panel denote theoretical dust masses at shock velocities $v_s$=5000 and 15\,000 km s$^{-1}$ assuming a shock heating scenario (see text for details); at the bottom right panel, the mentioned shock velocities are shown together with an upper limit of late-time ejecta velocities (black dashed line) expected in thermonuclear SNe \citep[$v_{ej,max}$=3000 km s$^{-1}$, based on][]{Silverman13}.}
\label{fig:param_Ia}
\end{center}
\end{figure}

\begin{figure}
\begin{center}
\includegraphics[width=15cm]{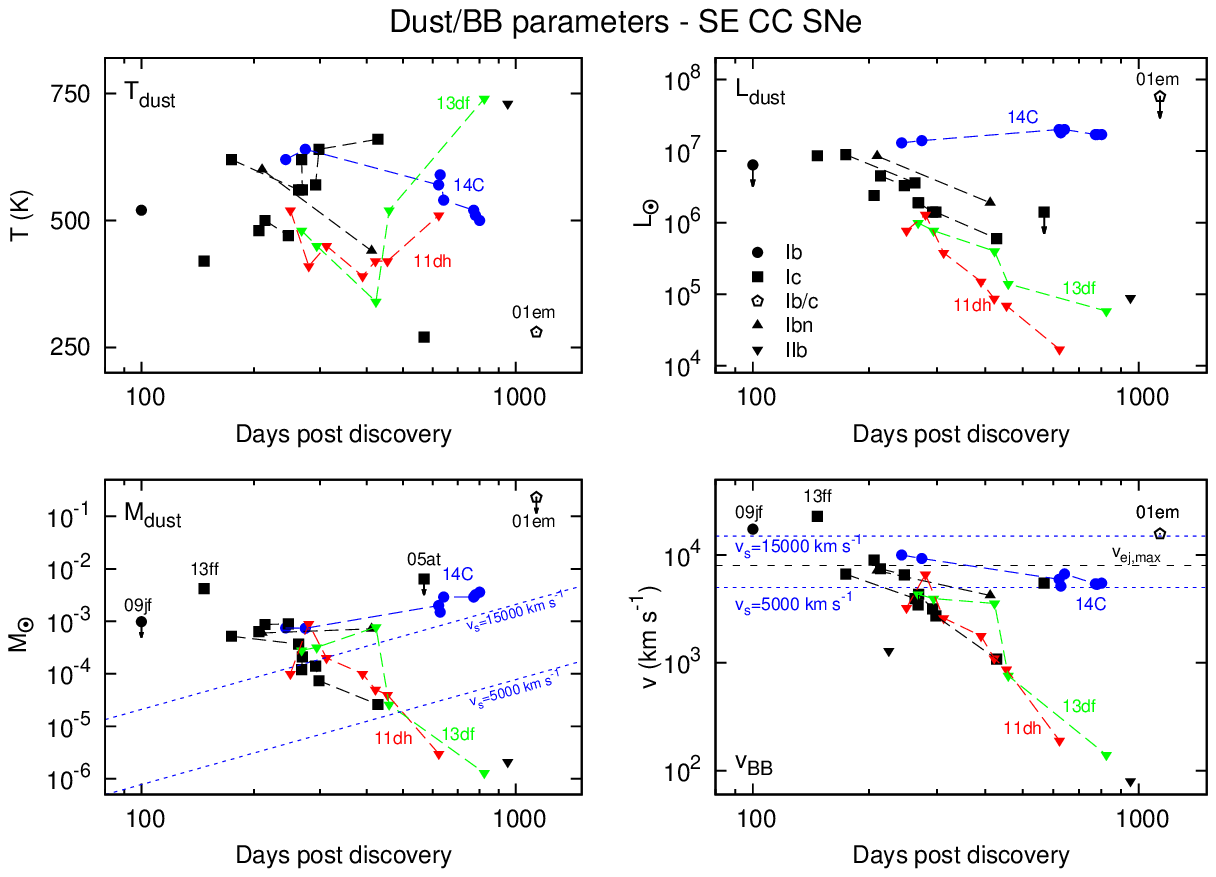}
\caption{Same as Figure \ref{fig:param_Ia}, except in this case for stripped-envelope CC SNe. Filled and empty symbols denote SNe whose absolute magnitudes were determined with or without image subtraction, respectively -- in the latter cases (and in some other ones), only upper limits can be determined for dust masses and luminosities (marked with arrows on both bottom left and top right panels). At the bottom right panel, black dashed line denotes an upper limit of late-time ejecta velocities expected in SE CC SNe \citep[$v_{ej,max}$=8000 km s$^{-1}$, based on][]{Taubenberger06}.}
\label{fig:param_Ibc}
\end{center}
\end{figure}

\begin{figure}
\begin{center}
\includegraphics[width=15cm]{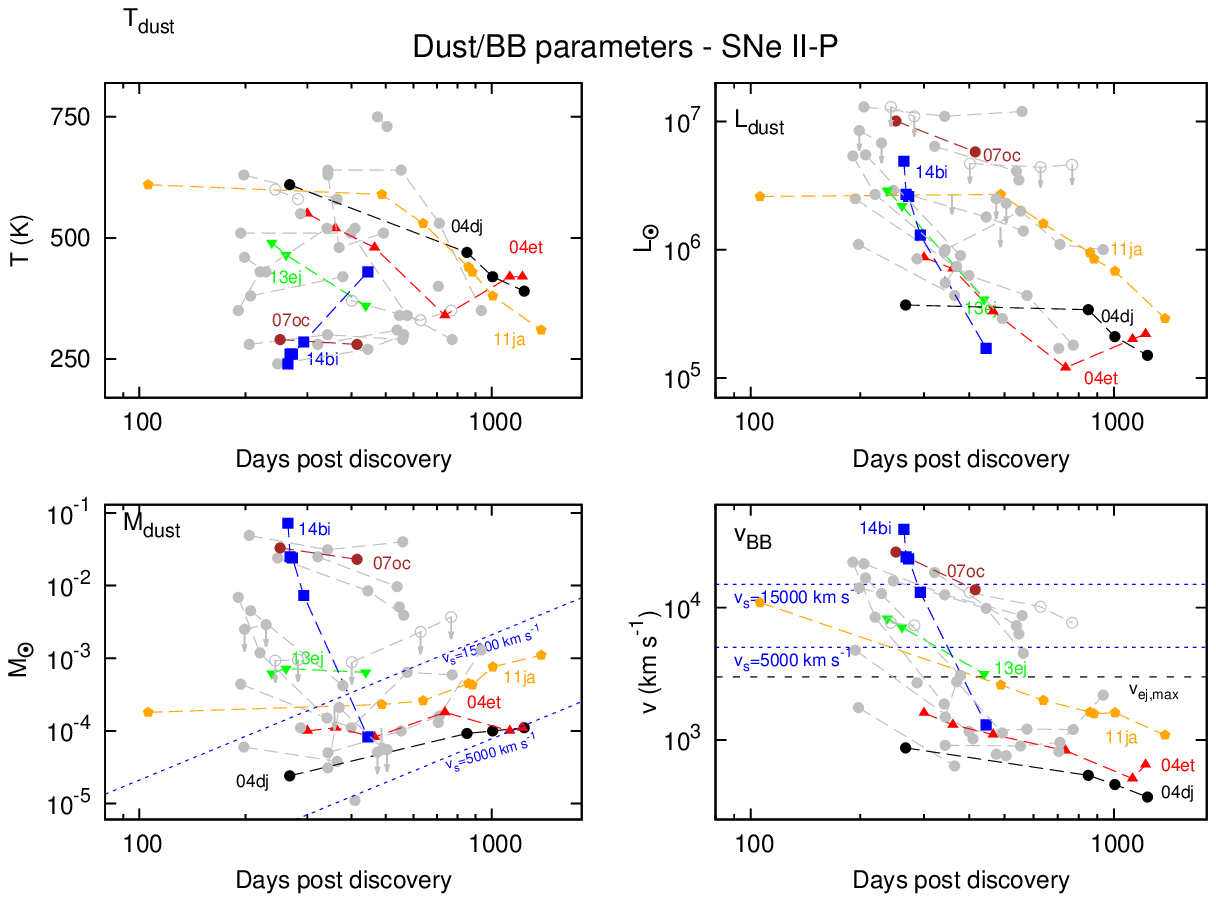}
\caption{ Same as Figure \ref{fig:param_Ia}, except in this case for SNe II-P. Filled and empty symbols denote SNe whose absolute magnitudes were determined with or without image subtraction, respectively -- in the latter cases (and in some other ones), only upper limits can be determined for dust masses and luminosities (marked with arrows on both bottom left and top right panels). At the bottom right panel, black dashed line denotes an upper limit of late-time ejecta velocities expected in SNe II-P \citep[$v_{ej,max}$=3000 km s$^{-1}$, based on][]{Szalai11}.}
\label{fig:param_IIP}
\end{center}
\end{figure}

\begin{figure}
\begin{center}
\includegraphics[width=15cm]{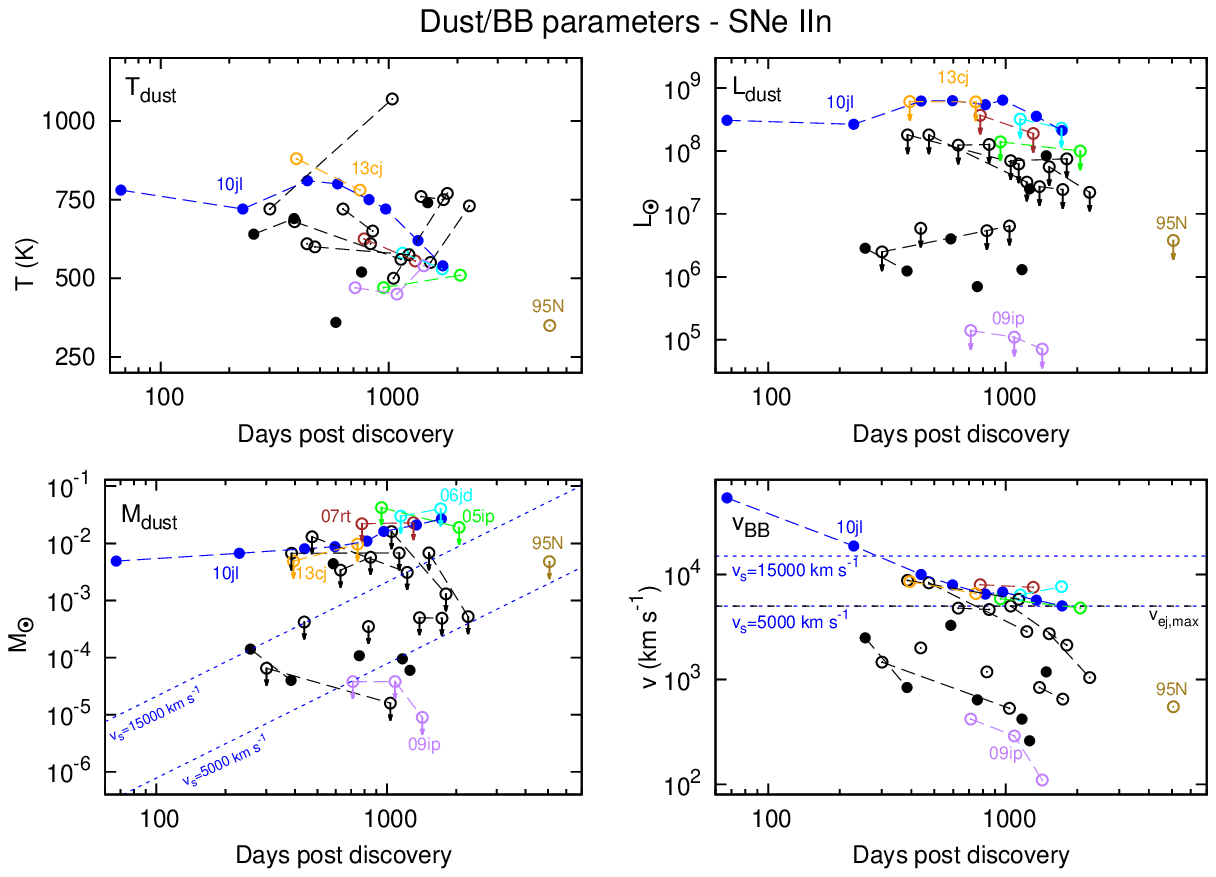}
\caption{Same as Figure \ref{fig:param_Ia}, except in this case for SNe IIn. Filled and empty symbols denote SNe whose absolute magnitudes were determined with or without image subtraction, respectively -- in the latter cases, only upper limits can be determined for dust masses and luminosities (marked with arrows on both bottom left and top right panels). At the bottom right panel, black dashed line denotes an upper limit of late-time ejecta velocities expected in SNe IIn \citep[$v_{ej,max}$=5000 km s$^{-1}$, based on][]{Patat01}.}
\label{fig:param_IIn}
\end{center}
\end{figure}

Table \ref{tab:dustparnew} lists and Figures \ref{fig:param_Ia}-\ref{fig:param_IIn} plot the best fit dust parameters (masses, temperatures, mid-IR luminosities). We fitted SEDs of all listed SNe using the dust model described above in order to generate a comparable set of dust parameters.
Nevertheless, in general, one cannot distinguish between dust compositions with only two IRAC filters. We include the results of silicate fits only in cases with spectroscopic evidence, such as SN 2004et \citep{Kotak09,Fabbri11} or SN 2005af \citep{Szalai13}. In some other cases, the temperature may provide guidance on the dust composition. For example, if $T_{dust} \gtrsim$1400 K, then the carbonaceous dust model makes the most sense since Si grains require lower temperatures for effective condensation \citep[see e.g.][]{Nozawa03}. 

Different dust models (composition, grain size) used in the literature result systematic uncertainties in dust parameters. After comparing our results with previously published ones, we draw the following conclusions: the uncertainties can be as large as $\sim$100-150K in dust temperature (which is also significantly influenced by number of SED points including additional optical and/or near-IR data), while dust masses and dust luminosities can vary within one order of magnitude and within a factor of $\sim$1-2, respectively.
We also note that choosing non-spherical geometry for the dust-forming region, or, assuming clumpy dust formation \citep[see e.g.][]{Meikle07,Ercolano07,Andrews16} may also lead to significantly (an order of magnitude lower/higher) different calculated dust masses.

While the SED fits in Table \ref{tab:dustparnew} have a number of uncertainties, we can still draw some useful conclusions.
The blackbody expansion velocities ($v_{BB}$), shown on the bottom right panels of Figs. \ref{fig:param_Ia}-\ref{fig:param_IIn}, can distinguish between newly-formed and pre-existing dust. In cases where $v_{BB}$ is quite low (several hundreds or a few thousands km s$^{-1}$), the dust likely formed in the ejecta. In these cases, cover many Type II-P and SE CC SNe, we find the estimated temperatures and dust masses ($\sim$10$^{-6}-10^{-2} M_{\odot}$) are in agreement with this scenario \citep[see e.g.][]{Fox11,Fox13,Szalai13,Tinyanont16}. 

In cases where $v_{BB}$ is a bit higher ($\sim$5000-15\,000 km s$^{-1}$), the velocities are consistent with the forward shock, suggesting new dust may be forming in the CDS behind the forward shock. Nevertheless, especially in the cases of SNe IIn or Ia-CSM, or other known interacting objects (e.g. SN Ib 2014C) with large ($>10^{-3} M_{\odot}$) observed dust masses, the presence of pre-existing dust should be invoked for explaining the amount of the observed mid-IR  luminosities. For the distinction between the collisional and radiative heating scenarios, we adopt the method presented in \citet{Fox11} \citep[and also used by e.g.][]{Tinyanont16}. Eq. \ref{eq:one}, assuming a dust to gas ratio of 0.01, gives the mass of dust processed by the forward shock of the SN:

\begin{equation}
\label{eq:one}
M_{\mathrm{d}}(M_{\odot}) \approx 0.0028 \left( \frac{v_{\mathrm{s}}}{15\,000 \,\mathrm{km} \,\mathrm{s}^{-1}} \right)^3 \left( \frac{t}{\mathrm{year}} \right)^2 \left( \frac{a}{\mathrm{\mu m}}\right),
\end{equation}

\noindent where $v_\mathrm{s}$ is the shock velocity, $t$ is the time post explosion, and $a$ is the grain size (assumed to be 0.1 $\mu$m). The calculated dust masses -- using $v_\mathrm{s}$ = 5\,000 km s$^{-1}$ and 15\,000 km s$^{-1}$ for the shock velocities assumed to be constant -- appear as straight lines on the bottom left panels of Figs. \ref{fig:param_Ia}-\ref{fig:param_IIn}. A large fraction of Type IIn and other strongly interacting SNe show much larger dust masses than what can be expected even at $v_\mathrm{s}$ = 15\,000 km s$^{-1}$; in these cases, radiative heating by the photons may emerge from the ongoing CSM interaction.


Finally, in cases where $v_{BB}$ is very high (over $\sim$15\,000 km s$^{-1}$), the dust is likely located beyond the forward shock, suggesting the dust is pre-existing at the time of the explosion and radiatively heated. Such high velocities can be seen mainly in early-time ($<$1yr) observations, found e.g. in the cases of Type IIn SN~2010jl \citep{Gall14,Fransson14} or Type II-P SN~2014bi \citep{Tinyanont16}.
In these cases, another possible source of mid-IR emission may be an IR echo in which the dust shell is heated by the peak luminosity of the SN \citep[e.g.][]{Bode80,Dwek83,Sugerman03,Meikle06}. At later epochs, however, this possibility probably can be ruled out \citep[see details in][]{Fox11}.

 
\section{Conclusion}\label{conc}

Here we presented a comprehensive study of all SNe (discovered before 2015) observed with {\it Spitzer}/IRAC, both targeted and untargeted. In total, we increased the published sample of {\it Spitzer} SNe by a factor of $\sim$5$\times$ (from $\sim$200 to $\sim$1100), including nearly $\sim$2$\times$~more detections ($\sim$70 to $\sim$120).

We carry out a thorough photometric analysis of the entire SN sample, including all previously published data. In general, we find good agreements with the published values ($\lesssim$10\% difference in fluxes), except in some cases that were captured in a very faint phase and/or with a complex sky background (however, for this reason, the uncertainties of their original fluxes are also implicitly large).

The results include both a detailed analysis of specific targets with unique behavior and a statistical analysis of the mid-IR evolution of the different SN subclasses. For each detection, we fit both black bodies and simple analytic dust models. Modeling the SEDs (even in cases with just two photometry points) can disentangle the dust origin and heating mechanism, and, in some cases, to determine the main physical parameters of the assumed dust. Large dust masses ($\gtrsim$10$^{-3} M_{\odot}$) are observed primarily in Type IIn and other strongly interacting SNe. The associated $v_{BB}$ is quite high in most of these cases, again consistent with pre-existing, radiatively heated grains.

The large data set allows us to draw some broad conclusions; nevertheless, we note that these are based on studying a quite heterogeneous sample with usually 1-2 epochs of data per object. In general, each subclass tends to fill its own region of phase space. Amongst thermonuclear explosions (looking over the late-time mid-IR data of several hundred objects, finding mostly non-detections), we see that i) SNe Ia-CSM may be rare indeed, and ii) only a very limited number of ``intermediate'' cases with moderately strong CSM interaction may exist (suggested by a $\sim$8-10 mag gap in late-time mid-IR brightness of strongly interacting and slightly- or nondetected objects). Secondly, in the heterogeneous group of stripped-envelope CC SNe, the length of mid-IR light-curve seems to correlate with the assumed size of the progenitor (the larger the progenitor was, the longer the mid-IR light-curve is, from Type eIIb SNe to Type cIIb and Type Ib/c ones); however, this finding is based on a not-so-large sample of objects. Finally, Type IIn SNe may remain bright for several years post-explosion or may fade more quickly.


Although this study has significantly enlarged the sample sizes for all subclasses, however, the cadence is quite under-sampled both spectrally and temporally.  Future observations with the {\it James Webb Space Telescope} (JWST) offer the sensitivity and spectral capabilities necessary to further constrain the dust geometry, mass, temperature, and composition.

\begin{acknowledgements}
We thank our anonymous referee for his/her valuable comments.
This work is part of the project ``Transient Astrophysical Objects'' GINOP-2-3-2-15-2016-00033 of the National Research, Development and Innovation Office (NKFIH), Hungary, funded by the European Union, and is also supported by the New National Excellence Program (UNKP-17-2, UNKP-17-4) of the Ministry of Human Capacities of Hungary.
TS has received funding from the Hungarian NKFIH/OTKA PD-112325 Grant.
OP is currently supported by award PRIMUS/SCI/17 from Charles University. TM was supported in part by the Ministry of Economy, Development, and Tourism’s Millennium Science Initiative through grant IC120009, awarded to the Millennium Institute of Astrophysics, MAS. TM thanks the LSSTC Data Science Fellowship Program, his time as a Fellow has benefited this work. TM was funded by the CONICYT PFCHA/DOCTORADO BECAS CHILE/2017 - 72180113.

This research has made use of the NASA/IPAC Infrared Science Archive, which is operated by the Jet Propulsion Laboratory, California Institute of Technology, under contract with the National Aeronautics and Space Administration; the NASA/IPAC Extragalactic Database (NED), which is operated by the Jet Propulsion Laboratory, California Institute of Technology, under contract with the National Aeronautics and Space Administration; and the SIMBAD database, operated at CDS, Strasbourg, France. This publication makes use of data products from the Two Micron All Sky Survey, which is a joint project of the University of Massachusetts and the Infrared Processing and Analysis Center/California Institute of Technology, funded by the National Aeronautics and Space Administration and the National Science Foundation.
We acknowledge the availability of NASA ADS services.
\end{acknowledgements}

\software{IRAF, HOTPANTS}

\appendix

\section{Basic data and mid-IR photometry of the studied SNe}
\setcounter{table}{0} 
\renewcommand{\thetable}{\Alph{section}\arabic{table}}

\startlongtable
\begin{deluxetable}{lcccccccc}
\tabletypesize{\scriptsize}
\tablecaption{\label{tab:sn} Basic parameters of the studied SNe with positive {\it Spitzer} detection}
\tablehead{
\colhead{Object} & \colhead{Type} & \colhead{Discovery} & \colhead{Host} & \colhead{RA (J2000)} & \colhead{Dec (J2000)} & \colhead{$d$} & \colhead{E(B-V)} & \colhead{Refe-} \\
\colhead{} & \colhead{} & \colhead{(MJD)} & \colhead{} & \colhead{} & \colhead{} & \colhead{(Mpc)} & \colhead{} & \colhead{rences}
}
\startdata
SN 1993J & IIb & 49074 & NGC 3031 & 09:55:24.77 & $+$69:01:13.7 & 3.6 $\pm$ 0.2 & 0.07 & 1 \\
SN 1995N & IIn & 49842 & MCG-02-38-17 & 14:49:28.29 & $-$10:10:14.0 & 24.0 $\pm$ 4.0 & 0.10 & 2 \\
$^*$SN 2001em & Ib/c & 52172 & UGC 11794 & 21:42:23.60 & $+$12:29:50.3 & 71.6 $\pm$ 0.7 & 0.10 & 3, 4 \\
SN 2001gd & IIb & 52237 & NGC 5033 & 13:13:23.89 & $+$36:38:17.7 & 16.5 $\pm$ 7.0 & 0.01 & 5 \\
SN 2002bu & IIn & 52361 & NGC 4242 & 12:17:37.18 & $+$45:38:47.4 & 5.8 $\pm$ 1.5 & 0.01 & 6 \\
SN 2002ed & II-P & 52482 & NGC 5468 & 14:06:38.22 & $-$05:27:28.0 & 43.8 $\pm$ 7.0 & 0.02 & 7, 8 \\
SN 2002hh & II-P & 52577 & NGC 6946 & 20:34:44.29 & $+$60:07:19.0 & 5.9 $\pm$ 0.4 & 0.30 & 9 \\
SN 2002ic & Ia-CSM & 52591 & A013002+2153 & 01:30:02.55 & $+$21:53:06.9 & 280.0$^{\dagger}$ & 0.05 & 10 \\
\enddata
\end{deluxetable}

\newpage
\startlongtable
\begin{deluxetable}{lccrccccc}
\tablecaption{ Mid-IR fluxes of SNe with previously unreported {\it Spitzer} photometry \label{tab:phot}}
\tabletypesize{\scriptsize}
\tablehead{
\colhead{Object} & \colhead{Type} & \colhead{MJD $-$} & \colhead{Epoch$^{\dagger}$} & \colhead{$F_{\nu,[3.6]}$} & \colhead{$F_{\nu,[4.5]}$} & \colhead{$F_{\nu,[5.8]}$} & \colhead{$F_{\nu,[8.0]}$} & \colhead{SN-targeted} \\
\colhead{} & \colhead{} & \colhead{2\,450\,000} & \colhead{(days)} & \colhead{($\mu$Jy)} & \colhead{($\mu$Jy)} & \colhead{($\mu$Jy)} & \colhead{($\mu$Jy)} & \colhead{observation?}
}
\startdata
$^*${\it SN 2001em}$^a$ & {\it Ib/c} & {\it 53307}$^e$ & {\it 1135} & {\it 303(41)} & {\it 349(36)} & {\it 481(50)} & {\it 859(105)} & {\it Yes} \\
SN 2001gd$^b$ & IIb & 53189$^f$ & 952 & 11(8) & 12(7) & non$^{\ddagger}$ & non$^{\ddagger}$ & No \\
SN 2002ed$^c$ & II-P & 53569$^g$ & 1086 & non$^{\ddagger}$ & 7(7) & non$^{\ddagger}$ & non$^{\ddagger}$ & No \\
$^*${\it SN 2004G}$^a$ & {\it II} & {\it 53047}$^h$ & {\it 24} & {\it 249(28)} & {\it 395(33)} & {\it 265(30)} & {\it 196(29)} & {\it No} \\
SN 2004W$^{a,b}$ & Ia & 53166$^h$ & 134 & non$^{\ddagger}$ & non$^{\ddagger}$ & non$^{\ddagger}$ & 212(29) & No \\
SN 2004bv$^b$ & Ia & 53308$^i$ & 159 & 22(17) & non$^{\ddagger}$ & non$^{\ddagger}$ & non$^{\ddagger}$ & No \\
SN 2005A$^{a,b}$ & Ia & 53387$^i$ & 12 & 154(112) & 64(64) & non$^{\ddagger}$ & non$^{\ddagger}$ & No \\
SN 2005P$^b$ & Iax & 53569$^g$ & 178  & non$^{\ddagger}$ & non$^{\ddagger}$ & non$^{\ddagger}$ & 54(32) & Yes \\
\enddata
\tablecomments{{\it Photometric methods}: $^a$aperture photometry on the original images; $^b$ap. photometry after subtracting a very late-time image + flux differences; $^c$ap. photometry after subtracting a pre-explosion image + flux differences; $^d$ ``Fox+11 method'' (see details in text). $^*$Positive detection based on single-epoch imaging. Data marked with italic denote cases where image subtraction can not be applied (in all these cases, measured fluxes can be handled only as upper limits).
{\it Program IDs}: $^e$3641 (PI van Dyk); $^f$159 (PI Kennicutt); $^g$20256 (PI Meikle); $^h$69 (PI Fazio); $^i$3672 (PI Mazzarella); $^j$80072 (PI Tully); $^k$30292 (PI Meikle); 
$^m$40619 (PI Kotak); $^n$50111 (PI McGaugh); $^o$80089 (PI Sanders); $^p$61060 (PI Sheth); $^q$30496 (PI Fisher); $^r$80254 (PI Stern); $^s$61065 (PI Sheth); $^t$61003 (PI Freedman); $^u$61009 (PI Freedman); $^v$60166 (PI Garnavich); $^w$11053 (PI Fox); $^x$70008 (PI Andrews); $^y$80131 (PI Andrews); $^z$70038 (PI Sanders); 
$^\varepsilon$10139 (PI Fox); $^\zeta$90137 (PI Fox);
$^\alpha$61066 (PI Sheth); $^\beta$61062 (PI Sheth); $^\gamma$10046 (PI Sanders); $^\delta$80023 (PI Fox); $^\varphi$90178 (PI Andrews); $^\eta$80069 (PI Beygu); $^\vartheta$10152 (PI Kasliwal); $^\iota$10136 (PI Kasliwal); $^\kappa$800025 (PI van Zee); $^\mu$80196 (PI Kasliwal); $^\nu$11063 (PI Kasliwal). {\it Additional notes}: $^{\dagger}$ days since discovery; $^{\ddagger}$no point source at the position of the SN.
}
\end{deluxetable}

\newpage
\startlongtable
\begin{deluxetable}{lccrccc}
\tablecaption{3.6 and 4.5 $\mu$m absolute Vega magnitudes of all SNe positively detected by {\it Spitzer}/IRAC. \label{tab:absmag_all}}
\tabletypesize{\scriptsize}
\tablehead{
\colhead{Object} & \colhead{Type} & \colhead{MJD $-$} & \colhead{Epoch} & \multicolumn{2}{c}{Absolute mag.} & \colhead{Source of} \\
\colhead{} & \colhead{} & \colhead{2\,450\,000} & \colhead{(days)} & \colhead{3.6 $\mu$m} & \colhead{4.5 $\mu$m} & \colhead{data$^{\dagger}$}
}
\startdata
SN 1993J & IIb & 52948 & 3874 & $-$11.67(0.10) & $-$12.79(0.10) & 1 \\
~ & ~ & 52976 & 3902 & $-$11.85(0.15) & $-$12.93(0.10) & 1 \\
~ & ~ & 53125 & 4051 & $-$11.48(0.10) & $-$12.55(0.10) & 1 \\
~ & ~ & 53357 & 4283 & $-$11.55(0.10) & $-$12.62(0.10) & 1 \\
~ & ~ & 53497 & 4423 & $-$11.71(0.10) & $-$12.58(0.10) & 1 \\
~ & ~ & 53665 & 4591 & $-$11.40(0.10) & $-$12.47(0.10) & 1 \\
~ & ~ & 54419 & 5345 & $-$11.15(0.10) & $-$12.00(0.10) & 1 \\
~ & ~ & 55736 & 6662 & $-$10.55(0.25) & $-$10.93(0.25) & 1 \\
\enddata
\tablecomments{$^{\dagger}$ Fluxes, distances, epochs and $E(B-V)$ values originate from our current work or are adopted from the papers listed here. $^{\ddagger}$. In the case of SN 2009ip, epochs are given regarding the SN-like outburst observed in 2012. Data marked with italic denote cases where image subtraction can not be applied.
References: 
$^1$\citet{Tinyanont16}; $^2$\citet{VD13};
$^3$\citet{Szczygiel12}; $^4$\citet{Meikle06}; $^5$\citet{FF13}; $^6$\citet{Meikle07}; $^7$\citet{Meikle05}; $^{8}$\citet{Szalai13}; $^{9}$\citet{Szalai11}; $^{10}$\citet{Meikle11}; $^{11}$\citet{Kotak09}; $^{12}$\citet{Fabbri11}; $^{13}$\citet{Kankare14}; $^{14}$\citet{Fox11};
$^{15}$\citet{Fox13}; $^{16}$\citet{Gerardy07}; $^{17}$\citet{Fox10}; $^{18}$\citet{Johansson17}; $^{19}$\citet{Gallagher12}; $^{20}$\citet{Mattila08}; 
$^{21}$\citet{Stritzinger12}; $^{22}$\citet{Kochanek11}; $^{23}$\citet{Andrews11b}; $^{24}$\citet{Andrews10}; $^{25}$\citet{McClelland13}; $^{26}$\citet{Prieto12}; $^{27}$\citet{Fraser15}; $^{28}$\citet{Gandhi13}; $^{29}$\citet{Andrews11a}; $^{30}$\citet{Fransson14}; $^{31}$\citet{Helou13}; $^{32}$\citet{Ergon15}; 
$^{33}$\citet{Andrews16}; $^{34}$\citet{Graham17}; $^{35}$\citet{Corsi14}; 
$^{36}$\citet{Andrews17}; $^{37}$\citet{Mauerhan17}; $^{38}$\citet{Fox16}.
Additional notes. Sources of new data of previously known positive {\it Spitzer} targets (PID): $^a$90174 (PI Fox); $^b$10139 (PI Fox); $^c$11053 (PI Fox); $^d$12099 (PI Fraser); $^e$13053 (PI Kasliwal); $^f$11063 (PI Kasliwal).}
\end{deluxetable}

\begin{longrotatetable}
\begin{deluxetable}{lcc|cc|ccc|l}
\tabletypesize{\scriptsize}
\tablecaption{\label{tab:dustparnew} Dust parameters of the studied SNe (pure graphite dust with a grain size of $a$=0.1 $\mu$m)}
\newcommand\T{\rule{0pt}{3.1ex}}
\newcommand\B{\rule[-1.7ex]{0pt}{0pt}}
\tablehead{
\colhead{Object} & \colhead{Type} & \colhead{Epoch} & \colhead{$R_{BB}$} & \colhead{$v_{BB}$} & \colhead{$T_{dust}$} & \colhead{$M_{dust}$} & \colhead{$L_{dust}$} & \colhead{Notes} \\
\colhead{} & \colhead{} & \colhead{(days)} & \colhead{(10$^{16}$ cm)} & \colhead{(km s$^{-1}$)} & \colhead{(K)} & \colhead{(10$^{-5}$ M$_{\odot}$)} & \colhead{($10^{6} L_{\odot}$)} & \colhead{}
}
\startdata
{\it SN 1995N} & {\it IIn} & {\it 5067} & {\it 2.4} & {\it 550} & {\it 350} & {\it $<$478} & {\it $<$3.8} & {\it Possible Si dust (M=1.4$\times10^{-3}$ M$_{\odot}$, T=410K)}\T\B \\
{\it SN 2001em} & {\it Ib/c} & {\it 1135} & {\it 15.4} & {\it 15\,700} & {\it 280} & {\it $<$22800} & {\it $<$58.4} & {\it + hot comp. (R=10$^{16}$ cm, T=1400K)}\B\T \\
SN 2001gd & IIb & 952 & 0.06 & 80 & 730 & 0.21 & 0.1 & ~\T\B \\
SN 2002bu & IIn & 759 & 0.4 & 640 & 520 & 10.8 & 0.7 & ~\T\B \\
SN 2002ic & Ia$-$CSM & 795 & 8.5 & 12\,370 & 590 & 1200 & 160 & ~\T \\
~ & ~ & 1179 & 8.8 & 8640 & 495 & 1600 & 87.0 & ~ \\
~ & ~ & 1370 & 8.0 & 6760 & 480 & 1400 & 63.0 & ~\B \\
SN 2003gd & II$-$P & 409 & 0.4 & 1020 & 520 & 1.1 & 0.1 & ~\T\B \\
\enddata
\end{deluxetable}
\end{longrotatetable}

\newpage
\section{Comparison of single-epoch Spitzer detections with pre-explosion 2MASS JHK data}
\setcounter{figure}{0} 
\setcounter{table}{0} 
\renewcommand{\thefigure}{\Alph{section}\arabic{figure}}
\renewcommand{\thetable}{\Alph{section}\arabic{table}}

\begin{figure*}[!h]
\begin{center}
\includegraphics[width=9cm]{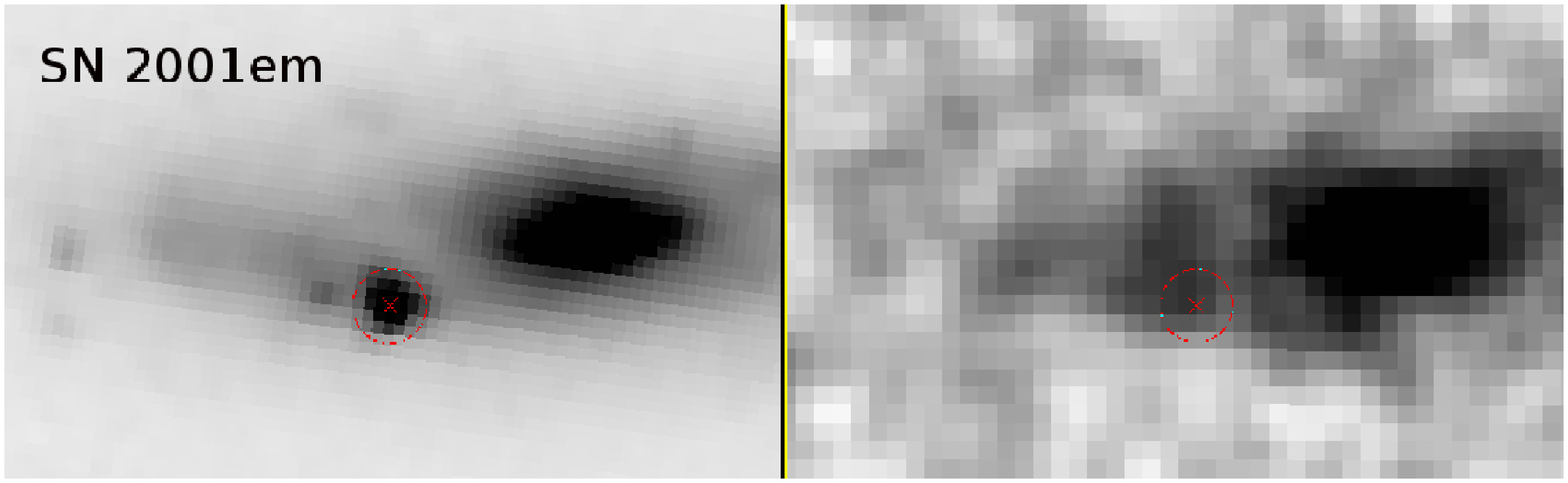}
\includegraphics[width=9cm]{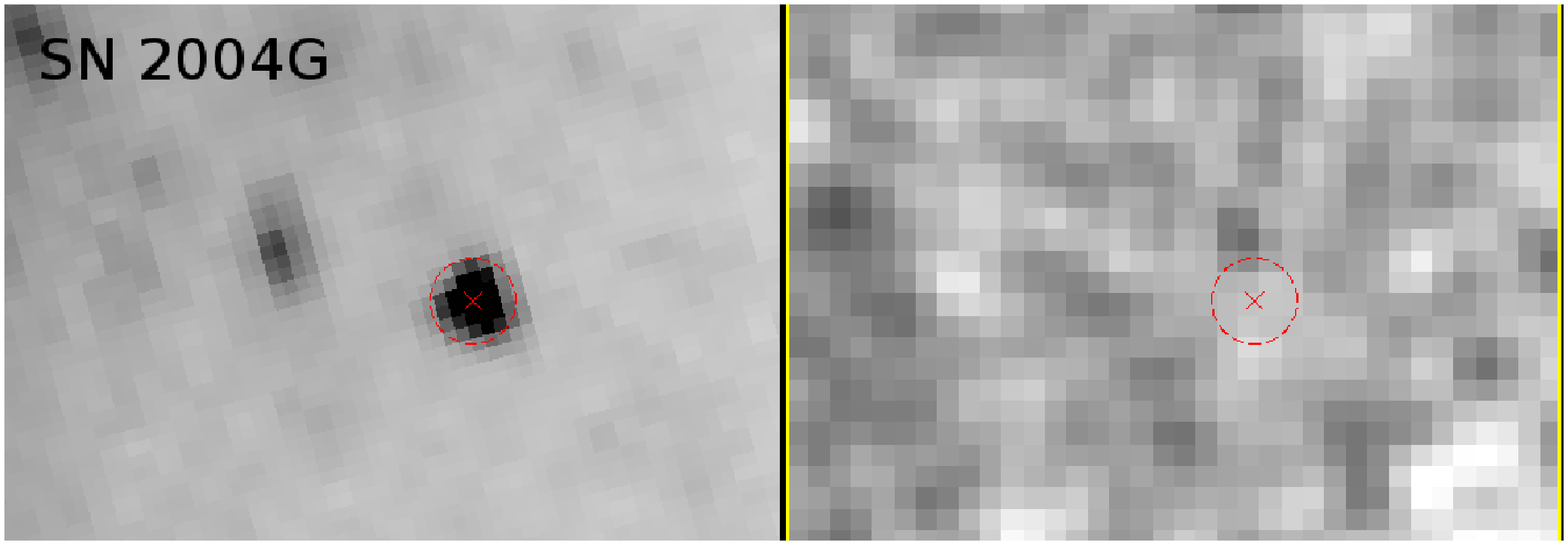}
\includegraphics[width=9cm]{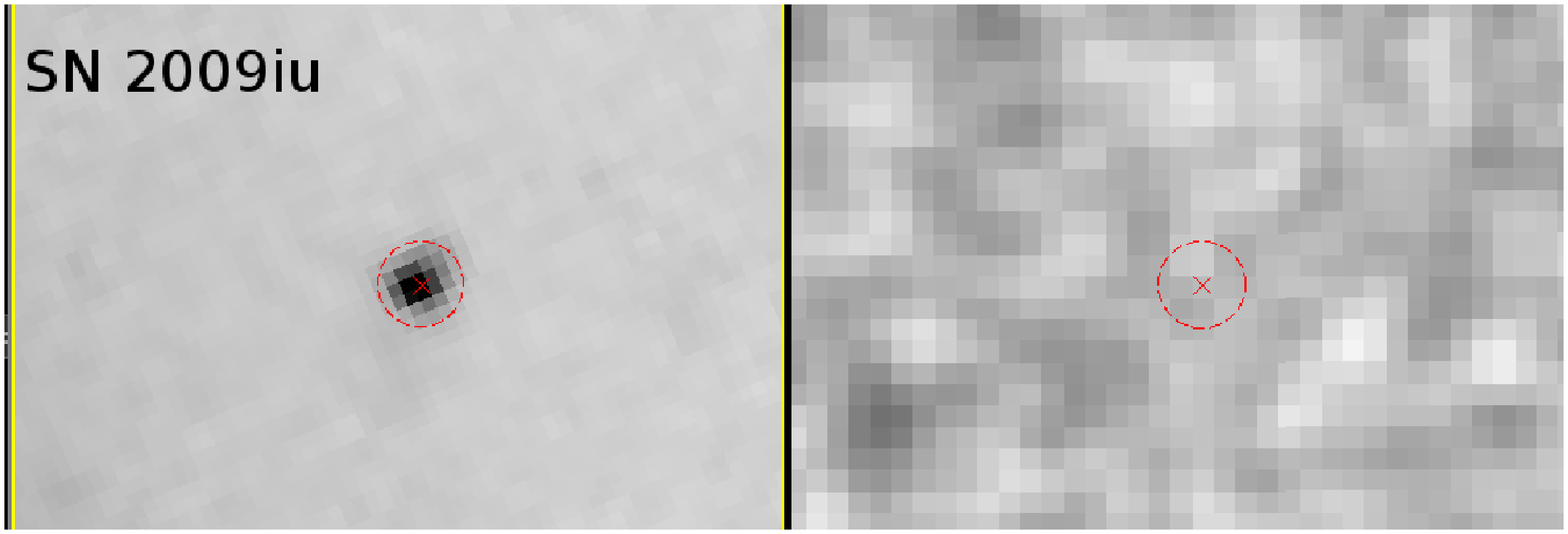}
\includegraphics[width=9cm]{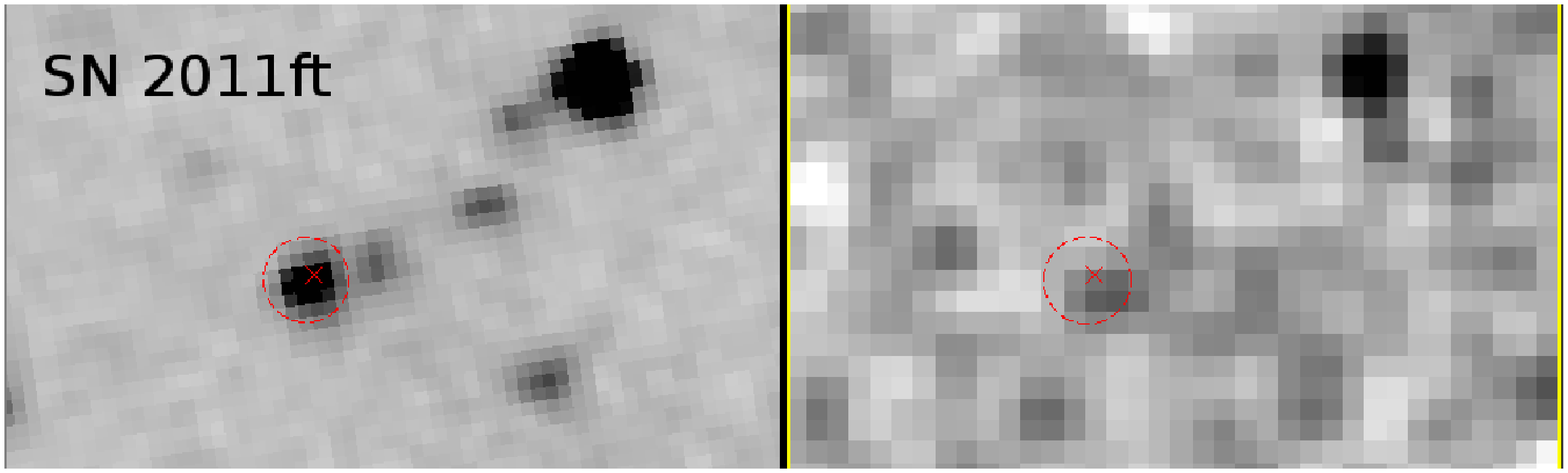}
\includegraphics[width=9cm]{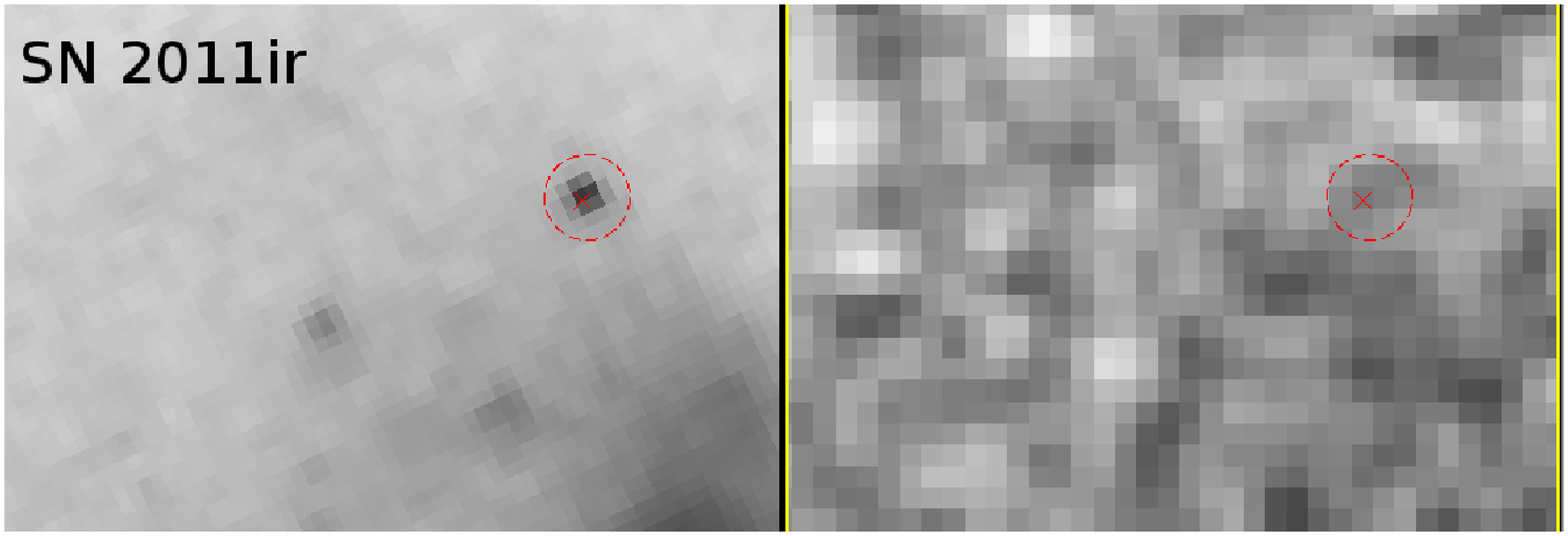}
\includegraphics[width=9cm]{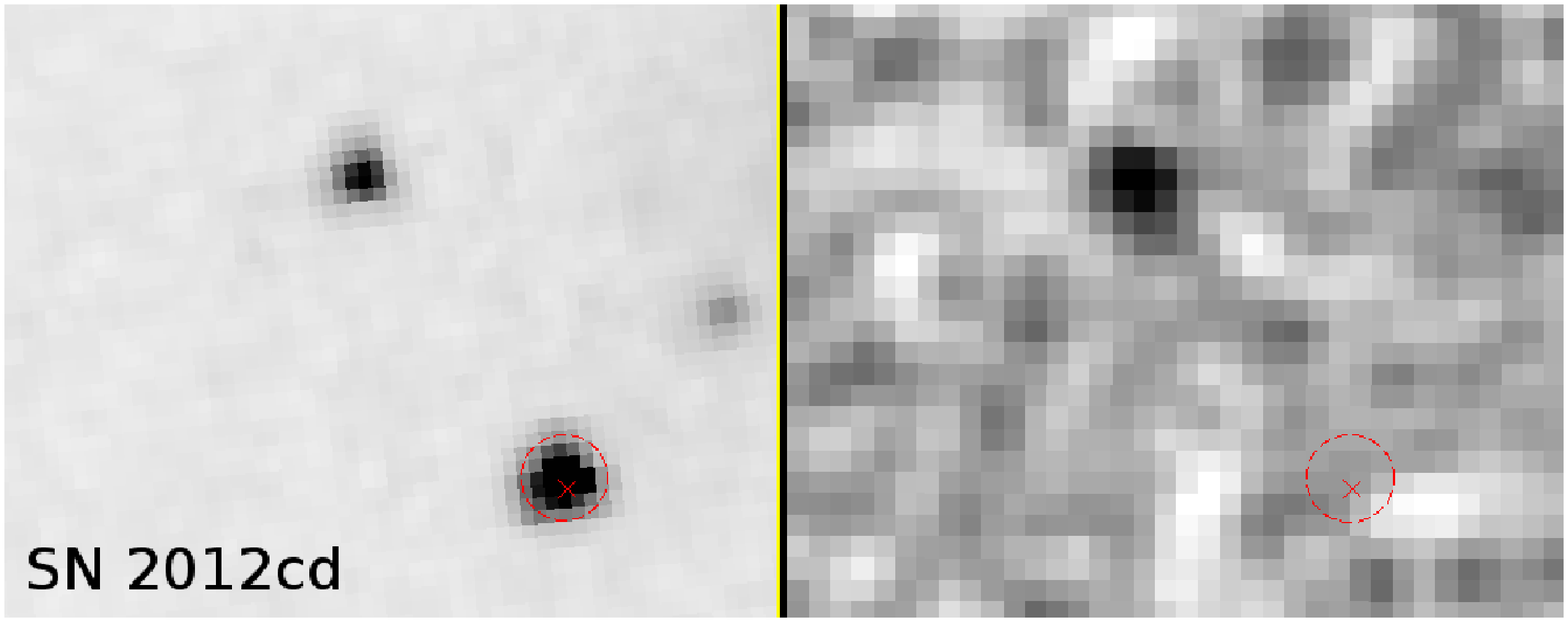}
\includegraphics[width=9cm]{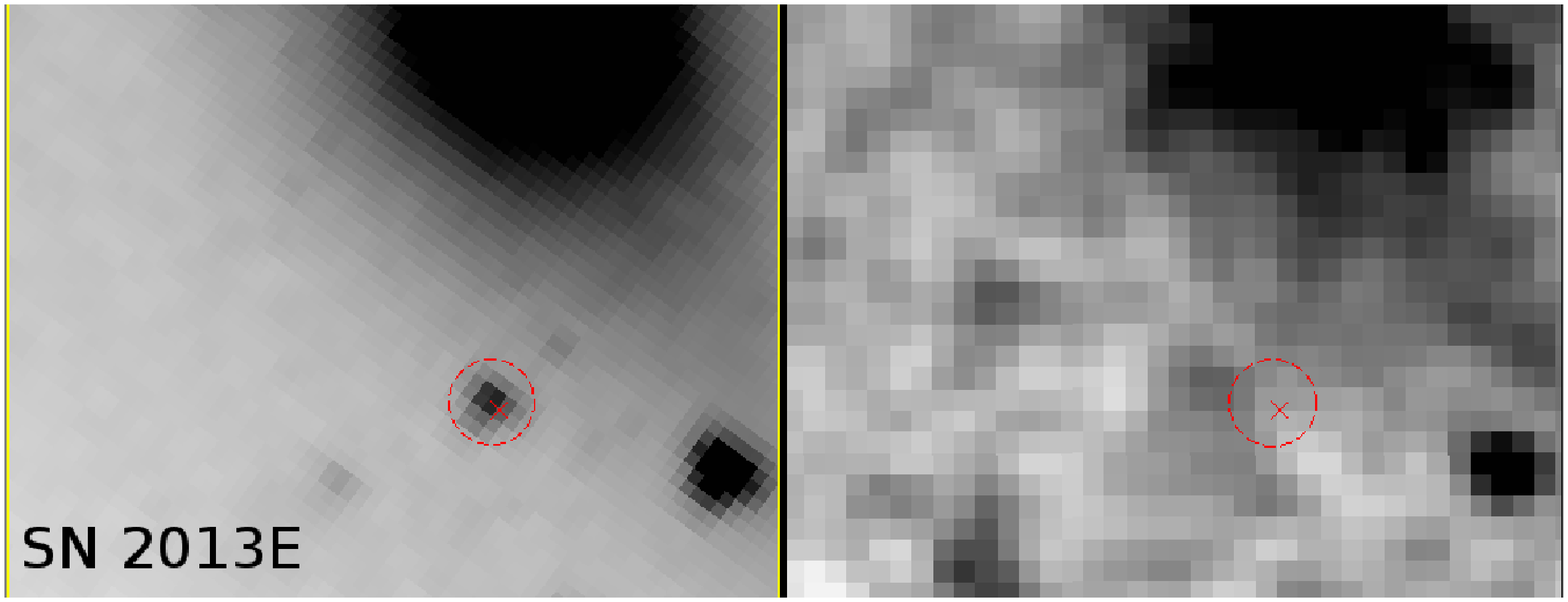}
\caption{Comparison of single-epoch {\it Spitzer}/IRAC 3.6 micron images (left) of SNe (classified as positive detections) with pre-explosion 2MASS K$_s$ images (right). Colors are inverted for better visibility. Red crosses denote the SN coordinates adopted from sources given in the Open Supernova Catalog, while red circles with radii of 2\arcsec show the typical PSF FWHM of point sources on IRAC images centered to the photometric centers of the mid-IR points sources.}
\label{fig:2mass_irac}
\end{center}
\end{figure*}

\newpage
\begin{table}
\footnotesize
\caption{\label{tab:2MASS} Uncertainties of absolute coordinates ($\sigma$RA, $\sigma$DEC), agreement of these coordinates and that of the photometric centers of mid-IR point sources ($\Delta$XY), and pre-explosion 2MASS JHK$_s$ magnitudes (estimating a general $\pm$0.4 magnitude error) regarding positively detected SNe based on single-epoch {\it Spitzer}/IRAC imaging.}
\newcommand\T{\rule{0pt}{3.1ex}}
\newcommand\B{\rule[-1.7ex]{0pt}{0pt}}
\begin{tabular}{c|ccc|c|ccc}
\hline
\hline
Object & $\sigma$RA & $\sigma$DEC & Ref. & $\Delta$XY & J (pre-exp.) & H (pre-exp.) & K$_s$ (pre-exp.) \\
\hline
SN~2001em & $<$0.001\arcsec & $<$0.001\arcsec & 1 & $<$0.6\arcsec & 18.38(0.40) & 17.16(0.40) & 16.58(0.40) \\
SN~2004G & unknown & unknown & -- & $<$0.6\arcsec & 17.05(0.40) & 16.66(0.40) & 16.15(0.40) \\
SN~2009iu & $<$0.1\arcsec & $<$0.1\arcsec & 2 & $<$0.6\arcsec & 18.15(0.40) & 18.50(0.40) & 19.54(0.40) \\
SN~2011ft & unknown & unknown & -- & $<$1.2\arcsec & 20.60(0.40) & 16.95(0.40) & 17.33(0.40) \\
SN~2011ir & unknown & unknown & -- & $<$1.2\arcsec & 19.25(0.40) & 20.03(0.40) & 17.78(0.40) \\
SN~2012cd & unknown & unknown & -- & $<$1.2\arcsec & 20.03(0.40) & 20.29(0.40) & 19.33(0.40) \\
SN~2013E & unknown & unknown & -- & $<$1.2\arcsec & 20.06(0.40) & 20.03(0.40) & 20.23(0.40) \\
\hline
\end{tabular}
\tablecomments{References: (1) \citet{BB05}; (2) \citet{Maza09}}
\end{table}

\begin{figure*}[!h]
\begin{center}
\includegraphics[width=7cm]{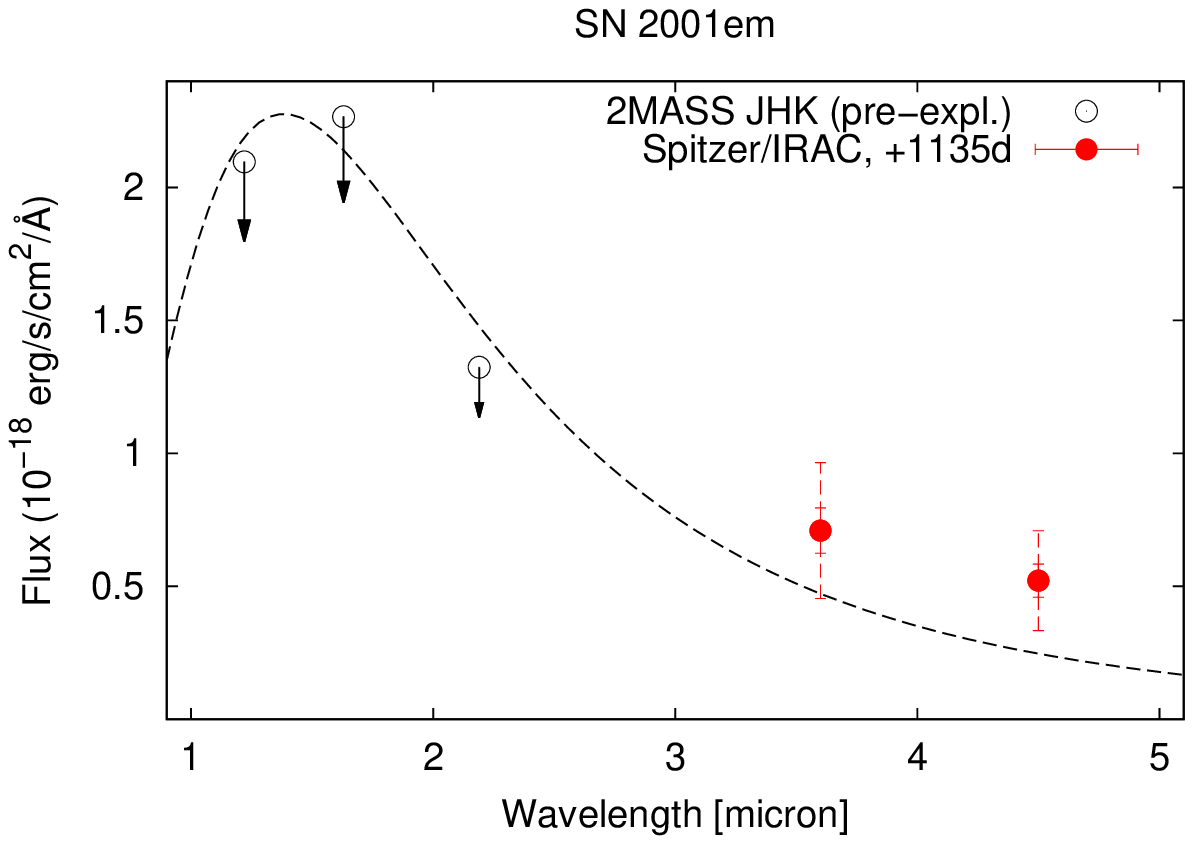}
\includegraphics[width=7cm]{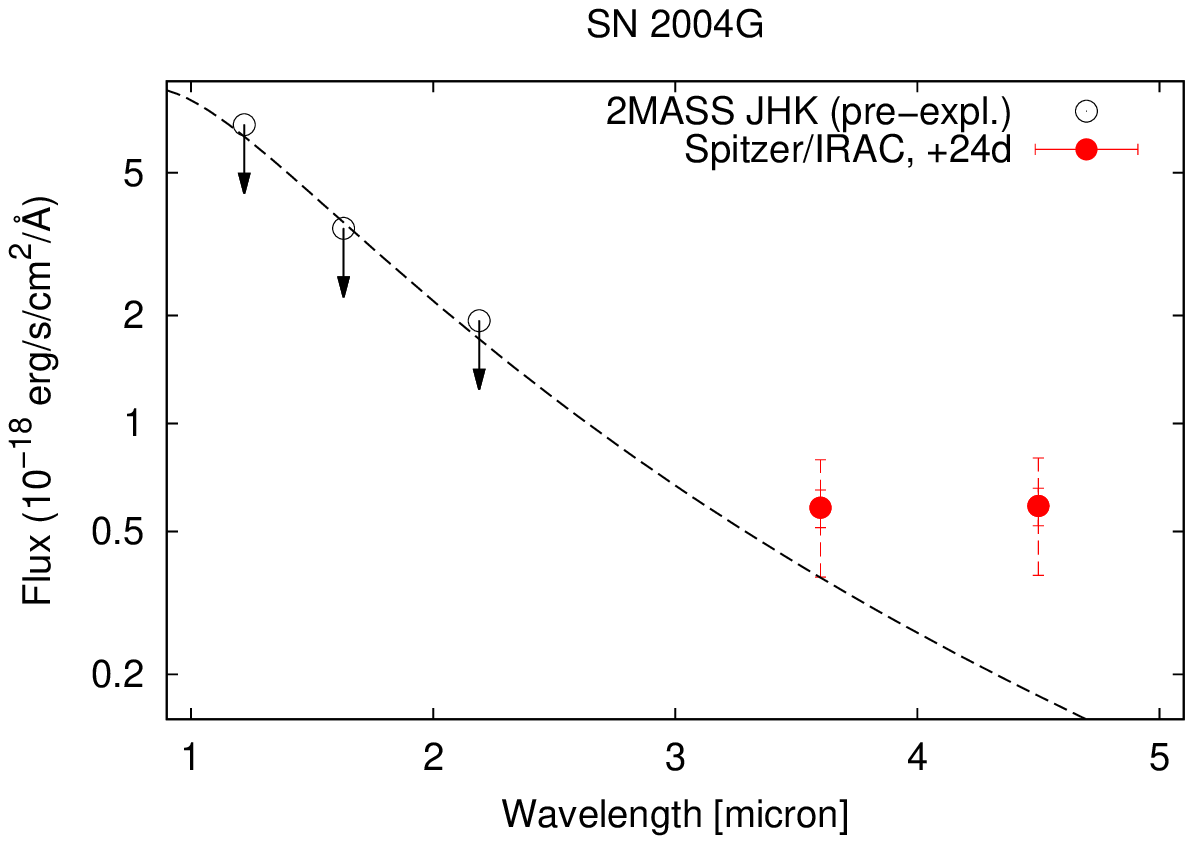}
\includegraphics[width=7cm]{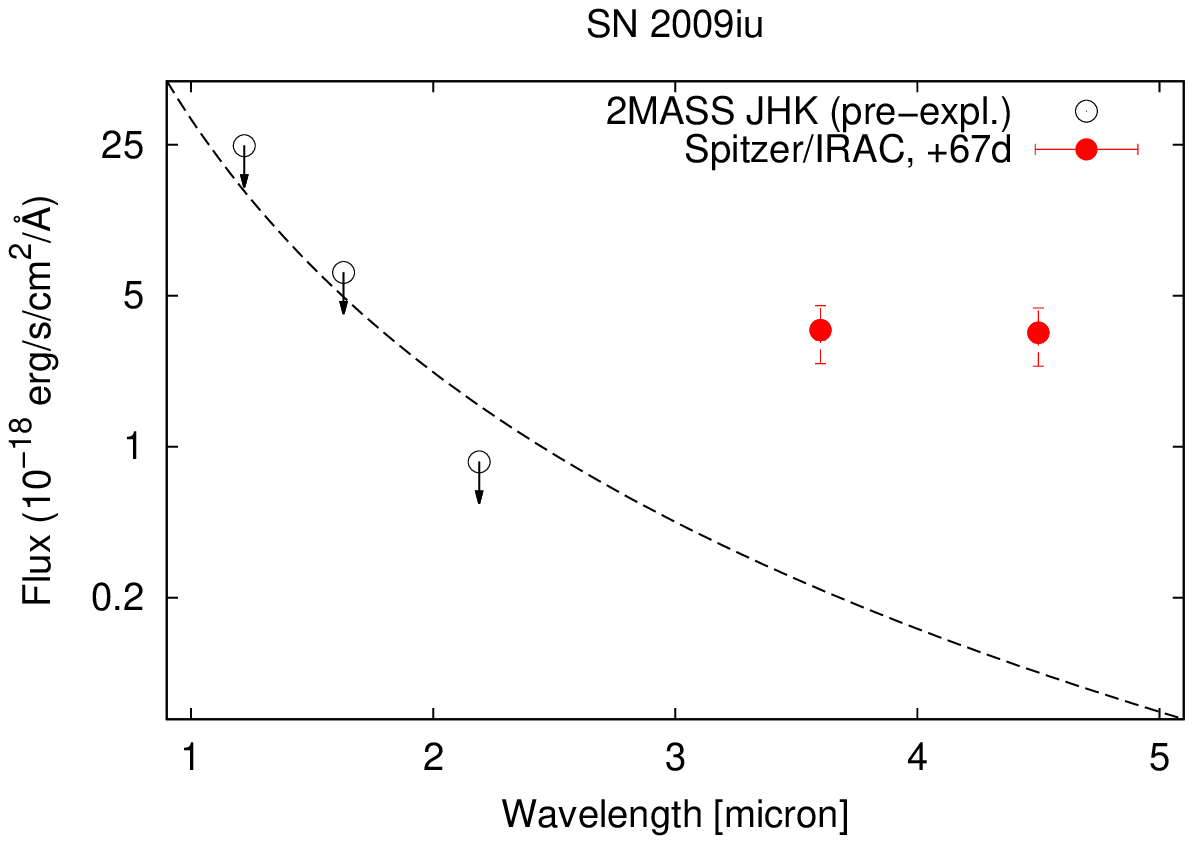}
\includegraphics[width=7cm]{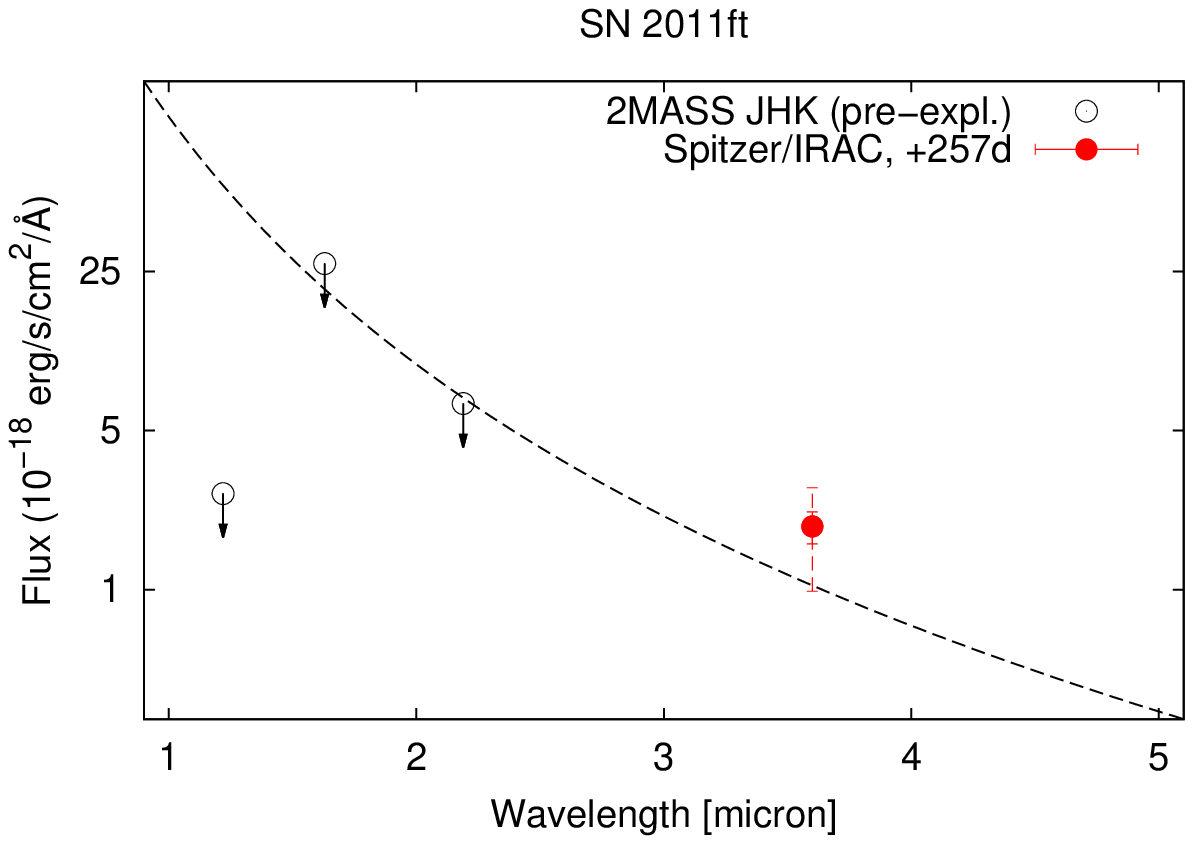}
\includegraphics[width=7cm]{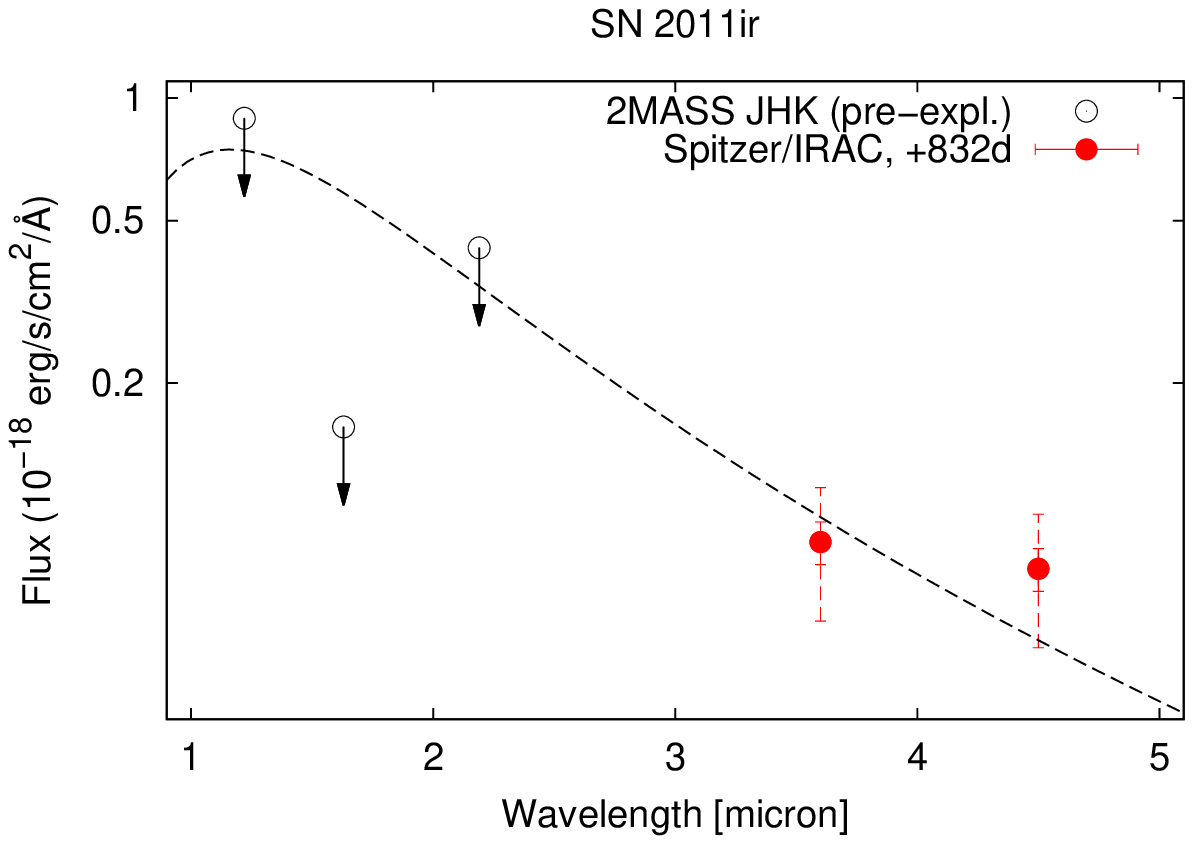}
\includegraphics[width=7cm]{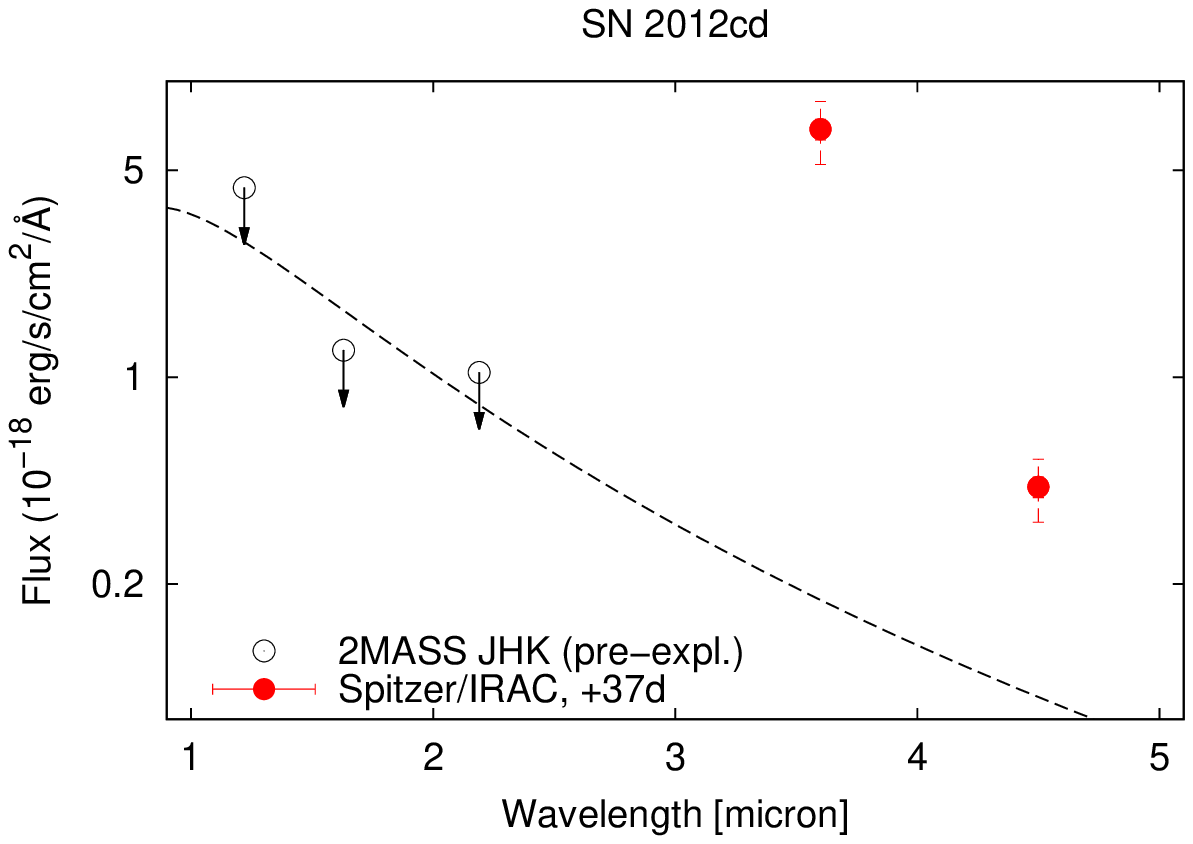}
\includegraphics[width=7cm]{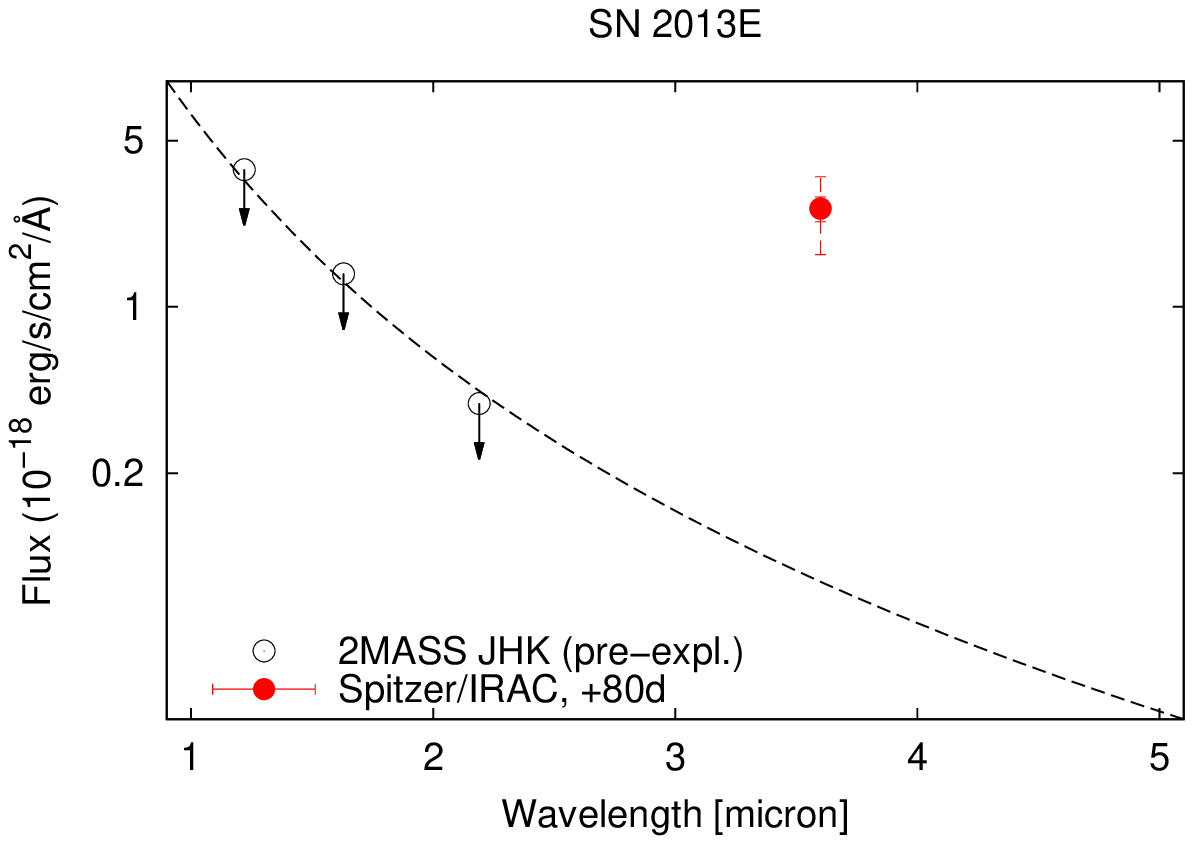}
\caption{Comparison of pre-explosion 2MASS JHK$_s$ (black open circles) and post-explosion mid-IR fluxes (red filled circles) in the cases of single-epoch {\it Spitzer}/IRAC SN observations classified as positive detections. Simple blackbodies are fitted to upper limits of JHK$_s$ fluxes in order to see whether there can be any 'real' mid-IR excess at post-explosion {\it Spitzer} images or not (see text for details). In the cases of IRAC fluxes, solid and dashed errorbars denote 1-$\sigma$ and 3-$\sigma$ photometric errors, respectively.}
\label{fig:2mass_sed}
\end{center}
\end{figure*}

\begin{figure*}[!h]
\begin{center}
\includegraphics[width=9cm]{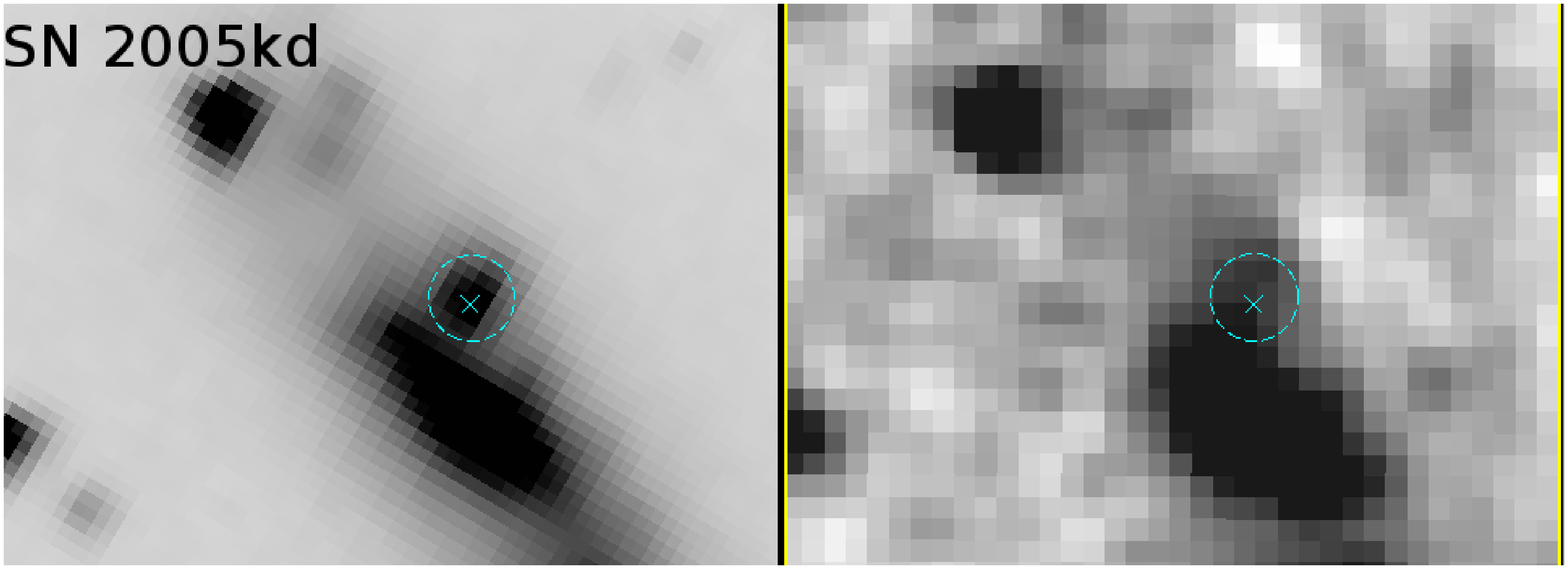}
\includegraphics[width=7cm]{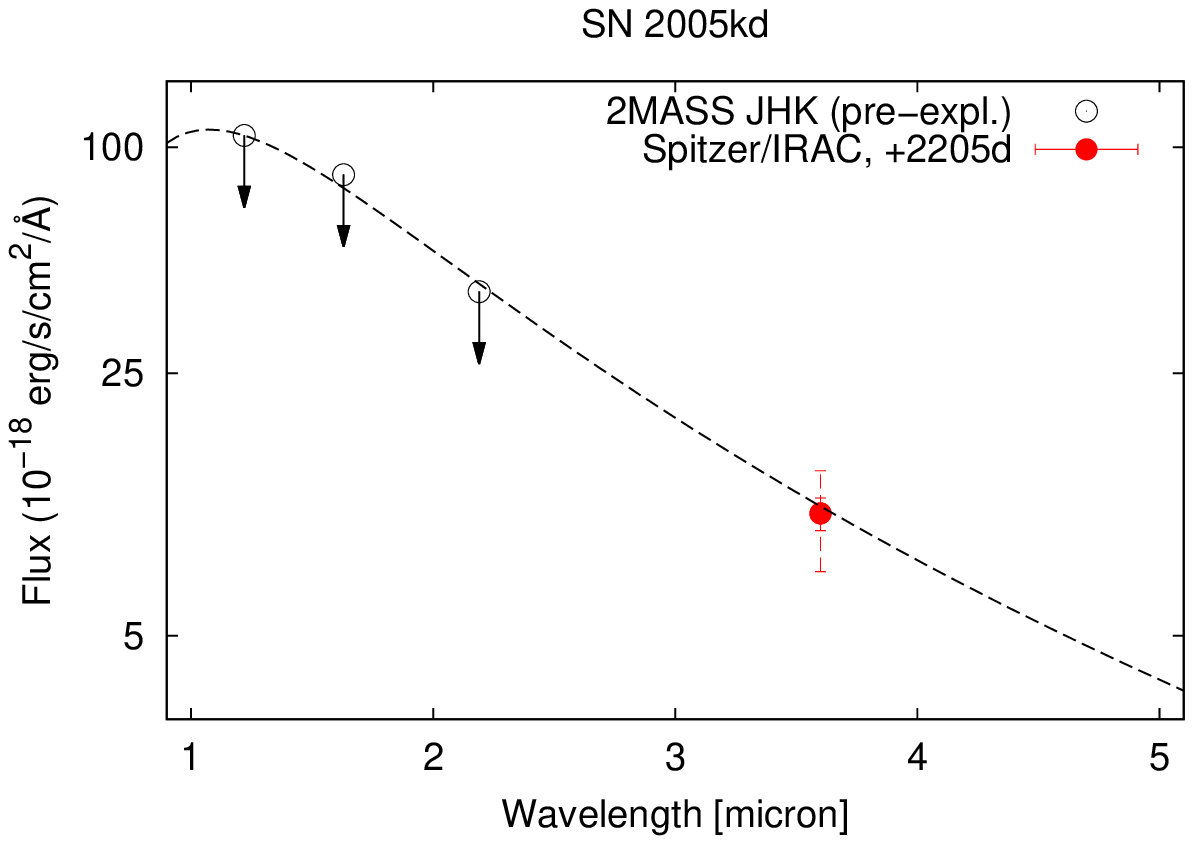}
\caption{SN~2005kd: an example of single-epoch {\it Spitzer}/IRAC supernova observations classified as negative detections. While one can see a mid-IR source at the position of the SN, some flux can be also detect on the pre-explosion 2MASS K$_s$ image. Doing the final step of checking, no significant excess could be revealed at the 3.6$\mu$m channel (the field was not observed in any other IRAC channels).}
\label{fig:05kd}
\end{center}
\end{figure*}


\begin{thebibliography}{}


\bibitem[\protect\citeauthoryear{Andrews et al.}{2010}]{Andrews10}
  Andrews, J. E., Gallagher, J. S., Clayton, G. C., et al. 2010, ApJ, 715, 541

\bibitem[\protect\citeauthoryear{Andrews et al.}{2011a}]{Andrews11a}
  Andrews, J. E., Clayton, G. C., Wesson, R., et al. 2011a, AJ, 142, 45  

\bibitem[\protect\citeauthoryear{Andrews et al.}{2011b}]{Andrews11b}
  Andrews, J. E., Sugerman, B. E. K., Clayton, G. C., et al. 2011b, ApJ, 731, 47 

\bibitem[\protect\citeauthoryear{Andrews et al.}{2015}]{Andrews15}  
  Andrews, J. E., Smith, N., Fong, W.-F., \& Milne, P. 2015, ATel, 7084, 1
  
\bibitem[\protect\citeauthoryear{Andrews et al.}{2016}]{Andrews16}  
  Andrews, J. E., Krafton, K. M., Clayton, G. C., et al. 2016, MNRAS, 457, 3241

\bibitem[\protect\citeauthoryear{Andrews et al.}{2017}]{Andrews17}  
  Andrews, J. E.,  Smith, N., McCully, C., et al. 2017, MNRAS, 471, 4047
  
\bibitem[\protect\citeauthoryear{Arbour \& Briggs}{2009}]{Arbour09}  
  Arbour, R., \& Briggs, D. 2009, CBET, 1964, 1
  
\bibitem[\protect\citeauthoryear{Balanutsa et al.}{2011}]{Balanutsa11}    
  Balanutsa, P., Parhomenko, A. V., Tlatov, A., et al. 2011, ATel, 3610, 1
  
\bibitem[\protect\citeauthoryear{Barna et al.}{2017}]{Barna17}
  Barna, B., Szalai, T., Kromer, M., et al., 2017, MNRAS, 471, 4865  
  
\bibitem[\protect\citeauthoryear{Benetti et al.}{2009}]{Benetti09}    
  Benetti, S.,Valenti, S., Magazzu, A., \& Harutyunyan, A. 2009, CBET, 1667, 1
  
%
%
\bibitem[\protect\citeauthoryear{Bietenholz \& Bartel}{2005}]{BB05}
  Bietenholz, M.~F., Bartel, N. 2005, ApJ, 625, L99

\bibitem[\protect\citeauthoryear{Blackman et al.}{2006}]{Blackman06}  
  Blackman, J., Schmidt, B., \& Kerzendorf, W. 2006, CBET, 541, 1
  
\bibitem[\protect\citeauthoryear{Bode \& Evans}{1980}]{Bode80}
  Bode, M. F., \& Evans, A. 1980, MNRAS, 193, 21

\bibitem[\protect\citeauthoryear{Boles}{2009}]{Boles09}  
  Boles, T. 2009, CBET, 1648, 1
  
\bibitem[\protect\citeauthoryear{Boles}{2011}]{Boles11}  
  Boles, T. 2011, CBET, 2851, 1  
  
\bibitem[\protect\citeauthoryear{Bose et al.}{2015}]{Bose15}  
  Bose, S., Sutaria, F., Kumar, B., et al. 2015, ApJ, 806, 160
  
\bibitem[\protect\citeauthoryear{Brimacombe et al.}{2013}]{Brimacombe13}    
  Brimacombe, J., Zaggia, S., Barbieri, M., et al. 2013, CBET, 3647, 1
  
\bibitem[\protect\citeauthoryear{Burket \& Li}{2005}]{Burket05}  
  Burket, J., \& Li, W. 2005, IAU Circ. 8472, 1
  
\bibitem[\protect\citeauthoryear{Casper et al.}{2013}]{Casper13}      
  Casper, C., Zheng, W., Li, W., Filippenko, A. V., \& Cenko, S. B. 2013, CBET, 3588, 1
  
\bibitem[\protect\citeauthoryear{Chakraborti et al.}{2016}]{Chakraborti16}        
  Chakraborti, S., Ray, A., Smith, R., et al. 2016, ApJ, 817, 22
  
\bibitem[\protect\citeauthoryear{Chen et al.}{2011}]{Chen11}        
  Chen, J., Wang, X.-F., Wang, X.-L., et al. 2011, CBET, 2943, 2
  
\bibitem[\protect\citeauthoryear{Chevalier \& Soderberg}{2010}]{Chevalier10}  
  Chevalier, R.~A., \& Soderberg, A.~M., 2010, ApJ, 711, L40  
  
\bibitem[\protect\citeauthoryear{Chomiuk et al.}{2016}]{Chomiuk16}  
  Chomiuk, L., Soderberg, A.~M., Chevalier, R.~A., et al. 2016, ApJ, 821, 119
  
\bibitem[\protect\citeauthoryear{Chornock \& Berger}{2009}]{Chornock09}    
  Chornock, R., \& Berger, E. 2009, CBET, 2086, 1
  
\bibitem[\protect\citeauthoryear{Chugai \& Chevalier}{2006}]{Chugai06}      
  Chugai, N.~N., \& Chevalier, R.~A. 2006, ApJ, 641, 1051
  
\bibitem[\protect\citeauthoryear{Ciroi et al.}{2009}]{Ciroi09}
  Ciroi, S., di Mille, F., Carco, M., et al. 2009, CBET, 1697, 2
  
%
\bibitem[\protect\citeauthoryear{Corsi et al.}{2014}]{Corsi14}
  Corsi, A., Ofek, E. O., Gal-Yam, A., et al. 2014, ApJ, 782, 42  
  
\bibitem[\protect\citeauthoryear{Cortini}{2009}]{Cortini09}  
  Cortini, G. 2009, CBET, 1697, 1
  
\bibitem[\protect\citeauthoryear{Cortini}{2013}]{Cortini13}  
  Cortini, G., Brimacombe, J., Tomasella, L., et al. 2013, CBET, 3597, 1
  
\bibitem[\protect\citeauthoryear{Dall'Ora et al.}{2014}]{DallOra14}   
  Dall'Ora, M., Botticella, M. T., Pumo, M. L., et al. 2014, ApJ, 787, 139
  
\bibitem[\protect\citeauthoryear{de Jaeger et al.}{2015}]{deJaeger15}   
  de Jaeger, T., Anderson, J. P., Pignata, G., et al. 2015, ApJ, 807, 63 
  
\bibitem[\protect\citeauthoryear{Dhungana et al.}{2016}]{Dhungana16}      
  Dhungana, G., Kehoe, R., Vink\'o, J., et al. 2016, ApJ, 822, 6
  
  
\bibitem[\protect\citeauthoryear{Drake et al.}{2013}]{Drake13}    Drake, A. J., Djorgovski, S. G., Graham, M. J., et al. 2013, CBET, 3570, 1
  
\bibitem[\protect\citeauthoryear{Drescher et al.}{2012}]{Drescher12}      
  Drescher, C., Parker, S., \& Brimacombe, J. 2012, CBET, 3101, 1

\bibitem[\protect\citeauthoryear{Dwek}{1983}]{Dwek83}  
  Dwek, E. 1983, ApJ, 274, 175
  
\bibitem[\protect\citeauthoryear{Dwek et al.}{2010}]{Dwek10}
  Dwek, E., Arendt, R. G., Bouchet, P., et al. 2010, ApJ, 722, 425
  
  
\bibitem[\protect\citeauthoryear{Elias-Rosa et al.}{2005}]{ER05}  
  Elias-Rosa, N., Navasardyan, H., Harutunyan, A., et al. 2005, IAU Circ., 8479, 3
  
\bibitem[\protect\citeauthoryear{Ercolano et al.}{2007}]{Ercolano07}  
  Ercolano, B., Barlow, M. J., \& Sugerman, B. E. K. 2007, MNRAS, 375, 753
  
\bibitem[\protect\citeauthoryear{Ergon et al.}{2015}]{Ergon15}
  Ergon, M., Jerkstrand, A., Sollerman, J., et al. 2015, A\&A, 580, 142  
  
\bibitem[\protect\citeauthoryear{Fabbri et al.}{2011}]{Fabbri11}
  Fabbri, J., Otsuka, M., Barlow, M. J., et al. 2011, MNRAS, 418, 1285

\bibitem[\protect\citeauthoryear{Filippenko et al.}{1993}]{Filippenko93}
  Filippenko, A.~V., Matheson, T., Ho, L.~C. 1993, ApJL, 415, L103

\bibitem[\protect\citeauthoryear{Fitzpatrick \& Massa}{2007}]{Fitzpatrick07}
  Fitzpatrick, E.~L., \& Massa, D. 2007, ApJ, 663, 320  

\bibitem[\protect\citeauthoryear{Folatelli et al.}{2010}]{Folatelli10}  
  Folatelli, G., Gutierrez, C., \& Pignata, G. 2010, CBET, 2390, 1
  
\bibitem[\protect\citeauthoryear{Foley}{2009}]{Foley09}  
  Foley, R. J. 2009, CBET, 1858, 1
  
\bibitem[\protect\citeauthoryear{Foley}{2010}]{Foley10}  
  Foley, R. J. 2010, CBET, 2137, 1  
  
\bibitem[\protect\citeauthoryear{Foley et al.}{2013}]{Foley13}
  Foley, R.~J., Challis, P.~J., Chornock, R., et al., 2013, ApJ, 767, 57
  
\bibitem[\protect\citeauthoryear{Foley et al.}{2016}]{Foley16}
  Foley, R.~J., Jha, S.~W., Pan, Y-C., et al. 2016, MNRAS, 461, 433

\bibitem[\protect\citeauthoryear{Fox \& Filippenko}{2013}]{FF13}  
  Fox, O. D., \& Filippenko, A. V. 2013, ApJ, 772, L6  
  
\bibitem[\protect\citeauthoryear{Fox et al.}{2010}]{Fox10}
  Fox, O. D., Chevalier, R. A., Dwek, E., et al. 2010, ApJ, 725, 1768 
  
\bibitem[\protect\citeauthoryear{Fox et al.}{2011}]{Fox11}
  Fox, O. D., Chevalier, R. A., Skrutskie, M. F., et al. 2011, ApJ, 741, 7

\bibitem[\protect\citeauthoryear{Fox et al.}{2013}]{Fox13}
  Fox, O. D., Filippenko, A. V., Skrutskie, M. F., et al. 2013, AJ, 146, 2    
  
\bibitem[\protect\citeauthoryear{Fox et al.}{2015}]{Fox15}
  Fox, O. D., Silverman, J. M., Filippenko, A. V., et al. 2015, MNRAS, 447, 772
  
\bibitem[\protect\citeauthoryear{Fox et al.}{2016}]{Fox16}  
  Fox, O. D., Johansson, J., Kasliwal, M., et al. 2016, ApJ, 2016, 816, L13  
  
\bibitem[\protect\citeauthoryear{Fransson et al.}{2014}]{Fransson14}  
  Fransson, C., Ergon, M., Challis, P.,  et al. 2014, ApJ, 797, 118  
  
\bibitem[\protect\citeauthoryear{Fraser et al.}{2013}]{Fraser13}  
  Fraser, M., Inserra, C., Jerkstrand, A., et al. 2013, MNRAS, 433, 1332
  
\bibitem[\protect\citeauthoryear{Fraser et al.}{2015}]{Fraser15}  
  Fraser, M., Kotak, R., Pastorello, A., et al. 2015, MNRAS, 453, 3886
  
  
\bibitem[\protect\citeauthoryear{Gall et al.}{2011}]{Gall11}
  Gall, C., Andersen, A. C., \& Hjorth, J., 2011, A\&A, 528, A13  

\bibitem[\protect\citeauthoryear{Gall et al.}{2014}]{Gall14}
  Gall, C., Hjorth, J., Watson, D., et al. 2014, Nature, 511, 326 

\bibitem[\protect\citeauthoryear{Gallagher et al.}{2012}]{Gallagher12}
  Gallagher, J. S., Sugerman, B. E. K., Clayton, G. C., et al. 2012, ApJ, 753, 109
  
\bibitem[\protect\citeauthoryear{Gandhi et al.}{2013}]{Gandhi13}
  Gandhi, P., Yamanaka, M., Tanaka, M., et al. 2013, ApJ, 767, 166
  
\bibitem[\protect\citeauthoryear{Gerardy et al.}{2002}]{Gerardy02}  
  Gerardy, C. L., Fesen, R. A., Nomoto, K., et al. 2002, ApJ, 575, 1007
  
\bibitem[\protect\citeauthoryear{Gerardy et al.}{2007}]{Gerardy07}  
  Gerardy, C. L., Meikle, W. P. S., Kotak, R., et al. 2007, ApJ, 661, 995
  
\bibitem[\protect\citeauthoryear{Guillochon et al.}{2017}]{Guillochon17} 
  Guillochon, J., Parrent, J., Kelley, L. Z., \& Margutti, R. 2017, ApJ, 835, 15
  
\bibitem[\protect\citeauthoryear{Graham \& Li}{2005}]{Graham05}  
  Graham, J., \& Li, W. 2005, IAU Circ., 8459, 3
  
\bibitem[\protect\citeauthoryear{Graham \& Meikle}{1986}]{Graham86}    
  Graham, J. R. \& Meikle, W. P. S. 1986, MNRAS, 221, 789
  
\bibitem[\protect\citeauthoryear{Graham et al.}{2017}]{Graham17}
  Graham, M. L., Harris, C. E., Fox, O. D., et al. 2017, ApJ, 843, 102 

\bibitem[\protect\citeauthoryear{Harutyunyan et al.}{2009}]{Harutyunyan09}  
  Harutyunyan, A., Bufano, F., \& Benetti, S. 2009, CBET, 1722, 1
  
\bibitem[\protect\citeauthoryear{Helou et al.}{2013}]{Helou13}  
  Helou, G., Kasliwal, M. M., Ofek E. O., et al., 2013, ApJ, 778, L19  
  
\bibitem[\protect\citeauthoryear{Holoien et al.}{2014}]{Holoien14}    
  Holoien, T. W.-S., Prieto, J. L., Kochanek, C. S., et al. 2014, ATel, 6436, 1

\bibitem[\protect\citeauthoryear{Holoien et al.}{2017a}]{Holoien17a}    
  Holoien, T. W.-S., Stanek, K. Z., Kochanek, C. S., et al. 2017a, MNRAS, 464, 2672
  
\bibitem[\protect\citeauthoryear{Holoien et al.}{2017b}]{Holoien17b}    
  Holoien, T. W.-S., Brown, J. S., Stanek, K. Z., et al. 2017b, MNRAS, 467, 1098
 
\bibitem[\protect\citeauthoryear{Holoien et al.}{2017c}]{Holoien17c}    
  Holoien, T. W.-S., Brown, J. S., Stanek, K. Z., et al. 2017c, MNRAS, 471, 4966
  
\bibitem[\protect\citeauthoryear{Howell \& Murray}{2010}]{Howell10}    
  Howell, D. A., Murray, D. 2010, CBET, 3313, 2
  
\bibitem[\protect\citeauthoryear{Howerton et al.}{2011}]{Howerton11}      
  Howerton, S., Drake, A. J., Djorgovski, S. G., et al. 2011, CBET, 2658, 1
  
\bibitem[\protect\citeauthoryear{Huang et al.}{2015}]{Huang15}          
  Huang, F., Wang, X., Zhang, J., et al. 2015, ApJ, 807, 59
  
\bibitem[\protect\citeauthoryear{Huang et al.}{2016}]{Huang16}  Huang, F., Wang, X., Zampieri, L., et al. 2016, ApJ, 832, 139
  
\bibitem[\protect\citeauthoryear{Indebetouw et al.}{2014}]{Indebetouw14}
  Indebetouw, R., Matsuura, M., Dwek, E., et al. 2014, ApJ, 782, L2

\bibitem[\protect\citeauthoryear{Inserra et al.}{2016}]{Inserra16}  
  Inserra, C., Fraser, M., Smartt, S. J., et al. 2016, MNRAS, 459, 2721  
  
\bibitem[\protect\citeauthoryear{Itagaki et al.}{2011}]{Itagaki11}    
  Itagaki, K., Brimacombe, J., Noguchi, T., \& Nakano, S. 2011, CBET, 2943, 1
  
\bibitem[\protect\citeauthoryear{Jencson et al.}{2017}]{Jencson17}
  Jencson, J. E., Kasliwal, M. M., Johansson, J., et al. 2017, ApJ, 837, 167  
  
\bibitem[\protect\citeauthoryear{Jencson et al.}{2018}]{Jencson18}
  Jencson, J. E., Kasliwal, M. M., Adams, S.~M., et al. 2018, ApJ, 863, 20
  
\bibitem[\protect\citeauthoryear{Jha et al.}{2006}]{Jha06}  
  Jha, S., Branch, D., Chornock, R., et al. 2006, AJ, 132, 189
  
 \bibitem[\protect\citeauthoryear{Jin et al.}{2013}]{Jin13}   	
  Jin, Z., Gao, X., McCully, C., \& Jha, S. W. 2013, CBET, 3520, 1
  
\bibitem[\protect\citeauthoryear{Johansson et al.}{2017}]{Johansson17}  
  Johansson, J., Goobar, A., Kasliwal, M. M., et al. 2017, MNRAS, 466, 3442	  

\bibitem[\protect\citeauthoryear{Kamble et al.}{2016}]{Kamble16}  
  Kamble, A., Margutti, R., Soderberg, A.~M., et al., 2016, ApJ, 818, 111  
  
\bibitem[\protect\citeauthoryear{Kankare et al.}{2014}]{Kankare14}    
  Kankare, E., Fraser, M., Ryder, S., et al. 2014, A\&A, 572, 75
  
\bibitem[\protect\citeauthoryear{Kasliwal}{2010}]{Kasliwal10}      
  Kasliwal, M. M. 2010, ATel, 2379, 1
  
\bibitem[\protect\citeauthoryear{Kasliwal et al.}{2009}]{Kasliwal09}    
  Kasliwal, M. M., Howell, J. L., Fox, D., Quimby, R., \& Gal-Yam, A. 2009, ATel, 2218, 1
  
\bibitem[\protect\citeauthoryear{Kasliwal et al.}{2017}]{Kasliwal17}  
  Kasliwal, M. M., Bally, J., Masci, F., et al. 2017, ApJ, 839, 88

\bibitem[\protect\citeauthoryear{Kochanek}{2017}]{Kochanek17} 
  Kochanek, C. S. 2017, MNRAS, 471, 3283
  
\bibitem[\protect\citeauthoryear{Kochanek et al.}{2011}]{Kochanek11} 
  Kochanek, C. S., Szczygiel, D. M., \& Stanek, K. Z. 2011, ApJ, 737, 76  
 
\bibitem[\protect\citeauthoryear{Kochanek et al.}{2012}]{Kochanek12} 
  Kochanek, C. S., Szczygiel, D. M., \& Stanek, K. Z. 2012, ApJ, 758, 142   
 
\bibitem[\protect\citeauthoryear{Kotak et al.}{2005}]{Kotak05} 
  Kotak, R., Meikle, W. P. S., van Dyk, S. D., H\"oflich, P. A., \& Mattila, S. 2005, ApJ, 628, L123 
 
\bibitem[\protect\citeauthoryear{Kotak et al.}{2006}]{Kotak06}
  Kotak, R., Meikle, W. P. S., Pozzo, M., et al. 2006, ApJ, 651, L117
 
\bibitem[\protect\citeauthoryear{Kotak et al.}{2009}]{Kotak09}
  Kotak, R., Meikle, W. P. S., Farrah, D., et al. 2009, ApJ, 704, 306

\bibitem[\protect\citeauthoryear{Kozasa et al.}{2009}]{Kozasa09}
  Kozasa, T., Nozawa, T., Tominaga, N., et al. 2009, in Cosmic Dust - Near and Far, ed. T. Henning, E. Gr\"un, \& J. Steinacker (San Francisco: ASP), ASP Conf. Ser., 414, 43  

\bibitem[\protect\citeauthoryear{Kumar et al.}{2016}]{Kumar16}  
  Kumar, B., Pandey, S. B., Eswaraiah, C., \& Kawabata, K. S. 2016, MNRAS, 456, 3157
  
\bibitem[\protect\citeauthoryear{Lennarz et al.}{2012}]{Lennarz12}
  Lennarz, D., Altmann, D., \& Wiebusch, C. 2012, A\&A, 538, A120

\bibitem[\protect\citeauthoryear{Leonard et al.}{2013}]{Leonard13}  
  Leonard, D. C., Pignata, G., Dessart, L., et al. 2013, ATel, 5275
  
\bibitem[\protect\citeauthoryear{Li et al.}{2004}]{Li04}  
  Li, W., Foley, R. J., \& Filippenko, A. V. 2004, IAU Circ., 8345, 4
  
\bibitem[\protect\citeauthoryear{Li et al.}{2009a}]{Li09a}    
  Li, W., Cenko, S. B., \& Filippenko, A. V. 2009a, CBET, 1656, 1
  
\bibitem[\protect\citeauthoryear{Li et al.}{2009b}]{Li09b}      
  Li, W., Cenko, S. B., \& Filippenko, A. V. 2009b, CBET, 1952, 1
  
\bibitem[\protect\citeauthoryear{Maeda et al.}{2015a}]{Maeda15a}  
  Maeda, K., Nozawa, T., Nagao, T., Motohara, K. 2015, MNRAS, 452, 3281
  
\bibitem[\protect\citeauthoryear{Maeda et al.}{2015b}]{Maeda15b}  
  Maeda, K., Hattori, T., Milisavljevic, D., et al., 2015, ApJ, 807, 35    
  
\bibitem[\protect\citeauthoryear{Margutti et al.}{2014}]{Margutti14}      
  Margutti, R., Milisavljevic, D., Soderberg, A. M., et al. 2014, ApJ, 780, 21
  
\bibitem[\protect\citeauthoryear{Margutti et al.}{2017}]{Margutti17}      
  Margutti, R., Kamble, A., Milisavljevic, D., et al. 2017, ApJ, 835, 140
  
\bibitem[\protect\citeauthoryear{Marion et al.}{2012}]{Marion12}          
  Marion, G. H., Foley, R. J., \& Calkins, M. 2012, CBET, 3106, 2
  
\bibitem[\protect\citeauthoryear{Marples \& Drescher}{2009}]{Marples09a}        
  Marples, P., \& Drescher, C. 2009, CBET, 2080, 1
  
\bibitem[\protect\citeauthoryear{Marples et al.}{2009}]{Marples09b}          
  Marples, P., Drescher, C., Quirk, S., \& Bock, G. 2009, CBET, 1856, 1
  
\bibitem[\protect\citeauthoryear{Matheson et al.}{2001}]{Matheson01}  
  Matheson, T., Jha, S., Challis, P.. Kirshner, R., \& Calkins, M. 2001, IAU Circ., 7761, 2
  
%
\bibitem[\protect\citeauthoryear{Matsuura et al.}{2011}]{Matsuura11}
  Matsuura, M., Dwek, E., Meixner, M., et al. 2011, Science, 333, 1258
  
\bibitem[\protect\citeauthoryear{Matsuura et al.}{2015}]{Matsuura15}
  Matsuura, M., Dwek, E., Barlow, M. J., et al. 2015, ApJ, 800, 50
  
\bibitem[\protect\citeauthoryear{Mattila et al.}{2008}]{Mattila08}  
  Mattila, S., Meikle, W. P. S., Lundqvist, P., et al. 2008, MNRAS, 389, 141  

\bibitem[\protect\citeauthoryear{Mauerhan et al.}{2013}]{Mauerhan13}  
  Mauerhan, J. C., Smith, N., Filippenko, A. V., et al. 2013, MNRAS, 430, 1801
  
\bibitem[\protect\citeauthoryear{Mauerhan et al.}{2017}]{Mauerhan17}  
  Mauerhan, J. C., Van Dyk, S. D., Johansson, J., et al. 2017, ApJ, 834, 118

\bibitem[\protect\citeauthoryear{Maza et al.}{2009}]{Maza09}    
  Maza, J., Hamuy, M., Antezana, R., et al. 2009, CBET, 1937, 1
  
\bibitem[\protect\citeauthoryear{Maza et al.}{2010a}]{Maza10a}      
  Maza, J., Hamuy, M., Antezana, R., et al. 2010a, CBET, 2125, 1  
  
\bibitem[\protect\citeauthoryear{Maza et al.}{2010b}]{Maza10b}      
  Maza, J., Hamuy, M., Antezana, R., et al. 2010b, CBET, 2388, 1
  
\bibitem[\protect\citeauthoryear{McClelland et al.}{2013}]{McClelland13}  
  McClelland, C. M., Garnavich, P. M., Milne, P. A., et al. 2013, ApJ, 767, 119  
\bibitem[\protect\citeauthoryear{McCully et al.}{2014}]{McCully14}
  McCully, C., Jha, S.~W., Foley, R.~J., et al., 2014, Nature, 512, 54  
  
\bibitem[\protect\citeauthoryear{Meikle et al.}{2005}]{Meikle05}  
  Meikle, W. P., Farrah, D., Fesen, R., et al. 2005, Spitzer Proposal, 20256
  
\bibitem[\protect\citeauthoryear{Meikle et al.}{2006}]{Meikle06}
  Meikle, W. P. S., Mattila, S., Gerardy, C. L., et al. 2006, ApJ, 649, 332 
  
\bibitem[\protect\citeauthoryear{Meikle et al.}{2007}]{Meikle07}
  Meikle, W. P. S., Mattila, S., Pastorello, A., et al. 2007, ApJ, 665, 608  

\bibitem[\protect\citeauthoryear{Meng \& Podsiadlowski}{2018}]{MP18}
  Meikle, W. P. S., Kotak, R., Farrah, D., et al. 2011, ApJ, 732, 109
 
\bibitem[\protect\citeauthoryear{Meikle et al.}{2011}]{Meikle11}
  Meng, X., Podsiadlowski, Ph. 2018, ApJ, 861, 127
  
\bibitem[\protect\citeauthoryear{Milisavljevic et al.}{2015}]{Milisavljevic15}  
  Milisavljevic, D., Margutti, R., Kamble, A., et al. 2015, ApJ, 815, 120
  
\bibitem[\protect\citeauthoryear{Modjaz et al.}{2005}]{Modjaz05}  
  Modjaz, M., Kirshner, R., Challis, P., et al. 2005, IAU Circ., 8461, 2  
  
\bibitem[\protect\citeauthoryear{Monard}{2009}]{Monard09}          
  Monard, L. A. G. 2009, CBET, 1798, 1
  
\bibitem[\protect\citeauthoryear{Monard \& Li}{2002}]{Monard02}              
  Monard, L. A. G., \& Li, W. 2002,  IAU Circ., 7940, 1
  
\bibitem[\protect\citeauthoryear{Monard \& Prieto}{2011}]{Monard11a}            
  Monard, L. A. G., \& Prieto, J. L. 2011, CBET, 2759, 1
  
\bibitem[\protect\citeauthoryear{Monard et al.}{2011a}]{Monard11b}              
  Monard, L. A. G., Prieto, J. L., \& Seth, K. 2011a, CBET, 2799, 1  
  
\bibitem[\protect\citeauthoryear{Monard et al.}{2011b}]{Monard11c}            
  Monard, L. A. G., Valenti, S., \& Benetti, S., 2011b, CBET, 2749, 1
  
\bibitem[\protect\citeauthoryear{Moore et al.}{2004}]{Moore04}    
  Moore, M., Li, W., Filippenko, A. V., Chornock, R., \& Foley, R. J., 2004, IAU Circ., 8286, 2
  
%
\bibitem[\protect\citeauthoryear{Morrell \& Phillips}{2009}]{Morrell09}      
  Morrell, N., \& Phillips, M. M. 2009, CBET, 1953, 1
  
\bibitem[\protect\citeauthoryear{Morrell et al.}{2006}]{Morrell13}        
  Morrell, N., Folatelli, G., Hamuy, M., \& Phillips, M. M. 2006, CBET, 368, 1
  
\bibitem[\protect\citeauthoryear{Morrell et al.}{2013}]{Morrell06}          
  Morrell, N., Hsiao, E. Y., Contreras, C., \& Phillips, M. M. 2013, ATel, 4707, 1

\bibitem[\protect\citeauthoryear{M\"uller et al.}{2017}]{Muller17}            
  M\"uller, T., Prieto, J. L., Pejcha, O., \& Clocchiatti, A. 2017, ApJ, 841, 127
  
\bibitem[\protect\citeauthoryear{Nagao et al.}{2017}]{Nagao17}      
  Nagao, T., Maeda, K., Yamanaka, M. 2017, ApJ, 835, 143  
  
\bibitem[\protect\citeauthoryear{Nakano et al.}{2004a}]{Nakano04a}      
  Nakano, S., Kushida, R.. Kushida, Y., \& Itagaki, K. 2004a, IAU Circ., 8272, 1

\bibitem[\protect\citeauthoryear{Nakano et al.}{2004b}]{Nakano04b}        
  Nakano, S., Kushida, R., \& Kushida, Y. 2004b, IAU Circ., 8344, 1
 
\bibitem[\protect\citeauthoryear{Nakano et al.}{2005}]{Nakano05}         
  Nakano, S., Hirose, Y., \& Li, W. 2005, IAU Circ., 8475, 1
  
\bibitem[\protect\citeauthoryear{Nakano et al.}{2009}]{Nakano09}           
  Nakano, S., Kadota, K., Ikari, Y., \& Itagaki, K. 2009, CBET, 1718, 1
  
\bibitem[\protect\citeauthoryear{Nakano et al.}{2010}]{Nakano10}             
  Nakano, S., Kadota, K., \& Itagaki, K. 2010, CBET, 2115, 1
  
\bibitem[\protect\citeauthoryear{Nakano et al.}{2012}]{Nakano12}               
  Nakano, S., Yusa, T., Yoshimoto, K., et al. 2012, CBET, 3263, 1
  
\bibitem[\protect\citeauthoryear{Navasardyan \& Benetti}{2009}]{Navasardyan09a}
  Navasardyan, H., \& Benetti, S. 2009, CBET, 1806, 1
  
\bibitem[\protect\citeauthoryear{Navasardyan et al.}{2009}]{Navasardyan09b}
  Navasardyan, H., Benetti, S., Bufano, F., \& Pastorello, A. 2009, CBET, 1738, 1
  
\bibitem[\protect\citeauthoryear{Nozawa et al.}{2003}]{Nozawa03}  
  Nozawa, T., Kozasa, T., Umeda, H., Maeda, K., Nomoto, K. 2003, ApJ, 598, 785
  
\bibitem[\protect\citeauthoryear{Nozawa et al.}{2011}]{Nozawa11}  
  Nozawa, T., Maeda, K., Kozasa, T., et al. 2011, ApJ, 736, 45   

  
\bibitem[\protect\citeauthoryear{Ochner et al.}{2014}]{Ochner14}      
  Ochner, P., Siviero, A., Tomasella, L., et al. 2014 ATel, 5767, 1
  
\bibitem[\protect\citeauthoryear{Ofek et al.}{2013}]{Ofek13}      
  Ofek, E. O., Sullivan, M., Cenko, S. B., et al. 2013, Nature, 494, 65
  
\bibitem[\protect\citeauthoryear{Papenkova et al.}{2001}]{Papenkova01}      
  Papenkova, M., Li, W. D., Wray, J., Chleborad, C. W., \& Schwartz, M. 2011, IAU Circ., 7722, 1
  

\bibitem[\protect\citeauthoryear{Pastorello et al.}{2013}]{Pastorello13}  
  Pastorello, A., Cappellaro, E., Inserra, C., et al. 2013, ApJ, 767, 1

\bibitem[\protect\citeauthoryear{Patat et al.}{2001}]{Patat01}  
  Patat, F., Cappellaro, E., Danziger, J., et al. 2001, ApJ, 555, 900
  
\bibitem[\protect\citeauthoryear{Pejcha \& Prieto}{2015}]{Pejcha15}    
  Pejcha, O., \& Prieto, J. L. 2015, ApJ, 806, 225

\bibitem[\protect\citeauthoryear{Phillips et al.}{2013}]{Phillips13}    
  Phillips, M.~M., Simon, J.~D., Morrell, N., et al. 2013, ApJ, 779, 38
  
\bibitem[\protect\citeauthoryear{Pignata et al.}{2011}]{Pignata11}        
  Pignata, G., Cifuentes, M., Maza, J., et al. 2011, CBET, 2623, 1
  
\bibitem[\protect\citeauthoryear{Ponticello et al.}{2006}]{Ponticello06}        Ponticello, N. J., Khandrika, H., Madison, D. R.,  et al. 2006, IAU Circ., 8709, 1
  
\bibitem[\protect\citeauthoryear{Pooley \& Levin}{2004}]{Pooley04}
  Pooley, D., \& Lewin, W.~H.~G. 2004, IAU Circ., 8323, 2  
  
\bibitem[\protect\citeauthoryear{Pozzo et al.}{2004}]{Pozzo04}
  Pozzo, M., Meikle, W.P.S., Fassia, A., Geballe, T., Lundqvist, P., Chugai, N. N., Sollerman, J. 2004, MNRAS, 352, 457
  
\bibitem[\protect\citeauthoryear{Prieto}{2011a}]{Prieto11a}   
  Prieto, J. L. 2011a, CBET, 2759, 2
  
\bibitem[\protect\citeauthoryear{Prieto}{2011b}]{Prieto11b}
  Prieto, J. L. 2011b, ATel, 3615, 1
  
\bibitem[\protect\citeauthoryear{Prieto \& Seth}{2011}]{Prieto11c}          
  Prieto, J. L., Seth, K. 2011, CBET, 2799, 2
  
\bibitem[\protect\citeauthoryear{Prieto et al.}{2011}]{Prieto11d}          
  Prieto, J. L., McMillan, R., Bakos, G., Grennan, D. 2011, CBET, 2903, 1  

  
\bibitem[\protect\citeauthoryear{Prieto et al.}{2012}]{Prieto12}        
  Prieto, J. L., Lee, J. C., Drake, A. J., et al. 2012, ApJ, 745, 70
  
\bibitem[\protect\citeauthoryear{Puckett et al.}{2006}]{Puckett06}        
  Puckett, T., Reddy, V., \& Li, W. 2006 CBET, 363, 1
  
\bibitem[\protect\citeauthoryear{Quinn et al.}{2008}]{Quinn08}
  Quinn, J., Baade, D., Clocchiatti, A., et al. 2008, CBET, 1510, 1
  
\bibitem[\protect\citeauthoryear{Reach et al.}{2006}]{Reach06a}
  Reach, W. T, Rho, J., Tappe, A., et al. 2006, AJ, 131, 1479
  

%
\bibitem[\protect\citeauthoryear{Sahu et al.}{2011a}]{Sahu11a}  
  Sahu, D. K., Arora, S., \& Anto, P. 2011a, CBET, 2658, 2
  
\bibitem[\protect\citeauthoryear{Sahu et al.}{2011b}]{Sahu11b}  
  Sahu, D. K., Gurugubelli, U. K., Anupama, G. C., \& Nomoto, K. 2011b, MNRAS, 413, 2583
  
\bibitem[\protect\citeauthoryear{Salvo \& Price}{2002}]{Salvo02}  
  Salvo, M., \& Price, P. 2002, IAU Circ., 7947, 2 
  
\bibitem[\protect\citeauthoryear{Sanders et al.}{2013}]{Sanders13}  
  Sanders, N. E., Soderberg, A. M., Foley, R. J., et al. 2013, ApJ, 769, 39

\bibitem[\protect\citeauthoryear{Shappee et al.}{2014}]{Shappee14}  
  Shappee, B. J., Prieto, J. L., Grupe, D., et al. 2014, ApJ, 788, 48
  
\bibitem[\protect\citeauthoryear{Shurpakov et al.}{2012}]{Shurpakov12}    
  Shurpakov, S., Balanutsa, P., Lipunov, V., et al. 2012, ATel, 4104, 1
  
\bibitem[\protect\citeauthoryear{Silverman et al.}{2009}]{Silverman09}  
  Silverman, J. M., Kandrashoff, M. T., \& Filippenko, A. V. 2009, CBET, 1968, 1
  
\bibitem[\protect\citeauthoryear{Silverman et al.}{2013}]{Silverman13}
  Silverman, J. M., Nugent, P. E., Gal-Yam, A., et al. 2013, ApJS, 207, 3  
  
\bibitem[\protect\citeauthoryear{Siviero et al.}{2012}]{Siviero12} 
  Siviero, A., Tomasella, L., Pastorello, A., et al. 2012, CBET, 3054, 4
  
\bibitem[\protect\citeauthoryear{Skrutskie et al.}{2006}]{Skrutskie06} 
  Skrutskie, M.~F., Cutri, R.~M., Stiening, R., et al. 2006, AJ, 131, 1163  
  
\bibitem[\protect\citeauthoryear{Smith et al.}{2009}]{Smith09}  
  Smith, N., Silverman, J. M., Chornock, R., et al. 2009, ApJ, 695, 1334
  
\bibitem[\protect\citeauthoryear{Soderberg et al.}{2004}]{Soderberg04}
  Soderberg, A.~M., Gal-Yam, A., \& Kulkarni, S.~R. 2004, GRB C. N., 2586, 1  
  
\bibitem[\protect\citeauthoryear{Stockdale et al.}{2004}]{Stockdale04}
  Stockdale, C.~J., Weiler, K. W., van Dyk, S.~D., et al. 2004, IAU Circ., 8282, 2  
  
\bibitem[\protect\citeauthoryear{Stritzinger et al.}{2011}]{Stritzinger11} 
  Stritzinger, M., Prieto, J. L., Morrell, N., \& Pignata, G. 2011, CBET, 2623, 2
  
\bibitem[\protect\citeauthoryear{Stritzinger et al.}{2012}]{Stritzinger12}    
  Stritzinger, M., Taddia, F., Fransson, C., et al. 2012, ApJ, 756, 173
  
\bibitem[\protect\citeauthoryear{Sugerman}{2003}]{Sugerman03}
  Sugerman, B. E. K. 2003, AJ, 126, 1939  
  
\bibitem[\protect\citeauthoryear{Sugerman et al.}{2004}]{Sugerman04}  
  Sugerman, B. E. K., Meixner, M., Fabbri, J., \& Barlow, M. 2004, IAU Circ., 8442, 1
  
\bibitem[\protect\citeauthoryear{Sugerman et al.}{2006}]{Sugerman06}
  Sugerman, B. E. K., Ercolano, B., Barlow, M. J., et al. 2006, Science, 313, 196
  
\bibitem[\protect\citeauthoryear{Szalai \& Vink\'o}{2013}]{Szalai13}  
  Szalai, T., \& Vink\'o, J. 2013, A\&A, 549, A79  
  
\bibitem[\protect\citeauthoryear{Szalai et al.}{2011}]{Szalai11}
  Szalai, T., Vink\'o, J., Balog, Z., et al. 2011, A\&A, 527, A61  
  
\bibitem[\protect\citeauthoryear{Szalai et al.}{2016}]{Szalai16}
  Szalai, T., Vink\'o, J., Nagy, A. P., et al. 2016, MNRAS, 460, 1500

\bibitem[\protect\citeauthoryear{Szczygiel et al.}{2012}]{Szczygiel12}
  Szczygiel, D. M., Kochanek, C. S., Dai, X. 2012, ApJ, 760, 20

\bibitem[\protect\citeauthoryear{Taubenberger et al.}{2006}]{Taubenberger06}
  Taubenberger, S., Pastorello, A., Mazzali, P. A., et al. 2006, MNRAS, 371, 1459

\bibitem[\protect\citeauthoryear{Thrasher et al.}{2008}]{Thrasher08}  
  Thrasher, P., Li, W., \& Filippenko, A. V. 2008, CBET, 1507, 1
  
\bibitem[\protect\citeauthoryear{Tinyanont et al.}{2016}]{Tinyanont16}  
  Tinyanont, S., Kasliwal, M. M., Fox, O. D., et al. 2016, ApJ, 833, 231

\bibitem[\protect\citeauthoryear{Tomasella et al.}{2011}]{Tomasella11} 
Tomasella, L.; Pastorello, A.; Valenti, S.; Benetti, S.

\bibitem[\protect\citeauthoryear{Xue et al.}{2016}]{Xue16}  
  Xue, M., Jiang, B.~W., Gao, J., Liu, J., Wang, S., Li, A. 2016, ApJS, 224, 23

\bibitem[\protect\citeauthoryear{Yamanaka et al.}{2016}]{Yamanaka16}  
  Yamanaka, M., Maeda, K., Tanaka, M., et al. 2016, PASJ, 68, 68
  
\bibitem[\protect\citeauthoryear{Valenti et al.}{2012}]{Valenti12}    
  Valenti, S., Pastorello, A., Cappellaro, E., et al. 2012, ATel, 4076, 1
  
\bibitem[\protect\citeauthoryear{Van Dyk}{2013}]{VD13}
  Van Dyk, S. 2013, AJ, 145, 118  
  
\bibitem[\protect\citeauthoryear{Vink\'o et al.}{2017}]{Vinko17}
  Vink\'o, J., Pooley, D., Silverman, J. M., et al. 2017, ApJ, 837, 62
  
\bibitem[\protect\citeauthoryear{Wenger et al.}{2000}]{Wenger00}
  Wenger, M., Ochsenbein, F., Egret, D., et al. 2000, A\&AS, 143, 9
  
\bibitem[\protect\citeauthoryear{Wesson et al.}{2015}]{Wesson15}
  Wesson, R., Barlow, M. J., Matsuura, M., \& Ercolano, B. 2015, MNRAS, 446, 2089 
  
\bibitem[\protect\citeauthoryear{Williams \& Fox}{2015}]{Williams15}
  Williams, B. J., \& Fox, O. D. 2015, ApJ, 808, L22
  
 \bibitem[\protect\citeauthoryear{Zheng et al.}{2014}]{Zheng14} 
  Zheng, W., Li, W., Filippenko, A. V., \& Cenko, S. B. 2014, ATel, 5770, 1
  
\end{thebibliography}
\end{document}